\documentclass[aps,physrev]{revtex4-2}
\usepackage{tikz}
\newcommand*\circled[1]{\tikz[baseline=(char.base)]{
            \node[shape=circle,draw,inner sep=2pt] (char) {#1};}}
\usepackage[utf8]{inputenc}
\usepackage[T1]{fontenc}
\usepackage{amsmath}
\usepackage{amsfonts}
\usepackage{graphicx}
\usepackage{subcaption}
\usepackage{float} 
\usepackage{setspace}
\usepackage[justification=raggedright,singlelinecheck=false]{caption}
\usepackage{epstopdf}

\begin{document}


\title{\textbf{Phase-Space Synchronization Driven by Moon-Magnetosphere Coupling in Gas Giants}}

\author{Adnane Osmane}
 \email{Contact author: adnane.osmane@helsinki.fi}
\affiliation{%
Department of Physics, University of Helsinki, Helsinki, Finland
}%

\author{Elias Roussos}
\affiliation{Max Planck Institute for Solar System Research, Goettingen, Germany}
\author{Peter Kollmann}
\affiliation{Applied Physics Laboratory, Johns Hopkins University, Laurel, MD, USA
}

\date{\today}

\begin{abstract}
We present a new theoretical framework to describe the rapid and spatially localized loss of energetic particles in planetary radiation belts, focusing on interactions between gas giant magnetospheres and their moons. Observations show that flux depletions—known as microsignatures—often refill on timescales comparable to a single drift period, which conflicts with traditional quasi-linear radial diffusion models that assume slow, gradual transport and predict refilling only over many drift periods. To resolve this inconsistency, we develop a drift-kinetic model that explicitly captures localized losses occurring on timescales similar to the azimuthal drift period. We demonstrate that such localized loss regions can synchronize the azimuthal Fourier modes of the particle distribution function, producing apparent refilling through phase-space synchronization rather than diffusion. The resulting governing equations are mathematically equivalent to a generalized Kuramoto model, widely used to describe synchronization phenomena. This framework provides a first-principles, non-diffusive explanation for the evolution of microsignatures near moons, highlighting synchronization as a fundamental yet overlooked mechanism in magnetized plasma environments.
\end{abstract}


\maketitle
\section{Introduction}
Planetary radiation belts are natural laboratories for studying weakly collisional, magnetized plasmas far from thermodynamic equilibrium. In these environments, magnetically trapped electrons can be accelerated to energies up to 10~MeV at Earth and as high as 50~MeV at gas giants such as Jupiter and Saturn, primarily through wave--particle interactions. Because these belts are open systems that exchange energy and momentum with external and internal sources, their particle populations are continually maintained in non-equilibrium states. At Earth, the dominant energy input comes from the solar wind and associated particle injections, which drive kinetic instabilities such as whistler-mode waves~\cite{Thorne10}. In contrast, at gas giants, rapid planetary rotation and internal plasma sources---including moons and rings---act simultaneously as reservoirs and sinks of energetic particles, fundamentally shaping the belts' dynamics~\cite{Roederer}. The resulting energy distributions reflect an ongoing interplay between particle injection, acceleration, and localized losses.

One longstanding puzzle in this context is the \emph{absorption of energetic particles by moons}, which produces persistent flux depletions known as \emph{microsignatures} (for energetic electrons) and \emph{macrosignatures} (for protons and very energetic electrons above 10~MeV)~\cite{Thomsen1977, vanAllen1980Mimas, Roussos07, Andri2011, Kollmann2011jgr, Roussos2018D}. These signatures appear as sharp dropouts in measured particle intensities due to irreversible loss to moons or rings within the magnetosphere. Unlike a local plasma wake, microsignatures and macrosignatures persist far downstream of the absorber and evolve as they drift, so their properties---depth, width, and refilling rate---are valuable diagnostics of the surrounding magnetospheric environment. Traditionally, their gradual refilling has been interpreted using radial diffusion models that invoke quasi-linear theory to estimate transport rates from the observed flux profiles~\cite{1983JGR....88.8947C, 2007JGRA..112.6214R}. However, the observed rapid refilling of microsignatures---on timescales comparable to or shorter than a single drift period---challenges the validity of this standard diffusion framework and exposes a gap in our understanding of transport in weakly collisional, magnetized plasmas with localized loss regions~\cite{2010JGRA..115.3202R, 2012Icar..220..503A, 2014Icar..229...57A}.
 
The difference between microsignatures and macrosignatures lies primarily in their spatial topology \cite{1993AdSpR..13j.221S}. Microsignatures are features whose profile (depth, width, position) depends strongly on their “age,” or equivalently, their angular separation from the absorbing object. Older microsignatures tend to appear more eroded, shallower, and broadened until they eventually dissipate. Macrosignatures, on the other hand, arise as a cumulative by-product of microsignatures when the erosion timescale is longer than it takes for the signature to drift entirely around the planet. In this case, the microsignature may re-encounter the original absorber (or merge with a younger microsignature), causing its depth to intensify again before it can refill fully. Depending on the balance of the refilling rate against the microsignature rencounter rate with a moon, particle flux losses build-up, resulting in a partially of fully charged particle depleted region in the vicinity of the moon's $L$-shell that persists over time and spreads at all longitudes or local times. For this reason, the profile of a macrosignature is independent of the angular separation between the spacecraft and the absorber.
 
Microsignatures and macrosignatures have been observed at all outer planets (Jupiter, Saturn, Uranus, Neptune) \cite{1989JGR....9415077H,1991JGR....9619137S,1994JGR....9919441M, 2024GeoRL..5104685H}, but they have been most thoroughly studied at Saturn \cite{2014Icar..229...57A, 2017NatAs...1..872K, 2018GeoRL..4510912K}. This is due to Saturn’s spin-aligned dipole and the presence of many large moons within its radiation belts (e.g., Janus, Epimetheus, Mimas, Enceladus, Tethys, Dione, Rhea). These conditions favor frequent and long-lasting formation of microsignatures and macrosignatures \cite{2007JGRA..112.6214R, 2013Icar..226.1595R}. Over 100 microsignatures events and various macrosignatures have been analyzed, many resolved across wide energy range. Saturn's conditions are so favorable that even asteroid-sized moons of the planet such as Methone, Anthe, Pallene, Telesto, and Helene have produced detectable absorption signatures \cite{2012Icar..220..503A, 2016Icar..274..272R}. Microsignatures are usually more prominent in energetic electrons (above 1 keV), while macrosignatures are more readily formed by protons above the low MeV range. At ultrarelativistic energies ($>$5-10 MeV), electron macrosignatures have also been observed at Saturn \cite{1993AdSpR..13j.221S, 2021GeoRL..4892690Y}, since these electrons have drift velocities as fast as MeV protons, with short re-encounter times at Saturn's inner moons.
 
Figures 1 and 2 display representative examples of microsignatures and macrosignatures observed by the Cassini spacecraft. The data come from the Magnetospheric Imaging Instrument (MIMI), specifically its Low Energy Magnetospheric Measurement System (LEMMS), which measures electrons from approximately 18 keV to 10 MeV and protons from about 20 keV to several hundred MeV \cite{2004SSRv..114..233K}. Figure 1 shows electron microsignatures from the Saturnian moons Dione ($L$ = 6.28) and Tethys ($L$ = 4.89) in four distinct energy channels spanning the 18–100 keV range. During these events, Cassini was located 1.31 degrees (8655 km) and 28.5 degrees (146600 km) downstream from Dione and Tethys, respectively. In both cases, pronounced dropouts are seen, with depths that decrease at higher electron energies (C0 being the lowest energy channel, C3 the highest). This is mainly due to the different signal-to-noise ratio SNR) across the energy spectrum: the SNR is highest relative to instrumental background in the 18–40 keV range, even though other factors are known to account for the energy dependent depth variation, such as the width of the energy channel in which they are detected. In addition, for any of the two events, the timings of the microsignature centers show a clear energy dependence, most clearly visible in the Tethys example. Each of the two microsignature groups(ie. all energies) is also offset as a whole from the corresponding moon’s nominal L-shell. These two characteristics have been interpreted as evidence for non-corotational plasma flows in Saturn’s magnetosphere \cite{2013Icar..226.1595R}, regularly attributed to a (quasi) uniform, global electric field with an approximate noon-midnight orientation \cite{2010JGRA..115.9214P, 2012Icar..220..503A}, an inference supported by numerous other evidence, not necessarily based on microsignature observations \cite{2012JGRA..117.9208T, 2019GeoRL..46.3590R, 2020ApJ...905L..10H, 2021GeoRL..4891595K, 2021JGRA..12629600S}. The process responsible for the microsignature refilling causes microsignatures to appear narrower at lower count-rate levels, and wider at count-rates closer to the ambient values.
 
Figure 2 compares electron microsignatures and proton macrosignatures, highlighting their different behavior. The dataset consists of electron channels E4–E6 (above 800 keV) and proton channel P8 (above 25 MeV), collected during Cassini's periapsis pass on day 353 of 2016. This pass crossed the L-shells of Enceladus ($L\sim$3.95), Mimas ($L\sim$3.1), and the Janus-Epimetheus pair($L\sim$2.5). The plot covers both the inbound and outbound segments of the pass, which allows for a comparison of the signatures at different longitudes (or local times). The proton macrosignatures (yellow profile) appear as broad, persistent depletions centered on the moons’ L-shells. These signatures are identical on both the inbound and outbound parts of the orbit. In contrast, electron microsignatures show strong variability, reflecting the changing angular separation from the absorbing objects. Two prominent microsignatures from Janus and/or Epimetheus appear around 21:20 and 21:35. Additional events at 21:53, 22:25, and later are also visible, showing how feature-rich and variable microsignatures are.

\begin{figure}[H]
     \centering
     \begin{subfigure}{0.49\textwidth}
         \centering
         \includegraphics[width=\textwidth]{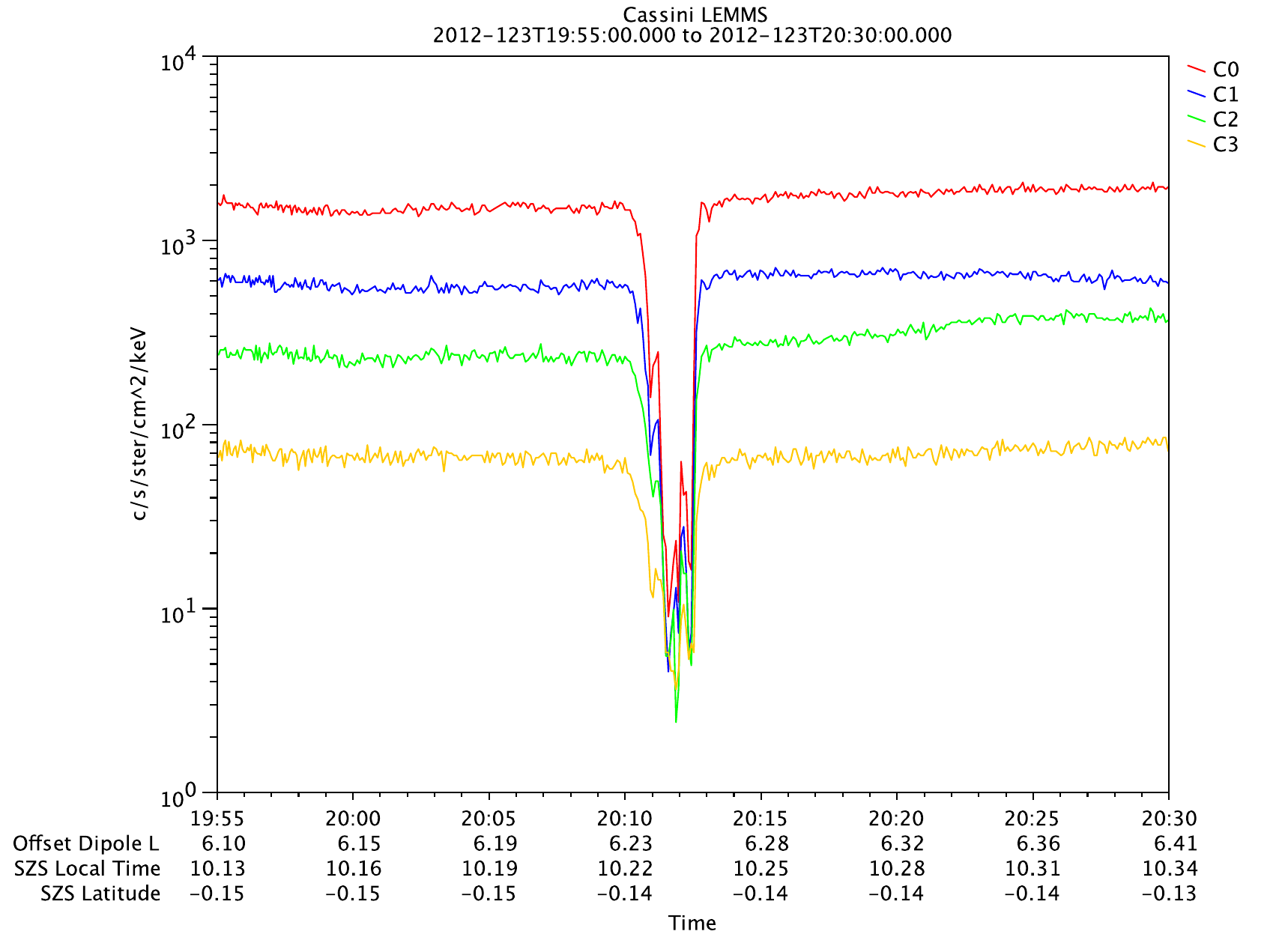}
     \end{subfigure}
     \hfill
     \begin{subfigure}{0.49\textwidth}
         \centering
         \includegraphics[width=\textwidth]{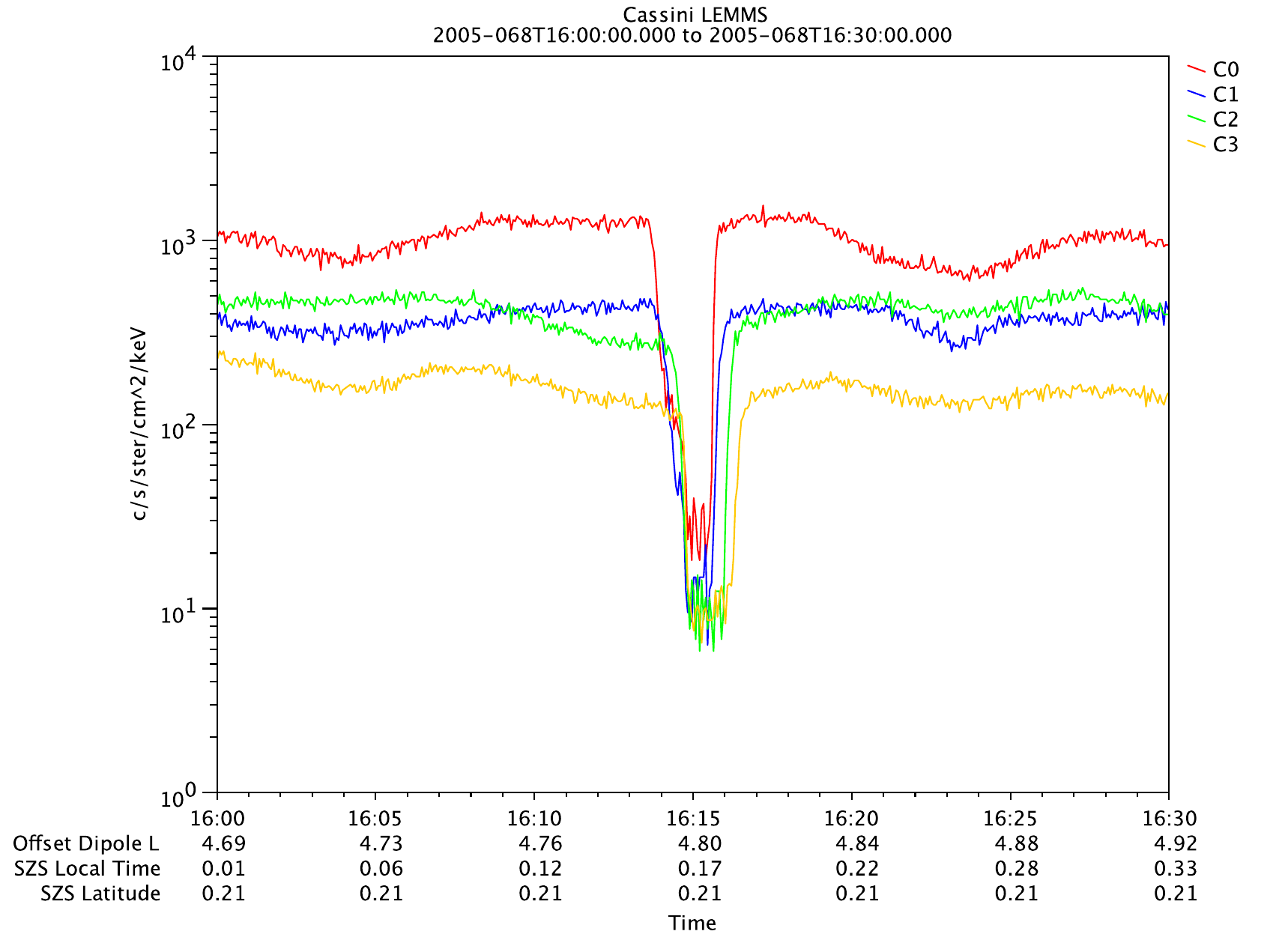}
     \end{subfigure}
     \caption{Two examples of microsignatures from Cassini's mission. On the left, a microsignature from Dione, and on the right from Tethys. Both figures show a smoothed shape due to the process of refilling with fluxes dropping by almost two orders of magnitude. The microsignatures also have an energy dependent position and an overall offset from their respective moon L-shell (4.89 for Tethys, 6.28 for Dione). The offset and energy dependent position is a geometric effect due to the convective (non-corotational) electric field drifts.}
     \label{Fig:microsignature}
\end{figure}

Over the past decades, both micro- and macro-signature features \cite{vanAllen1980Mimas, Roussos07} have been primarily interpreted within the framework of radial diffusion \cite{Falthammar65, Parker60, Elkington99, Lejosne20, osmane2023radial}, implicitly assuming the validity of the quasi-linear theory originally developed by \citet{Parker60} and \citet{Falthammar65}. As the azimuthal distance from the absorbing moon increases, the depleted region, in the case of microsignatures, can be observed to gradually refill. \citet{vanAllen1980Mimas} quantitatively described this refilling process through radial diffusion. Readers unfamiliar with the refilling of microsignatures in terms of radial diffusion can find a detailed calculation and discussion in the Appendix. There, we present the solution of \citet{vanAllen1980Mimas} for explaining microsignatures in terms of radial diffusion, as well as a generalization for cases where the absorption region is not negligible in size.

\begin{figure}
    \centering
    \includegraphics[width=0.75\linewidth]{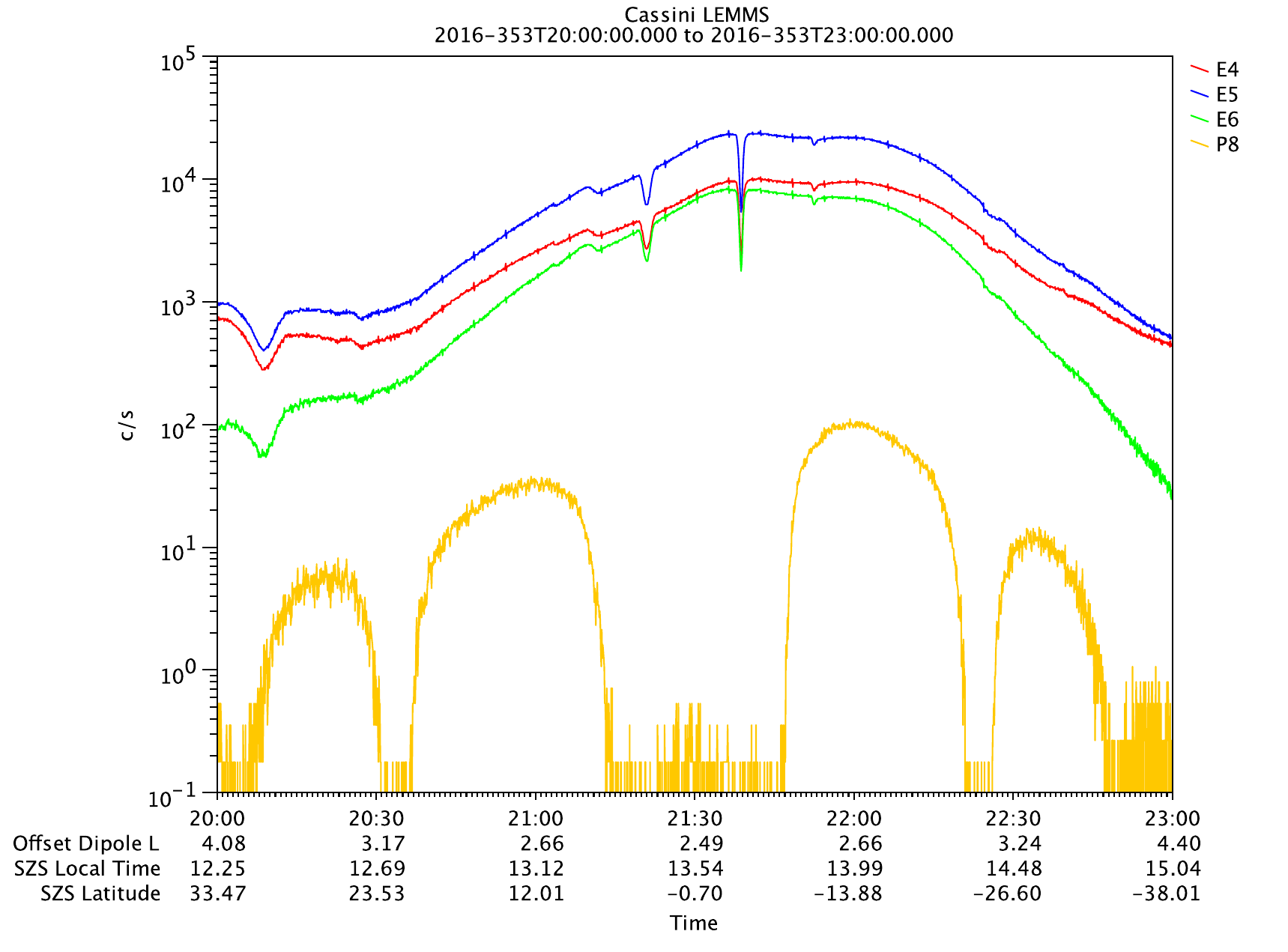}
    \caption{Example of micro- and macro-signatures in the stronger magnetic field regions of Saturn, seen in integral electron channels (>800 keV for E4, E5, $>$1.6 MeV for E6), embedded within the proton macrosignature (empty fluxes at all longitudes and local times) of Janus and Epimetheus (P8 proton channel $>$25 MeV in yellow).}
    \label{Fig:Macrosignature}
\end{figure}

This diffusion interpretation relies on a specific physical regime: radial diffusion is driven by electromagnetic fluctuations with spatial and temporal scales much larger than the particles’ Larmor radius and period, respectively. Since fluctuations at kinetic scales are generally too weak in rotation-driven magnetospheres of gas giants, radial transport driven by large-scale electromagnetic fields remains the most plausible mechanism. Furthermore, because microsignatures and their associated absorption primarily affect energetic electrons in the far tail of the distribution, the resulting depletions are unlikely to generate self-consistently significant electrostatic fluctuations at large spatial scales that could contribute to refilling the depletions. 

However, while radial diffusion provides a conceptually and quantitatively appealing framework for explaining the refilling of microsignatures, it has notable limitations that make its application to this phenomenon, at best, incomplete. Existing radial diffusion models, originally developed during the early space age \cite{Parker60, Falthammar65}, are grounded in quasi-linear theory. These quasi-linear models—whether describing radial transport or local wave-particle interactions—assume that particles experience a large number of small-amplitude perturbations over timescales much longer than the rapid timescales characteristic of their motion \cite{Kennel66, Schulz74, Eijnden, Kulsrud}. In the context of radial transport, the relevant fast timescale is the azimuthal drift period. As a result, observable signatures of radial diffusion are expected to appear only on timescales significantly longer than the drift period, since it takes time for a collection of particles to lose memory of the electromagnetic fields that perturb their radial positions.

It is important to emphasize that this requirement is not merely a theoretical inconsequential subtlety. \citet{osmane2023radial} presents a derivation of quasi-linear radial diffusion that closely mirrors the approach of \citet{Kennel66} for local wave–particle interactions. This work illustrates more clearly than \citet{Falthammar65} how the separation of timescales—between fast particle motion and slow diffusion—and the resulting loss of memory of the electromagnetic field are essential to the \textit{quasi-linear} radial diffusion process. Moreover, these assumptions are supported by numerical experiments that explicitly examine the relevant timescales for radial diffusion of magnetically trapped particles. For instance, early particle-tracing studies by \citet{Riley92} demonstrated that features consistent with radial diffusion only emerge on timescales much longer than the duration of geomagnetic storms at Earth, implying that radial diffusion for energetic electrons operates over thousands of drift periods. Further test-particle simulations by \citet{Sasha_poops_on_radial_diffusion}, examining 1 MeV electrons interacting with ULF waves generated by strong solar wind dynamic pressure in a global MHD code, found the resulting radial transport to be primarily non-diffusive and inconclusive in terms of supporting rapid diffusion.

More recently, \citet{Li_Mann24} showed that radial diffusion of phase-space density can indeed be recovered in the presence of a broad spectrum of ULF waves, but it still requires hundreds of drift periods for these effects to become manifest and consistent with radial diffusion. The findings of \citet{Li_Mann24} are consistent with theoretical requirements of quasi-linear radial diffusion, underscoring that observable diffusion signatures necessitate waiting several drift periods for particles to lose memory of their initial conditions. This fundamental feature of quasi-linear radial diffusion thus places significant constraints on the applicability of these models to the specific context of micro- and macro-signature refilling \footnote{While the numerical tests for radial diffusion discussed above have been primarily applied to Earth’s radiation belts, the fundamental assumptions underlying quasi-linear radial diffusion are not specific to any particular source of electromagnetic fluctuations. Whether these fluctuations are electromagnetic in nature and driven by solar wind interactions—as in Earth’s case—or sustained by electrostatic processes in the ionosphere, the key condition remains: transport must occur on timescales much longer than the drift period. The conclusions drawn by \citet{Sasha_poops_on_radial_diffusion} and \citet{Li_Mann24} extend beyond Earth’s environment and are equally applicable to radial diffusion processes at Jupiter, Saturn, or any planetary system where particles are magnetically trapped in a dipolar field and influenced by electromagnetic perturbations.}.

How can we reconcile observations showing that microsignature refilling occurs on timescales as short as a fraction of the drift period for energetic electrons with existing radial transport models? One possibility is that if the refilling is governed by radial diffusion, standard quasi-linear models cannot account for the rapid timescales observed, as they predict much slower evolution. This suggests the need for a new theoretical framework that moves beyond quasi-linear assumptions and accurately captures the rapid response of the distribution function to large-amplitude fluctuations. However, such a theory has not yet been developed, and addressing the limitations of quasi-linear models remains an open challenge in plasma physics, often leading to reliance on phenomenological approaches \cite{Dupree66, Dupree72,Diamond10, scheko2016}. Alternatively, the refilling of microsignatures may not be a diffusive process at all, and could instead result from inherently fast, non-diffusive mechanisms that remain to be fully understood and quantified. 

In this communication, rather than developing a radial transport equation that incorporates large-amplitude fluctuations and rapid diffusive processes, we use a drift kinetic formalism \cite{Hazeltine73, Hazeltine18, Parra} to construct a model for magnetically trapped particles interacting with localized sink regions such as moons. This approach considers processes on timescales comparable to a fraction of the azimuthal drift period. In our model, the absorption region is localized in magnetic local time (MLT) and can have any spatial extent. The framework is versatile and can be applied to different environments, including modeling particle losses at Earth resulting from magnetopause shadowing, as well as absorption by moons or rings in the radiation belts of gas giants. 

The report is organized as follows. In section \ref{Sec:methodology}, we introduce the drift kinetic model, which includes corotation drift, guiding center trajectories, and a loss term that operates on timescales comparable to the drift period. This loss term can be confined to a specific segment of the drift orbit in magnetic local time, which reflects the fact that moons typically occupy only a small region along the particle's trajectory. When losses are incorporated, the drift kinetic model can be reduced to a system of linearly coupled oscillators for the Fourier modes of the distribution function. In section \ref{Sec:Linear}, we demonstrate that this system is linearly unstable to localized sink regions with absorption rates comparable to the particle drift period. This finding is particularly relevant for the study of microsignatures, as it suggests that in gas giants, energetic electrons—which have longer azimuthal drift periods due to the co-rotation electric field—will enter the unstable regime, while energetic protons with shorter drift periods will not. 

In Section \ref{Sec:Linear}, we also present numerical solutions for the system of linearly coupled equations. Our results indicate that when the system is marginally or linearly stable, the various Fourier modes of the distribution function interact, preventing secular damping of the phase-space density. In other words, although the moons serve as particle sinks, the coupling and synchronization of these modes produce a phase-space density profile that displays apparent refilling. This apparent refilling arises from synchronization among the Fourier modes, not from a classical plasma instability, since the system remains dissipative and particles are not created. Furthermore, we show that the linearly coupled equations describing magnetic local time–localized absorption regions correspond to the Kuramoto model, which is a well-established framework for understanding synchronization in complex systems \cite{Acebron05}. We therefore provide a new explanation for microsignatures in energetic electrons that does not require radial diffusion, nor the presence of a large-amplitude electromagnetic fluctuations. In Section \ref{Sec:Conclusion}, we summarize our results as well limitations and future work.

\section{Methodological approach}\label{Sec:methodology}
\subsection{Kinetic radial transport without absorptions or losses}

We first proceed with a derivation of the guiding center drifts for the special case of an electrostatic poloidal electric field superposed onto a dipolar magnetic field, which serves as a good first approximation for gas giants. The background magnetic field is written as: 
\begin{equation}
\label{Magnetic_field_eq_1}
    \boldsymbol{B}=-\frac{B_P R_P^3}{r^3} {\boldsymbol{\hat{z}}},
\end{equation}
in terms of the planetary magnetic moment $B_P$ and the planetary radius $R_P$. We ignore the latitude dependence of the background magnetic field because we derive the radial transport equation for the special case of equatorially trapped particles, and thus the unit magnetic field vector $\boldsymbol{b}=\frac{\boldsymbol{B}}{|B|}=-{\boldsymbol{\hat{z}}}$ in our model points in the negative z direction for an orthogonal cylindrical system of coordinates written in terms of $(r, \varphi, z)$, where $\varphi$ is the azimuthal angle. The poloidal electric field due to ULF waves at Earth or dusk-dawn co-rotation field in gas giants is written as: 
\begin{eqnarray}
\label{electric_field_eq2}
\boldsymbol{E}&=&   {\delta E}_\varphi (r, \varphi, t) \boldsymbol{\hat{\varphi}} \nonumber \\
&=&\frac{R_P}{r}  \sum_m {\delta E}_{\varphi, m}e^{im\varphi}  {\boldsymbol{\hat{\varphi}}}, 
\end{eqnarray}
in terms of the time-dependent Fourier coefficients $\delta E_{\varphi, m} (t)$ with the $1/r$ dependence of the electric field coming from Faraday's equation for an electrostatic field that satisfies $\nabla \times \mathbf{E}=0$. The above electrostatic field representation in terms of Fourier modes can be simplified to a day–night or dawn–dusk asymmetric configuration, as appropriate for gas giants and Earth. However, it does not include the corotation electric field, which is introduced separately in the following section.

With the electric and magnetic field given by Equations (\ref{Magnetic_field_eq_1}) and (\ref{electric_field_eq2}), we can compute the guiding center drift velocities \cite[]{Hazeltine18}. Particles trapped in the equatorial plane experience two drifts: the $\mu \nabla B$ drift, for a particle of charge $q$, relativistic Lorentz factor $\gamma$ and first adiabatic invariant $\mu$: 
\begin{equation}\label{mugradBdrift}
    -\frac{\mu\nabla B\times \boldsymbol{b}}{q\gamma B} =\frac{3 \mu}{q\gamma r} \boldsymbol{\hat{\varphi}}
\end{equation}
and the $E$ cross $B$ drift: 
\begin{eqnarray}\label{EcrossBequation8}
\frac{\boldsymbol{E} \times \boldsymbol{b}}{B}&=&-\boldsymbol{\hat{r}}\sum_m \frac{\delta E_{\varphi, m}}{B_P} \frac{r^2}{R_P}e^{im\varphi}  
\end{eqnarray}
The drift velocity $\boldsymbol{v}_d$ can therefore be written as: 
\begin{equation}\label{drift_velocity}
    \boldsymbol{v}_d=\frac{3 \mu}{q\gamma r} \boldsymbol{\hat{\varphi}}
-\boldsymbol{\hat{r}} \sum_{m}\frac{\delta E_{\varphi, m}}{ B_P}\frac{r^2}{R_P} e^{im\varphi}
\end{equation}
In the absence of loss processes, the kinetic equation for equatorially trapped guiding centers in planetary radiation belts (see \cite{osmane2023radial} and references therein) can be expressed in conservative form as:
\begin{equation}\label{kinetic1}
  \frac{\partial}{\partial t}  (B f)+\nabla\cdot (\boldsymbol{v}_d Bf)=0.
\end{equation}
The magnetic field amplitude $B$ is included in Equation (\ref{kinetic1}) because it represents the Jacobian of the transformation when the particle’s velocity is expressed in field-aligned coordinates. The distribution function of guiding centers is denoted as  $f=f(r, \varphi, t)$. Utilizing Liouville’s theorem
\begin{equation}
     \frac{\partial}{\partial t}  (B)+\nabla\cdot (\boldsymbol{v}_d B)=0,
\end{equation}
the evolution of the distribution function is given by:
\begin{equation}\label{kinetic1_}
  B\frac{\partial f}{\partial t}+B\boldsymbol{v}_d \cdot\nabla f=0,
\end{equation}
or more explicitly for the drift velocity in Equation (\ref{drift_velocity}) by:
\begin{equation}\label{kinetic_2}
\frac{\partial f}{\partial t} +\frac{3 \mu}{q\gamma r^2} \frac{\partial f}{\partial \varphi}-\sum_m \frac{\delta E_{\varphi, m}}{B_P}\frac{r^2}{R_E^2}e^{im\varphi} \frac{\partial f}{\partial r}=0, 
\end{equation}
We note that the azimuthal drift frequency $\Omega_d=3\mu /q\gamma r^2$ on the left-hand side of Equation (\ref{kinetic_2}) provides the drift along magnetic local time and the last term of Equation (\ref{kinetic_2}) is responsible for the radial transport. Unlike standard radial diffusion treatment, the above equation accounts for azimuthal dependence of the distribution function. The right-hand side of equation (\ref{kinetic_2}) is set to zero in the absence of losses but in the next section we add a term to account for spatially localized losses. 

\subsection{Accounting for the corotation drift}
In gas giants such as Jupiter or Saturn, the rapid planetary rotation induces a strong corotation electric field in the inertial frame, which must be included when computing the guiding center motion of charged particles. This electric field can be determined from Faraday's equation and by the requirement that in the rotating planetary frame, the plasma is sufficiently conductive to render a zero electric field. Thus, by a Lorentz transformation of the electric field, a corotation electric field in a magnetic dipole field given 
\begin{equation}
    \boldsymbol{E}=-\boldsymbol{v_\Omega}\times\boldsymbol{B}
\end{equation}
in terms of the angular velocity $\boldsymbol{v}=\omega_z \boldsymbol{r}$, and the corotation frequency $\omega_z$, must be sustained \footnote{In the case of Saturn, corotation frequency scales as $\omega_z\simeq -2.1 \times 10^{-5} L E$ rad/s, where $L$ is the drift shell and $E$ is the relativistic kinetic energy.}. This corotation electric field, results in an additional $E\times B$ azimuthal drift given by: 
\begin{equation}
    \frac{\boldsymbol{E}\times \boldsymbol{b}}{B}=\omega_z r \boldsymbol{\hat{\varphi}}
\end{equation}
The net drift velocity $\boldsymbol{v}_d$ in Equation (\ref{drift_velocity}) now contains an additional azimuthal component: 
\begin{equation}\label{drift_velocity2}
    \boldsymbol{v}_d=\frac{3 \mu}{q\gamma r} \boldsymbol{\hat{\varphi}}+\omega_r r\boldsymbol{\hat{\varphi}}
-\boldsymbol{\hat{r}} \sum_{m}\frac{\delta E_{\varphi, m}}{ B_P}\frac{r^2}{R_P} e^{im\varphi}
\end{equation}
and the kinetic equation is now given by: 
\begin{equation}\label{kinetic_2_}
\frac{\partial f}{\partial t} +\left(\frac{3 \mu}{q\gamma r^2}+\omega_r\right) \frac{\partial f}{\partial \varphi}-\sum_m \frac{\delta E_{\varphi, m}}{B_P}\frac{r^2}{R_E^2}e^{im\varphi} \frac{\partial f}{\partial r}=0, 
\end{equation}
When the corotational electric field is taken into account, the net azimuthal drift frequency decreases for negatively charged particles ($q < 0$) and increases for positively charged ones ($q > 0$). As a result, electrons and protons with equal magnetic moments experience fundamentally different drift periods in the environment of a rapidly rotating planet, making them susceptible to distinct radial transport mechanisms.  In the following analysis, we demonstrate how this charge-dependent drift asymmetry can play a crucial role in shaping the dynamical evolution of particle distribution functions—and in interpreting microsignatures. However, Equation (\ref{kinetic_2_}) still does not account for localized losses, which is discussed and added in the next section.

\subsection{Kinetic radial transport with MLT localized losses}
Our aim is to decompose the drift kinetic equation for the guiding centers into Fourier modes. It is therefore convenient, when incorporating a loss or source term, to express it in terms of a Fourier representations. In this communication we choose to represent MLT localised losses in terms of the von Mises distribution $W(\varphi, \kappa)$, which is defined as:
\begin{eqnarray}\label{Eq:vonMises}
    W(\varphi; \kappa)&=&\frac{e^{\kappa \cos(\varphi)}}{2\pi I_0(\kappa)}, \nonumber\\
&=&\frac{1}{2\pi}\left(1+2\sum_{n=1}^\infty \psi_n(\kappa) \cos(n\varphi)\right)   
\end{eqnarray}
 where the parameter $\kappa \in [0, \infty )$ can be shown to quantify the spread in MLT of the loss region, and the coefficient $\psi_n(\kappa)$ is the ratio of the modified Bessel function of the first kind with order $n$, $I_n(\kappa)$ divided by the modified Bessel function of the first kind with degree zero, $I_0(\kappa)$ \citep{abramowitz1968handbook}. Hence,  $\psi_n(\kappa)= I_n/I_0$ and is always less than unity for $\kappa>0$. The von Mises distribution is the analogue to the Maxwellian distribution on a circle. Written in the above form, it is also normalised as can be shown when integrated along $\varphi$ between $-\pi$ and $\pi$.

 The use of the von Mises distribution allows us to confine losses to any range of magnetic local times we deem appropriate for a given problem. The range of MLT localisation is parameterised by the coefficient $\kappa$ and has a simple interpretation when $\kappa \gg 1$. If $\kappa=0$, the loss region is uniformly spread across all magnetic local times. But as $\kappa$ becomes much greater than one, the von Mises distribution converges to a Maxwellian: 
\begin{equation}
   \lim_{\kappa \longrightarrow \infty} f(\varphi;\kappa) = \sqrt{\frac{\kappa}{2\pi}} e^{-\kappa \varphi ^2/2}. 
\end{equation}   
Thus, the parameter $\kappa$ is equivalent to the inverse of the variance of a Maxwellian, and the von Mises distribution confines losses around $\varphi= \pm 2 \sqrt{1/\kappa}$. For very large values of $\kappa$, the model can be used to quantify losses at the moons, whereas values of $\kappa\simeq 1$ can be used for magnetopause shadowing or perhaps grazing of planetary rings. Now the next step is to use the von Mises distribution to represent localised losses in terms of a Fourier representation of the form:
\begin{equation}
    \label{EQ:Fourier_decomposition}
    \nu=\sum_m c_m e^{i m\varphi}
\end{equation}
with $\nu=\nu(\varphi, L, t)$ a loss frequency that can be made dependent on the magnetic local time $\varphi$, the normalised radial distance $L$, and time. In this report, since we are introducing this model for the first time, we are not going into the details of the loss at the moons, which affect energies, phases and pitch-angle differently. Instead, we choose a generic treatment, that can be generalised for various pitch-angles or energies, and which can be used for moons in gas giants or for magnetopause shadowing at Earth. The loss term can then be inserted in the drift kinetic equation as: 
\begin{equation}
    \frac{d f}{dt}= -\nu (f-f_0)
\end{equation}
where $f_0$ is the slow part of the distribution function which can be extracted by the following averaging procedure: 
\begin{equation}\label{averaging_procedure}
    f_0(r, t)=\langle f \rangle_\varphi=\frac{1}{2\pi T} \int_{-\pi}^{\pi} \mathrm{d}\varphi \int_0^T \mathrm{d}t \hspace{1mm}f(r, \varphi, t),
\end{equation}
where $T$ is the drift period timescale. It should be emphasized that $f_0$ is not some initial value of the distribution function, before the magnetically trapped particles encounter the loss region, but a drift average value. In short, we want to quantify losses that are MLT localised, and thus, if the losses are significant on timescales comparable to several drift periods (or less), than it should be visible within the MLT dependent part of the distribution function $f-f_0$. 

\subsection{Fourier representation of MLT localised losses}
We seek a Fourier representation to MLT localised losses that can be written in terms of the von Mises distribution as: 
\begin{equation}
   \nu=\beta \ W(\kappa, \varphi) 
\end{equation}
where $\beta$ is a parameter that is independent of the magnetic local time $\varphi$ and can be made dependent on the radial distance, the magnetic latitude or even the particle's properties, i.e., the pitch-angle or adiabatic invariant. In order to find the find the Fourier coefficients $c_m$ in Equation (\ref{EQ:Fourier_decomposition}) we simply need to solve the following integral: 
\begin{eqnarray}
   c_m&=& \frac{1}{2\pi} \int_{-\pi}^{\pi} \mathrm{d}\varphi \  \nu e^{-im\varphi} \nonumber \\
   &=&\frac{1}{2\pi} \int_{-\pi}^{\pi} \mathrm{d}\varphi \ \beta \ W(\kappa, \varphi)  e^{-im\varphi}
\end{eqnarray}
which gives the following Fourier coefficients for a fixed azimuthal number $m$:
\begin{eqnarray}
   c_m&=& \frac{\beta}{2\pi} \left(\delta_{m0}+2\psi_m\right),  
\end{eqnarray}
in terms of the Kronecker delta function $\delta_{ij}$. The Fourier representation of MLT localised losses can therefore be written as: 
\begin{eqnarray}
    \label{EQ:Fourier_coefficient_Bessel}
    \nu&=&\frac{\beta}{2\pi} \sum_m \left(\delta_{m0}+2\psi_m\right)e^{im\varphi} \nonumber \\ 
    &=&\frac{\beta}{2\pi}\left(1+2\sum_m \psi_me^{im\varphi}\right).
\end{eqnarray}
Thus, the Fourier coefficients for $m>0$ are the ratio of the modified Bessel function of degree $n$ and $0$ and are function of the parameter $\kappa$. If the loss source is higly localised in MLT, than $\kappa$ is large and one needs a large number of Fourier coefficients, but if $\kappa$ is small, a few Fourier coefficient might suffice. 

Even though the von Mises function is a normalised distribution, the loss term peak grows by a factor of 10 between $\kappa=0$ and $\kappa=30$. We therefore need to ensure that trapped particles sample loss sources of comparable magnitude over a total drift orbit. We therefore rescale the loss term in a manner that is not affected by the $\kappa$ parameter. In order to do so, we compute the root mean square value $\sqrt{\langle \nu^2\rangle}$, where the brackets denote the drift average, and add the additional requirement that a particle samples the same root mean square fluctuation on a drift orbit. This procedure was also used in \citet{Osmane25} to ensure that the MLT ULF localised wave power was independent of the parameter $\kappa$. We find that the root mean square value $ \langle \nu^2\rangle$ needs to be rescaled according to: 
\begin{eqnarray}
\beta = \frac{2\pi \sqrt{\langle \nu^2\rangle}}{\left(1+4\sum_m \psi_m^2\right)^{1/2}},
\end{eqnarray}
and thus, we write the Fourier representation of the loss term as: 
\begin{eqnarray}
    \label{EQ:Final_form_loss_term}
    \nu(\kappa)&=&\frac{\sqrt{\langle \nu^2 \rangle}}{\sqrt{1+4\sum_m \psi_m^2}} \left({1+2\sum_m \psi_m e^{im\varphi}}\right) \nonumber \\
    &=&\sigma(\kappa)\left({1+2\sum_m \psi_m e^{im\varphi}}\right)
\end{eqnarray}
where in the last line of Equation (\ref{EQ:Final_form_loss_term}) we wrote the normalised root-mean square fluctuation as $\sigma=\frac{\sqrt{\langle \nu^2 \rangle}}{\sqrt{1+4\sum_m \psi_m^2}}$. The loss term is therefore a function of two parameters: 
\begin{enumerate}
    \item[1.] $\langle \nu \rangle^{1/2}$ which quantifies the root mean square loss rate over a drift orbit. 
    \item[2.] $\kappa$ which quantifies the MLT localization of our loss region. 
\end{enumerate}
After inserting Equation (\ref{EQ:Final_form_loss_term}) into the right-hand side of the kinetic equation (\ref{kinetic_2_}) we find : 
\begin{equation}\label{kinetic_with_losses}
\frac{\partial f}{\partial t} +\left(\frac{3 \mu}{q\gamma r^2}+\omega_r\right) \frac{\partial f}{\partial \varphi}-\sum_m \frac{\delta E_{\varphi, m}}{B_P}\frac{r^2}{R_E^2}e^{im\varphi} \frac{\partial f}{\partial r}=-\sigma(\kappa)\left({1+2\sum_m \psi_m e^{im\varphi}}\right) (f-f_0)
\end{equation}
If we now apply a Fourier decomposition of the distribution function along the magnetic local time to Equation (\ref{kinetic_with_losses}), i.e., 
\begin{eqnarray}\label{decomposition_f}
    f(r, \varphi, t)&=&f_0(r, t) + \delta f (r,\varphi, t) \nonumber  \\ 
        &=& f_0(r, t) + \sum_m \delta f_m(r, t) e^{im\varphi}, 
\end{eqnarray}
we find the following equation for the drift average distribution function $f_0$:
\begin{eqnarray}
    \label{EQ:quasi_linear_with_loss}
  \boxed{\frac{\partial f_0}{\partial t} =\sum_m L^2 \frac{\partial}{\partial L} \bigg{\langle} \frac{\delta E_{\varphi, m}^* \delta f_m}{B_p R_p} \bigg{\rangle}-2\sigma \sum_m \psi_m \langle \delta f_m \rangle.} 
\end{eqnarray}
Equation (\ref{EQ:quasi_linear_with_loss}) is, to the best of our knowledge, the first quasilinear equation that incorporates losses which can occur on timescales comparable to the drift period. The first term on the right-hand side of Equation (\ref{EQ:quasi_linear_with_loss}) is responsible for radial diffusion, and the second term accounts for irreversible losses on long timescales much larger than the drift period of the guiding centres \footnote{A priori $\langle \delta f_m \rangle$ would be expected to be negligeable, but when one accounts for the evolution of the distribution function due to MLT localised losses detailed in the remaining section, or for higher order effects, as discussed in \citet{osmane2023radial}, this additional term can result in a net loss term. If we solve $\delta f_m$ linearly, due to the interaction with electromagnetic fluctuations, the resulting loss term will be proportional to $\frac{\partial f_0}{\partial L}$ times the drift averaged expression for the electromagnetic guiding center radial drift velocity. Thus, the loss term that enters into this quasilinear generalisation for MLT localized losses behaves like a drift transport coefficients that grows with large gradients in the distribution functions. This effect would be null in the inner magnetosphere of the Earth, far from the boundaries, but could prove significant near the magnetopause and result in enhanced losses, even in the absence of a large value of $D_{LL}$.}. 

In order to solve full the distribution function we also need an evolution equation for each perturbed Fourier modes $\delta f_m$. Following the same procedure as described in Appendix B of \citet{osmane2023radial}, we find: 
\begin{eqnarray}
\label{EQ:Perturbed_part}
   \boxed{\frac{\partial \delta f_m}{\partial t}+\underbrace{im\Omega_d \delta f_m}_{\circled{1}}+\underbrace{\sigma \delta f_m}_{\circled{2}}=\underbrace{\frac{\delta E_{\varphi, m}}{B_p R_p}L^2\frac{\partial f_0}{\partial L}}_{\circled{3}}+\underbrace{\sum_{m'}\frac{\delta E_{\varphi, m'-m}}{B_p R_p}L^2\frac{\partial \delta f_{m'}}{\partial L}}_{\circled{4}}-\underbrace{2\sigma\sum_{m'\neq m}\psi_{m'-m} \delta f_{m'}}_{\circled{5}}.} \nonumber
\end{eqnarray}
where the drift frequncy $\Omega_d$, now accounts for the corotation drift, i.e., $\Omega_d=\frac{3\mu}{q\gamma r^2}+\omega_r$. Thus, the rate of change of the perturbed part coefficient $\delta f_m$ is a function of ballistic motion responsible for drift echoes ${\circled{1}}$, the MLT localised damping of the mode $m$ ${\circled{2}}$, linear wave-particle interaction between the ULF fluctuations and the particles ${\circled{3}}$, higher order wave-particle interactions ${\circled{4}}$ and a linear coupling between modes $m$ and $m'$ arising due to localized losses ${\circled{5}}$. We note that if the loss term is MLT independent and thus $\kappa=0$, different Fourier modes of the perturbed distribution function are uncoupled and will produce homogeneous solutions of the form:
\begin{equation}
    \delta f_m(r, t)=\delta f_m(r, t=0) e^{-im\Omega_d t-\sigma t} \nonumber
\end{equation}
 However, if $\kappa \neq 0$, even if higher order effects quantified by term  ${\circled{4}}$ are negligeable, the different Fourier modes $m$ and $m'$ will be linearly coupled to one another, with a coupling constant given by $\psi_{m'-m}$.

In the following, we will ignore higher order effects due to linear wave-particle interactions, assume that the background distribution is time independent\footnote{We also focus hereafter on changes in the distribution function that are taking place on the order of a few drift periods. And if the quasi-linear equation given by Equation (\ref{EQ:quasi_linear_with_loss}) holds at all, it is expected to keep $f_0$ constant on timescales comparable to the drift period.}, and focus instead on the effect of MLT localised losses on the distribution function. We rewrite Equation (\ref{EQ:Perturbed_part}) in terms of the following rescaled quantities: 
\begin{equation}
   \boxed{a_m=\frac{\delta f_m}{f_0}; \hspace{3mm} t\Rightarrow \langle \nu^2\rangle^{1/2}t;  \hspace{3mm}\omega_m =\frac{m\Omega_d}{\langle \nu^2\rangle^{1/2}}\hspace{3mm} \sigma \Rightarrow  \sigma/\langle \nu^2\rangle^{1/2};  \hspace{3mm}  \eta_m=\frac{\delta E_{\varphi, m}}{B_p R_p}\frac{L^2}{\langle \nu^2\rangle^{1/2}}\frac{\partial \log f_0}{\partial L};} 
\end{equation}
where \( a_m \) is the normalized Fourier mode \( m \) of the distribution function that responds to the moon-induced losses. Time and the drift frequency are both normalized by the root mean square of the loss term. The driving term \( \eta_m \) is nonzero whenever the drift-averaged distribution function has a gradient and the electric field contains a nonzero Fourier component at mode \( m \). The more compact form for the evolution of the distribution function on fast timescales is therefore given by
\begin{equation}
   \label{EQ:oscillators}
   \boxed{\frac{\partial a_m}{\partial t}+(i\omega_m+\sigma)a_m=\eta_m-2\sigma\sum_{m'\neq m} \psi_{m'-m} a_{m'}}
\end{equation}
Equation (\ref{EQ:oscillators}) indicate that the Fourier modes of the perturbed distribution function are linearly coupled through one another. Note that the MLT dependence of the loss region is contained within the parameter $\sigma$ and the coupling terms $\psi_{m'-m}$. The more narrow the loss regions, i.e. the higher the value of $\kappa$, the more terms one needs to keep in the sum on the right-hand side of Equation (\ref{EQ:oscillators}). Based on the above equation we expect the mode numbers to oscillate with frequency $\omega_m$ and decay on timescales of $\sigma t\simeq 1$ in the absence of driving ($\eta_m(t)=0$) and MLT localised loss sources ($\kappa=0$). This behavior of the distribution function is known as drift echoes, and responsible for the appearance of zebra stripes in planetary radiation belts (see, e.g., \cite{osmane2023radial} and references therein). However, let's determine if the decaying of drift echoes is altered by the presence of driving due to wave-particle interactions or MLT localised sources. 

\section{Linear solutions and stability analysis}\label{Sec:Linear}
In this section, we look at solutions to the linearly coupled system of Equation (\ref{EQ:oscillators}) for a fixed radial distance or drift shell. Equation (\ref{EQ:oscillators}) reduces to a set of linearly coupled ordinary differential equations for fixed radial distances:

\begin{equation}
   \label{EQ:oscillators2}
   {\frac{d a_m}{dt}+(i\omega_m+\sigma)a_m=\eta_m-2\sigma\sum_{m'\neq m} \psi_{m'-m} a_{m'}}
\end{equation}
which can be solved exactly in driven ($\eta_m\neq 0$) and undriven case ($\eta_m = 0$) for any value of the parameter $\kappa$ as a linear algebra problem \footnote{It should be emphasised that the driving term $\eta_m$ is non-zero for non-radial electric fields that results in radial $E\times B$ drifts. But the driving can be treated as stochastic, as commonly done for radial diffusion derivations \cite{Falkovich, osmane2023radial} or as deterministic, when looking for coherent responses to narrow band ULF fluctuations.}. We therefore consider the vector \( \boldsymbol{A}(t) = [a_1(t), a_2(t), \dots, a_N(t)]^T \), of each individual Fourier modes \( m \in \mathbb{Z} \)  such that $1 \leq m \leq N$, and write the coupled system:
\begin{eqnarray}
\label{EQ:oscillators3}
\frac{d\boldsymbol{A}}{dt} &=& -(\boldsymbol{\Omega} + 2\sigma \Psi)\boldsymbol{A} + \boldsymbol{\eta} \\
    &=& \mathbf{M} \boldsymbol{A} + \boldsymbol{\eta}
\end{eqnarray}

with matrices $\boldsymbol{\Omega}$, $\boldsymbol{\Psi}$ and $\mathbf{M}$ defined as:
\[
\boldsymbol{\Omega} =
\begin{pmatrix}
i\omega_1 + \sigma & 0 & \cdots & 0 \\
0 & i\omega_2 + \sigma & \cdots & 0 \\
\vdots & \vdots & \ddots & \vdots \\
0 & 0 & \cdots & i\omega_N + \sigma
\end{pmatrix}
\]

\[
\Psi =
\begin{pmatrix}
0 & \psi_{2-1} & \psi_{3-1} & \cdots & \psi_{N-1} \\
\psi_{1-2} & 0 & \psi_{3-2} & \cdots & \psi_{N-2} \\
\psi_{1-3} & \psi_{2-3} & 0 & \cdots & \psi_{N-3} \\
\vdots & \vdots & \vdots & \ddots & \vdots \\
\psi_{1-N} & \psi_{2-N} & \psi_{3-N} & \cdots & 0
\end{pmatrix}
\]

\begin{eqnarray}
    \mathbf{M} = -\boldsymbol{\Omega} - 2\sigma \Psi, \nonumber
\end{eqnarray}

and the forcing vector given by:
\[
\boldsymbol{\eta} =
\begin{pmatrix}
\eta_1 \\
\eta_2 \\
\vdots \\
\eta_N
\end{pmatrix}.
\]

Putting it all together, for example for $N=4$, the system of equation can be written as:
\[
\frac{d}{dt}
\begin{pmatrix}
a_1 \\
a_2 \\
a_3 \\
a_4
\end{pmatrix}
=
- 
\begin{pmatrix}
i\omega_1 + \sigma & 2\sigma \psi_{1} & 2\sigma \psi_{2} & 2\sigma \psi_{3} \\
2\sigma \psi_{-1} & i\omega_2 + \sigma & 2\sigma \psi_{1} & 2\sigma \psi_{2} \\
2\sigma \psi_{-2} & 2\sigma \psi_{-1} & i\omega_3 + \sigma & 2\sigma \psi_{1} \\
2\sigma \psi_{-3} & 2\sigma \psi_{-2} & 2\sigma \psi_{-1} & i\omega_4 + \sigma
\end{pmatrix}
\begin{pmatrix}
a_1 \\
a_2 \\
a_3 \\
a_4
\end{pmatrix}
+
\begin{pmatrix}
\eta_1 \\
\eta_2 \\
\eta_3 \\
\eta_4
\end{pmatrix}
\]

With specified initial conditions \( \boldsymbol{A}(0)= [a_1(0), a_20t), \dots, a_N(0)]^T \) the general solution for the Fourier coefficients $a_m$ is:
\[
\boldsymbol{A}(t) = e^{\mathbf{M} t} \boldsymbol{A}(0) + \int_0^t e^{\mathbf{M}(t - \tau)} \boldsymbol{\eta} \, d\tau
\]

\begin{figure}[H]
    \centering
    \includegraphics[width=0.5\linewidth]{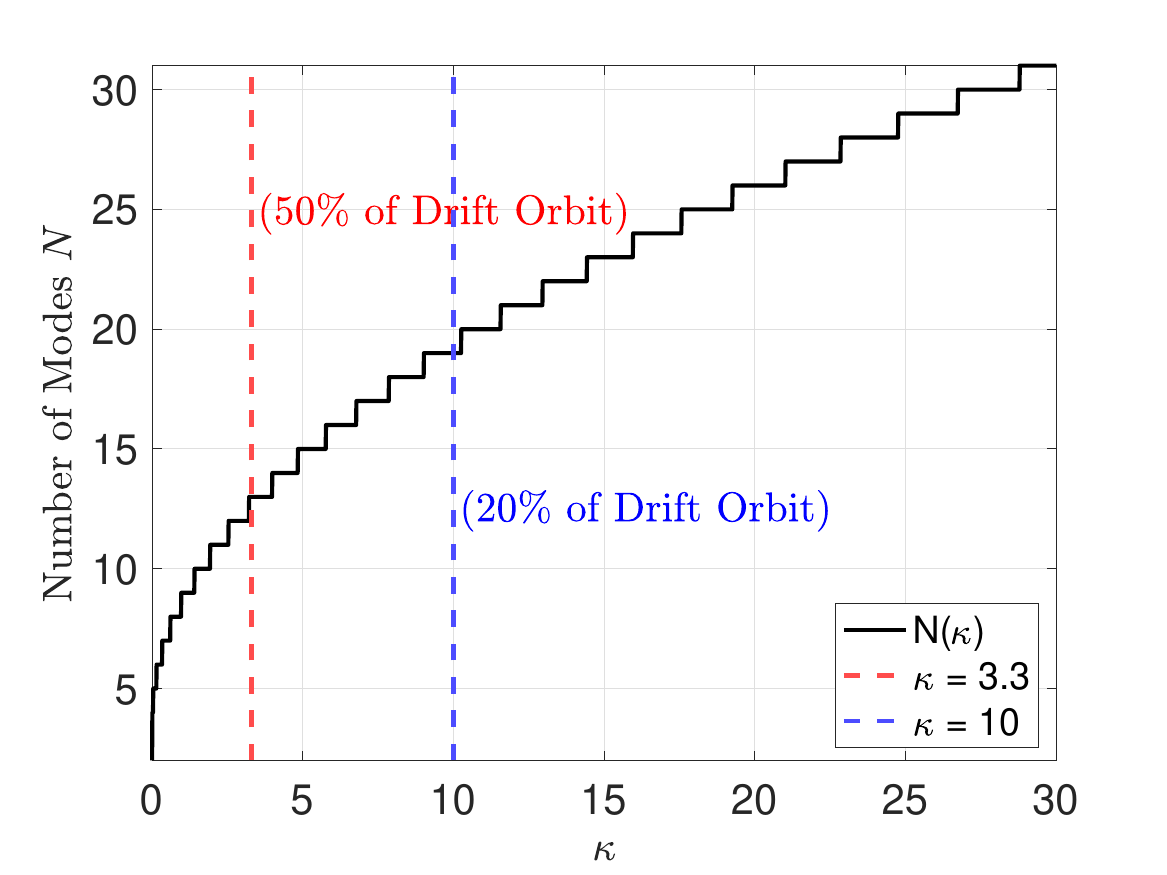}
    \caption{Number of modes $N$ required as a function of coupling parameter $\kappa$ for a threshold $\varepsilon = 10^{-6}$. The values of $\kappa$ corresponding to 50\% and 20\% of the drift orbit are shown in dashed vertical lines.}
    \label{fig:0}
\end{figure}

If $\boldsymbol{\eta}$ is constant (time independent) and $\mathbf{M}$ is diagonalizable, this integral can be computed analytically. This representation of the problem using linear algebra helps for the case where absorption region is very localized, and thus when many Fourier modes $a_m$ need to be computed. But before solving the set of linear equations, we must determine the number of terms we need to keep in $\boldsymbol{A}$ and thus the size of the $N\times N$ matrix $\mathbf{M}$. We note that the coefficients $\psi_{m-m'}$ are always less than 1 and that for large values of $|m-m'|$, $\psi_{m-m'}\rightarrow 0$. However, the exact number of terms needed is a function of the parameter $\kappa$. The more MLT localized the loss terms, the more the number of modes $a_m$ we need to compute. So in order to address this, we determine a cut-off threshold such that $\psi_{m_{\text{max}}}<10^{-6}$ to determine the number of $m$ terms needed for a given value of $\kappa$. Once we determine the value of $m_{\text{max}}$ for a given $\kappa$ that satisfies the cutoff threshold, we set $N=m_{\text{max}}+1$. The result is shown in Figure (\ref{fig:0}) for values of $\kappa$ ranging between 0 and 30, with marks showing the value of $\kappa$ for which 50\% and 20\% of the drift orbit are occupied by the loss region. Figure (\ref{fig:0}), shows that we typically need less than 30 terms in order to account for the couplings between a given mode. 

\subsection{Stability analysis}\label{section:stability}
The system of Equation (\ref{EQ:oscillators3}) can be solved exactly in terms of a linear combination of modes of the form: 
\begin{equation}
    \boldsymbol{A}(t)=\sum_k \alpha_k(t=0) \boldsymbol{p}_k e^{-\lambda_k t}
\end{equation}
where $\alpha_k(t=0)$ depends on the initial condition, and $\boldsymbol{p}_k$ are the eigenvectors corresponding to the eigenvalues $\lambda_k$ of the matrix $\mathbf{M}$. Thus, for a value of $\kappa$, we can determine the exact solutions for the Fourier modes $\delta f_m(L,t)$ of MLT dependent distribution function by computing the eigenvalues and eigenvectors of the matrix $\mathbf{M}$.  

\begin{figure}
  \includegraphics[width=0.75\textwidth]{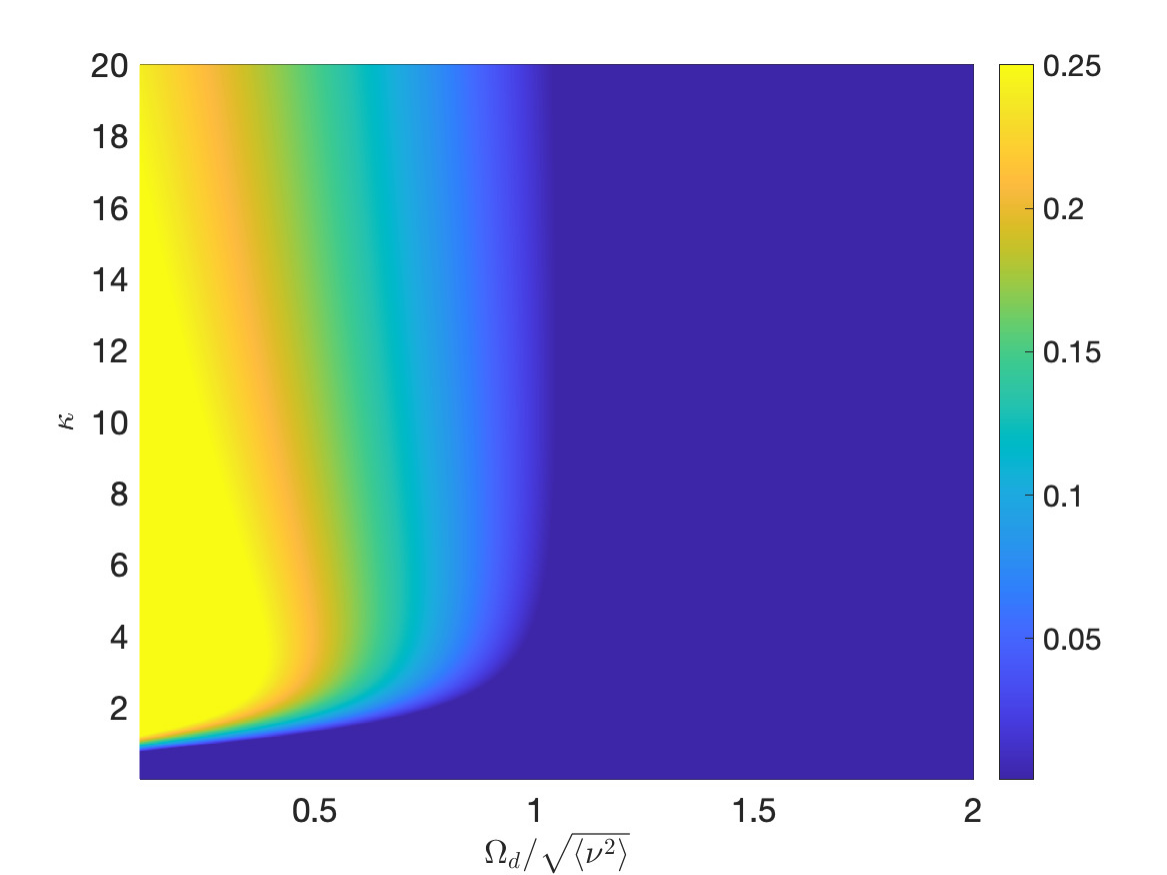}
  \caption{Dependence of the fastest growing linear mode as a function of the $\kappa$ parameter and the ratio of the drift frequency $\Omega_d$ to the root-mean-square loss rate $\sqrt{\langle \nu^2 \rangle}$. The color bar shows the transition from stable (dark blue), marginally stable (light blue) to unstable (green and yellow) eigenvalues.}
\label{fig:ka}
\end{figure}

In Figure (\ref{fig:ka}), we show the dependence of the fastest growing eigenvalue, or in the case where all eigenvalues are negative, the slowest damping eigenvalue, as a function  of the $\kappa$ parameter and the ratio of the drift frequency $\Omega_d$ to the root-mean-square loss rate $\sqrt{\langle \nu^2 \rangle}$. The most interesting observation from Figure (\ref{fig:ka}) is that when the effective drift frequency becomes comparable or less than the loss rate $\sqrt{\langle \nu^2 \rangle}$, and with $\kappa\geq 1$ (spatially localised loss region), that the system can become unstable. And thus, even though the system is dissipative, localised losses in MLT can result in some azimuthal wave numbers $m$ to grow at the expense of others. Physically, the requirement for the instability to be triggered for $\Omega_d/\sqrt{\langle \nu^2 \rangle}<1$ means that the effective drift period must be comparable or longer than the root mean square loss rate. This regime is consistent with the microsignature case, as observations sometimes demonstrate that fluxes are almost completely depleted following interaction with moon—indicating that the loss timescale is therefore much shorter than the drift period. In short, losses must be sufficiently fast when compared to the effective drift periods, for some modes to grow. However, the Fourier modes $a_m$ cannot grow indefinitely, as they are constrained by the requirement to conserve the total number of particles along a drift shell. In this context, stability means that localized losses can sustain the growth of Fourier modes at various azimuthal wave numbers $m$, but only to a limited extent. This is mathematically analogous to linearly unstable plasma systems, where specific configurations allow for the growth of electromagnetic field fluctuations. Yet such growth must eventually saturate—either because the energy source is depleted or due to nonlinear processes (e.g., wave–wave interactions) that act as sinks. We leave a discussion of the saturation of these Fourier modes in the unstable regime to future studies and focus on cases where the linear approximation remains valid, but we can already assume that the neglect of the linear coupling term between modes $m'-m$ and $m'$ in the kinetic equation for $a_m$ can not be justified when a given mode $m$ grows to large amplitude. 

\begin{figure}
     \begin{subfigure}{0.45\textwidth}
         \centering
         \includegraphics[width=\textwidth]{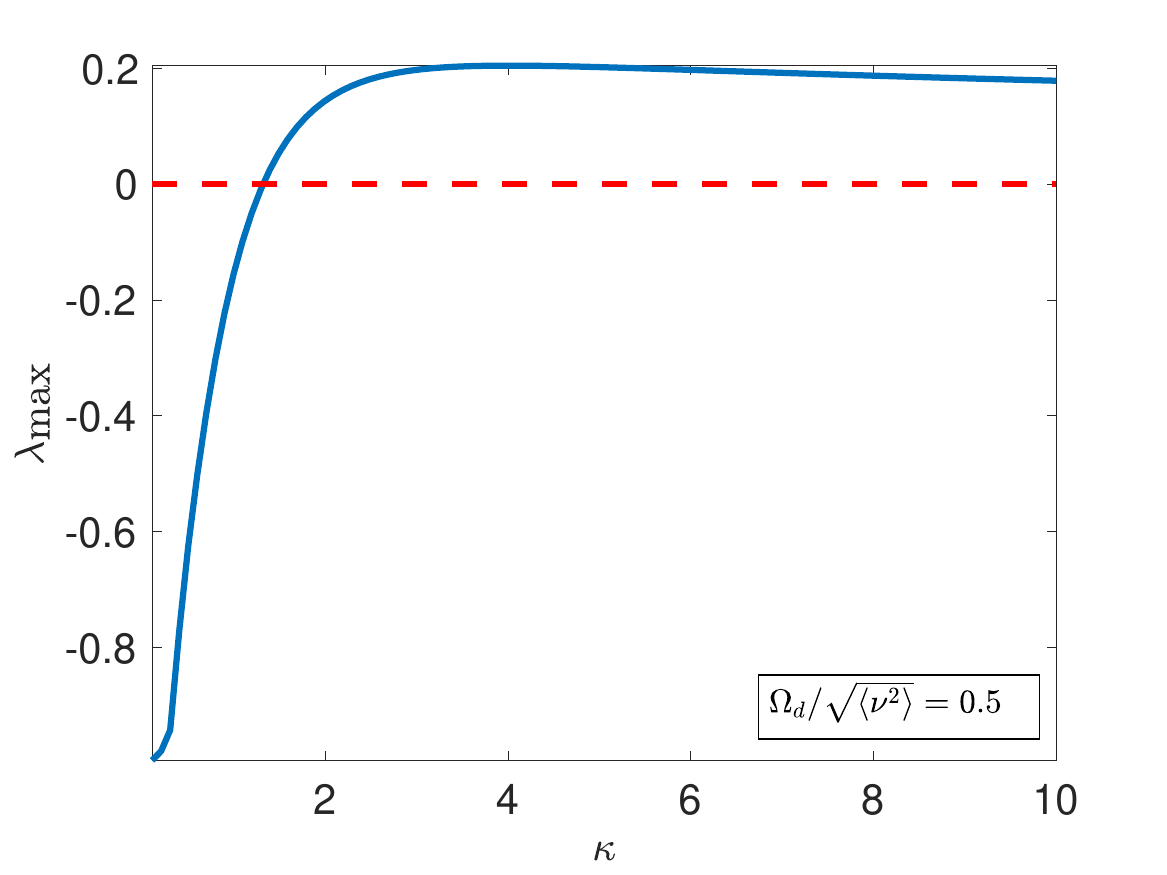}
         \caption{}
         \label{fig:1a}
     \end{subfigure}
     \hfill
     \begin{subfigure}{0.45\textwidth}
         \centering
         \includegraphics[width=\textwidth]{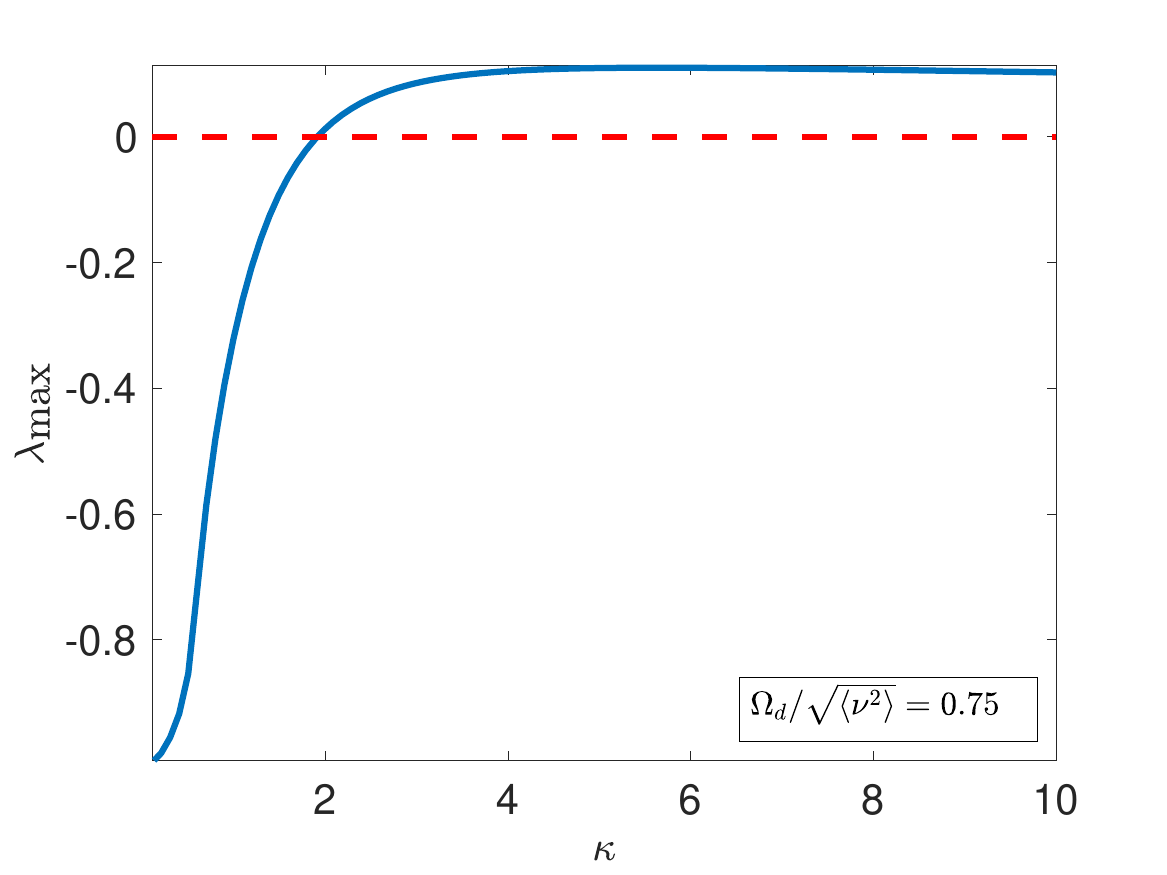}
         \caption{}
         \label{fig:1b}
     \end{subfigure}
     \hfill
     \begin{subfigure}{0.45\textwidth}
         \centering
         \includegraphics[width=\textwidth]{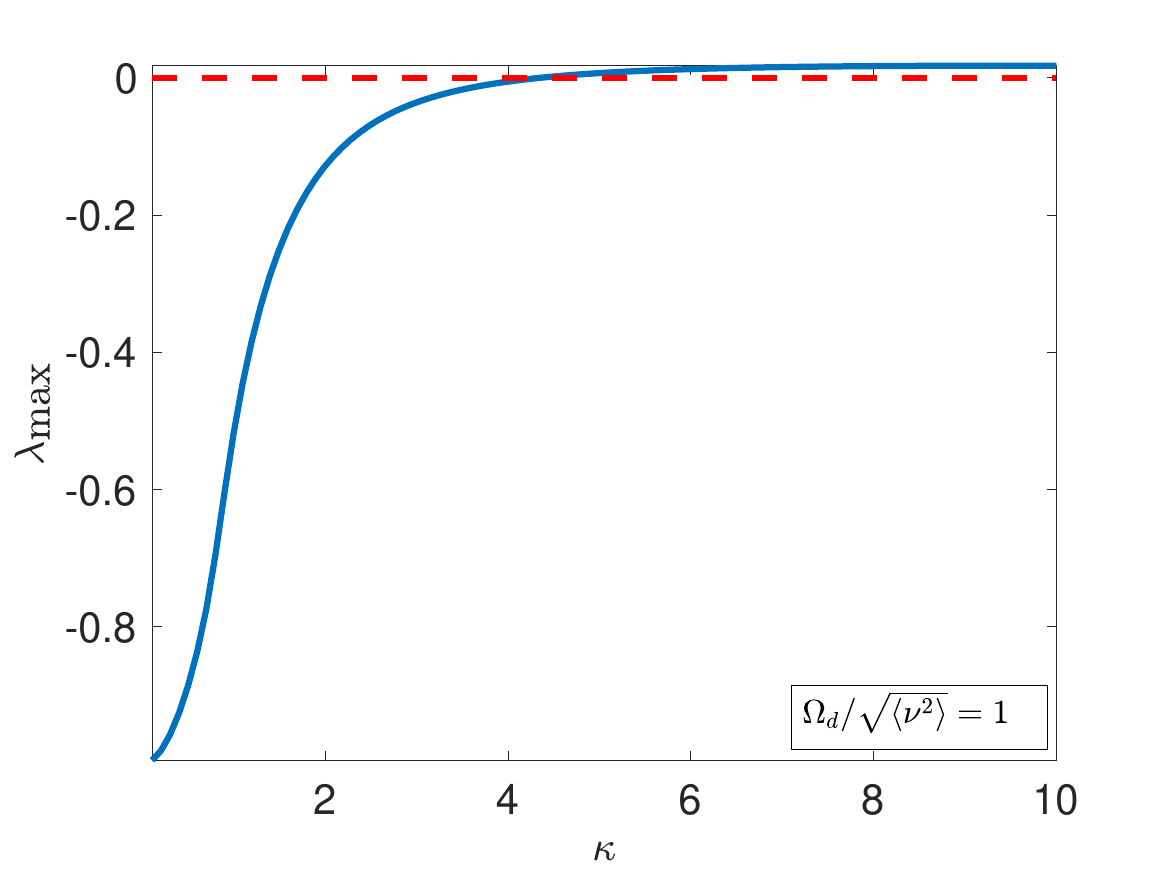}
         \caption{}
      \label{fig:1c}
     \end{subfigure}
     \hfill
      \begin{subfigure}{0.45\textwidth}
         \centering
         \includegraphics[width=\textwidth]{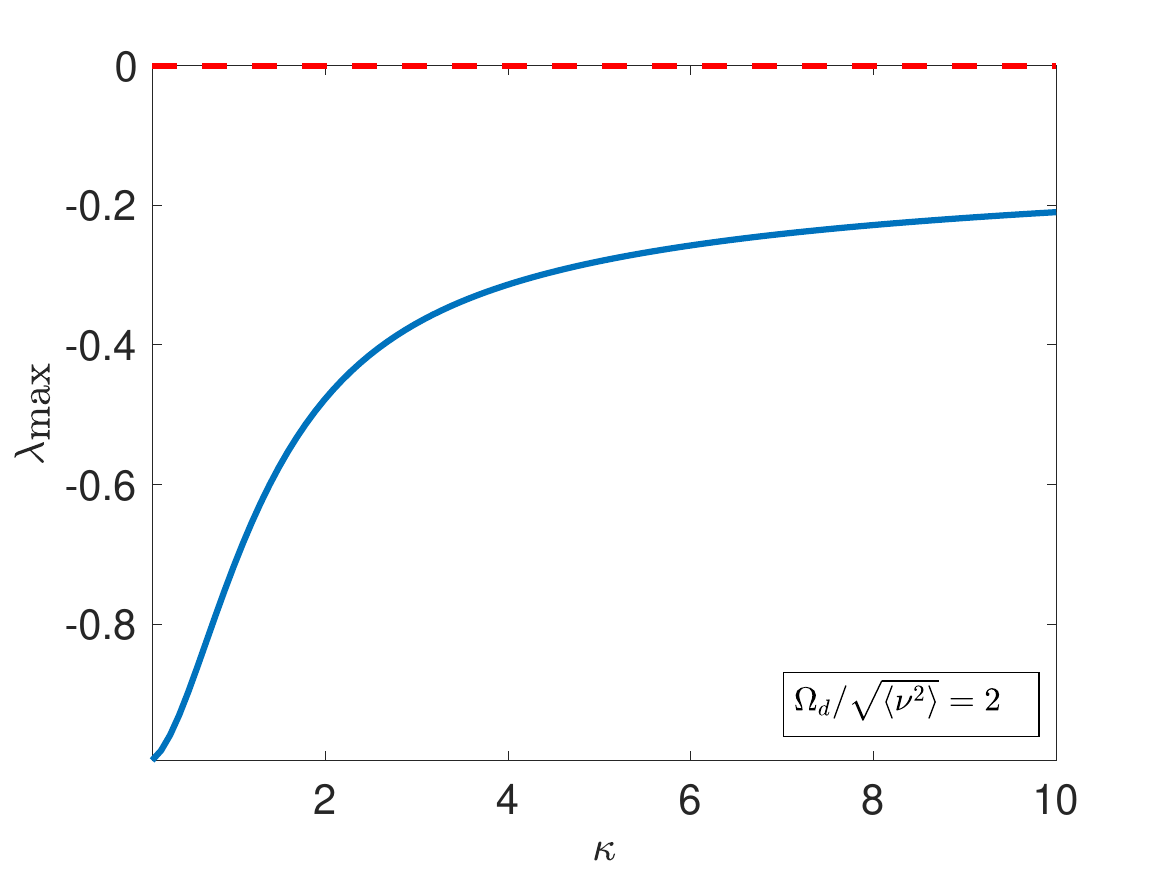}
         \caption{}
         \label{fig:1d}
     \end{subfigure}
        \caption{Dependence of the fastest growing mode $\lambda_m$ as a function of the parameter $\kappa$ for ratio of the drift frequency $\Omega_d$ to the root-mean-square loss frequency $\sqrt{\langle \nu^2 \rangle}$  equal to (a) 1/2, (b) 0.75, (c) 1 and (4) 2, respectively. When losses are fast, i.e., when $1/\Omega_d \geq 1/ \sqrt{\langle \nu^2 \rangle}$ the distribution function is linearly unstable for values of $\kappa>1$.}
        \label{fig:1}
\end{figure}

The parametric dependence of the faster growing eigenmode $\lambda_m$ as a function of the $\kappa$ parameter and the ratio of the drift frequency $\Omega_d$ to the root-mean-square loss rate $\sqrt{\langle \nu^2 \rangle}$ is also shown in Figure (\ref{fig:1}). When losses are fast, i.e., when $1/\Omega_d \ll 1/ \sqrt{\langle \nu^2 \rangle}$ the distribution function is linearly unstable for values of $\kappa>1$, whereas one needs very localised loss regions of $\kappa>5$ when $1/\Omega_d \simeq 1/ \sqrt{\langle \nu^2 \rangle}$. 

Based on this stability analysis, we make two remarks: (1) if we consider two particle populations with the same energy on the same drift shell, where one population (e.g., electrons due to the co-rotation drift) has a longer drift period than the other (e.g., proton), the population with the lower drift frequency (i.e., longer drift period) is more likely to enter an unstable regime under identical interactions with moons. The features we will see unfolding when we come close to the unstable parameter space is therefore more likely to take place for electrons than protons of comparable adiabatic invariant $\mu$ and absorption frequency $\sqrt{\langle\nu^2\rangle}$. (2) If the homogeneous solution is unstable, the particular solution which accounts for the forcing term $\boldsymbol{\eta}$ is also unstable. The forcing term at a given $m$ values is responsible for triggering fluctuations in the distribution function at a given $m$ azimuthal number. But the instability of the distribution function to localised is independent of the forcing. In the next section, we solve the linear system for an initial perturbation concentrated among the first two modes, that is, $m=1$ and $m=2$, for the driven and undriven case, and for parameter regimes where the equations are both stable and unstable. 

\subsection{Numerical solutions for driven and undriven case}
\begin{figure}[H]
     \centering
     \begin{subfigure}[b]{0.8\textwidth}
         \centering
         \includegraphics[width=\textwidth]{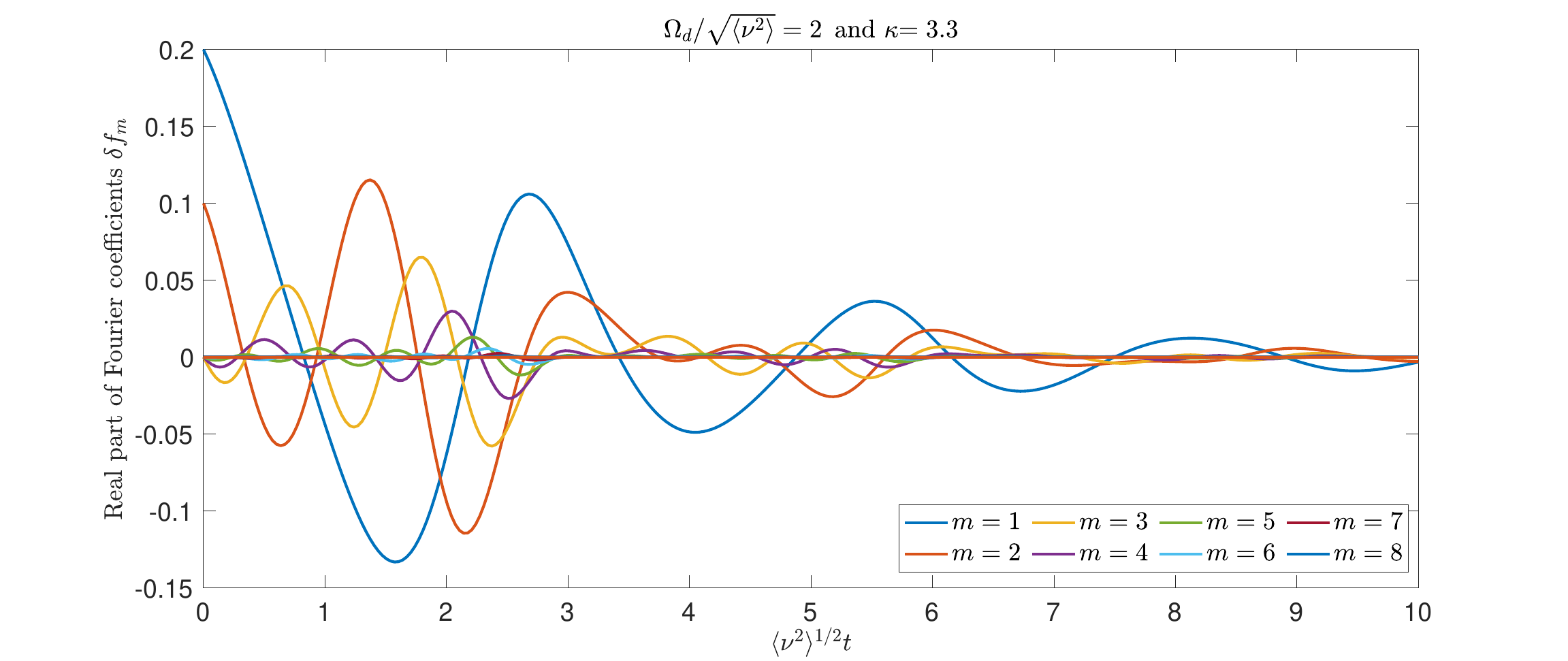}
         \caption{}
         \label{fig:3a}
     \end{subfigure}
     \hfill
     \begin{subfigure}[b]{0.8\textwidth}
         \centering
         \includegraphics[width=\textwidth]{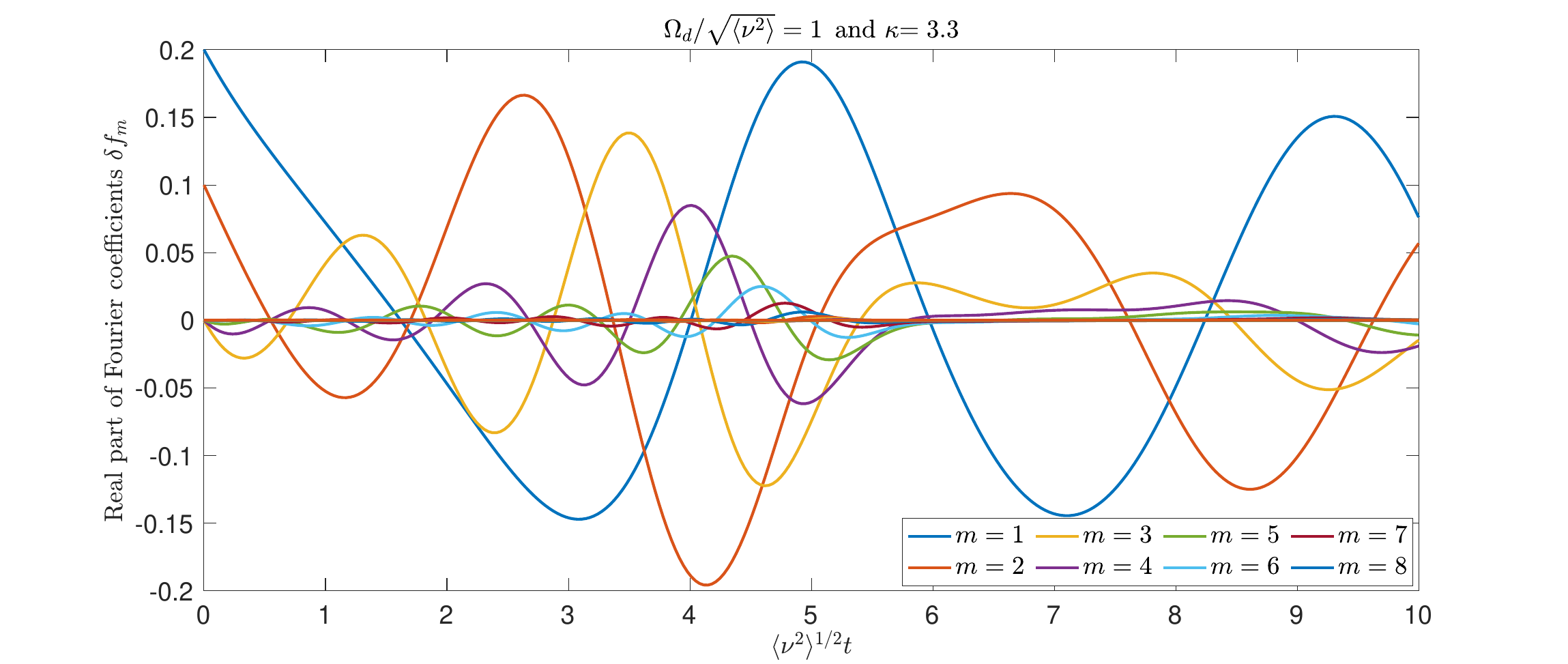}
         \caption{}
      \label{fig:3b}
     \end{subfigure}
     \hfill
      \begin{subfigure}[b]{0.8\textwidth}
         \centering
         \includegraphics[width=\textwidth]{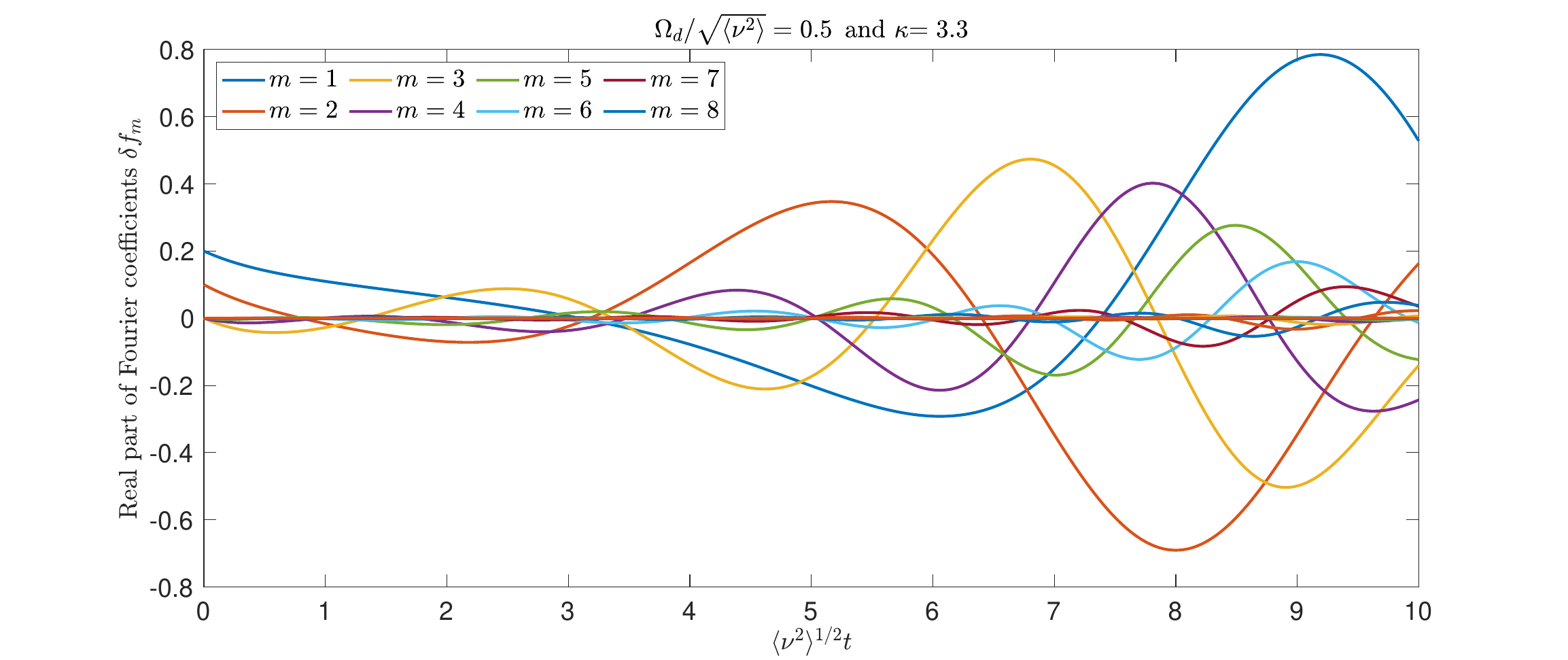}
         \caption{}
         \label{fig:3c}
     \end{subfigure}
        \caption{Time evolution of the Fourier coefficients $\delta f_m$ for stable parameters \textbf{(a)}, marginally stable parameters \textbf{(b)} and unstable parameters \textbf{(c)}. The $\kappa$ parameter is equal to 3.3 for all panels, which corresponds to a loss region localized along 50\% of the drift orbit.}
        \label{fig:3}
\end{figure}

In this section we solve the coupled Equations (\ref{EQ:oscillators2}) for the driven and undriven cases and for parameters where the system is both linearly stable and unstable. In the first examples, we initialise the system with $\delta f_{m=1}=0.2$ and $\delta f_{m=2}=0.1$. Figure (\ref{fig:3}) shows the time evolution of the Fourier coefficients $\delta f_m$ with $m$ ranging between 1 and 8 for stable parameters, panel (a), marginally stable parameters, panel (b), and unstable parameters, in panel (c). The $\kappa$ parameter for Figure (\ref{fig:3}) is equal to 3.3, which corresponds to a loss region localized along 50\% of the drift orbit. For linearly stable parameters, as shown in panel (a), the various Fourier modes eventually decay on a timescale comparable to 20 drift periods. For marginally stable parameters, that is, the region where at least one eigenvalue of the linear system of equations is negative but close to zero, some of the modes of the distribution functions do not decay on timescales comparable to ten drift periods. Panel (b) Figure (\ref{fig:3}) shows that the different modes oscillate out of phase with one another, allowing them to retain their initial amplitudes for several drift periods. Panel (c) in Figure (\ref{fig:3}) shows an example of a linearly unstable regime with $\Omega_d/ \sqrt{\langle \nu^2 \rangle}=1/2$. For the unstable case, the $m=1$ mode grows secularly on timescales comparable to 5 drift periods, which demonstrate that the instability in the distribution function operates on fast non-diffusive timescales.  

\begin{figure}[H]
     \centering
     \begin{subfigure}[b]{0.8\textwidth}
         \centering
         \includegraphics[width=\textwidth]{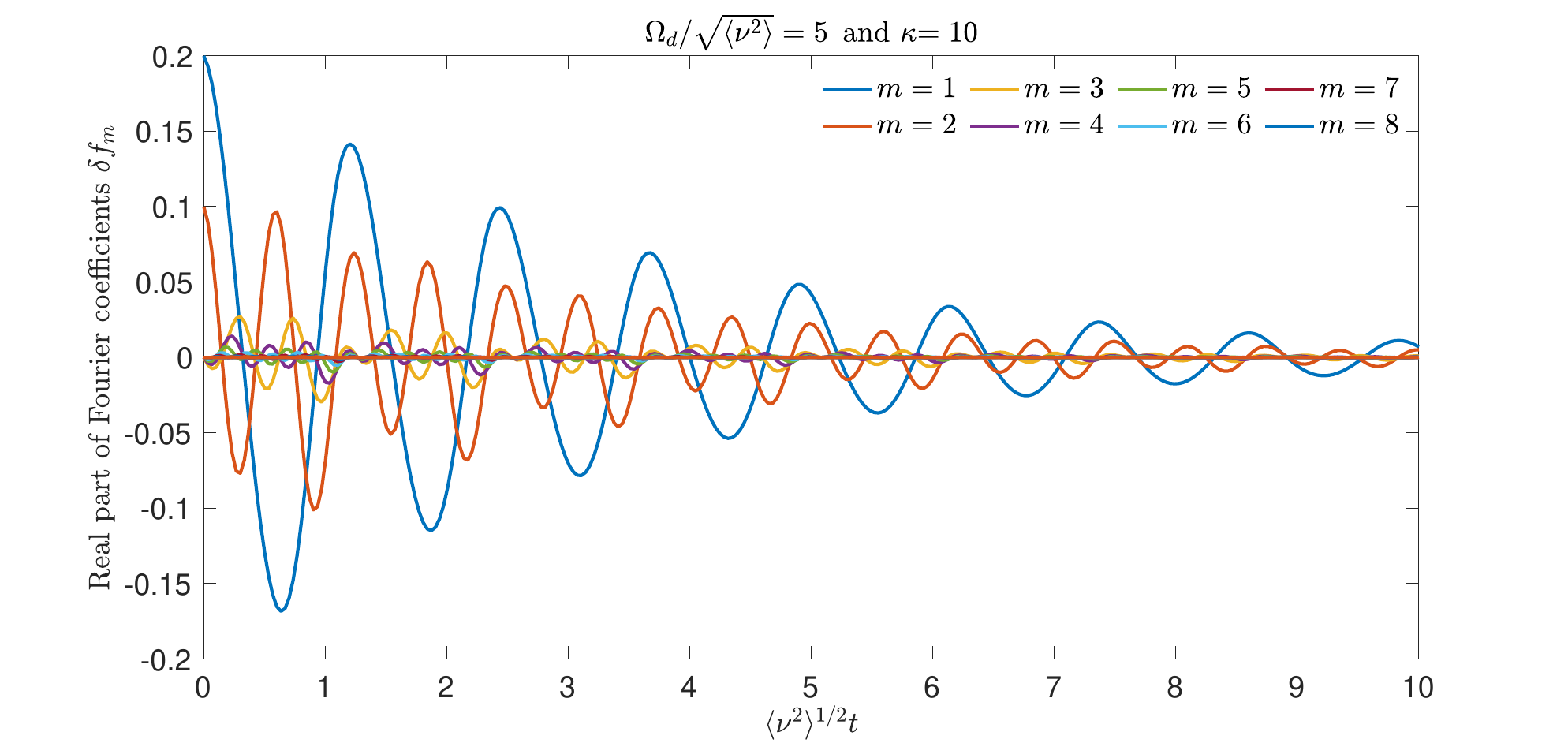}
         \caption{}
         \label{fig:4a}
     \end{subfigure}
     \hfill
     \begin{subfigure}[b]{0.8\textwidth}
         \centering
         \includegraphics[width=\textwidth]{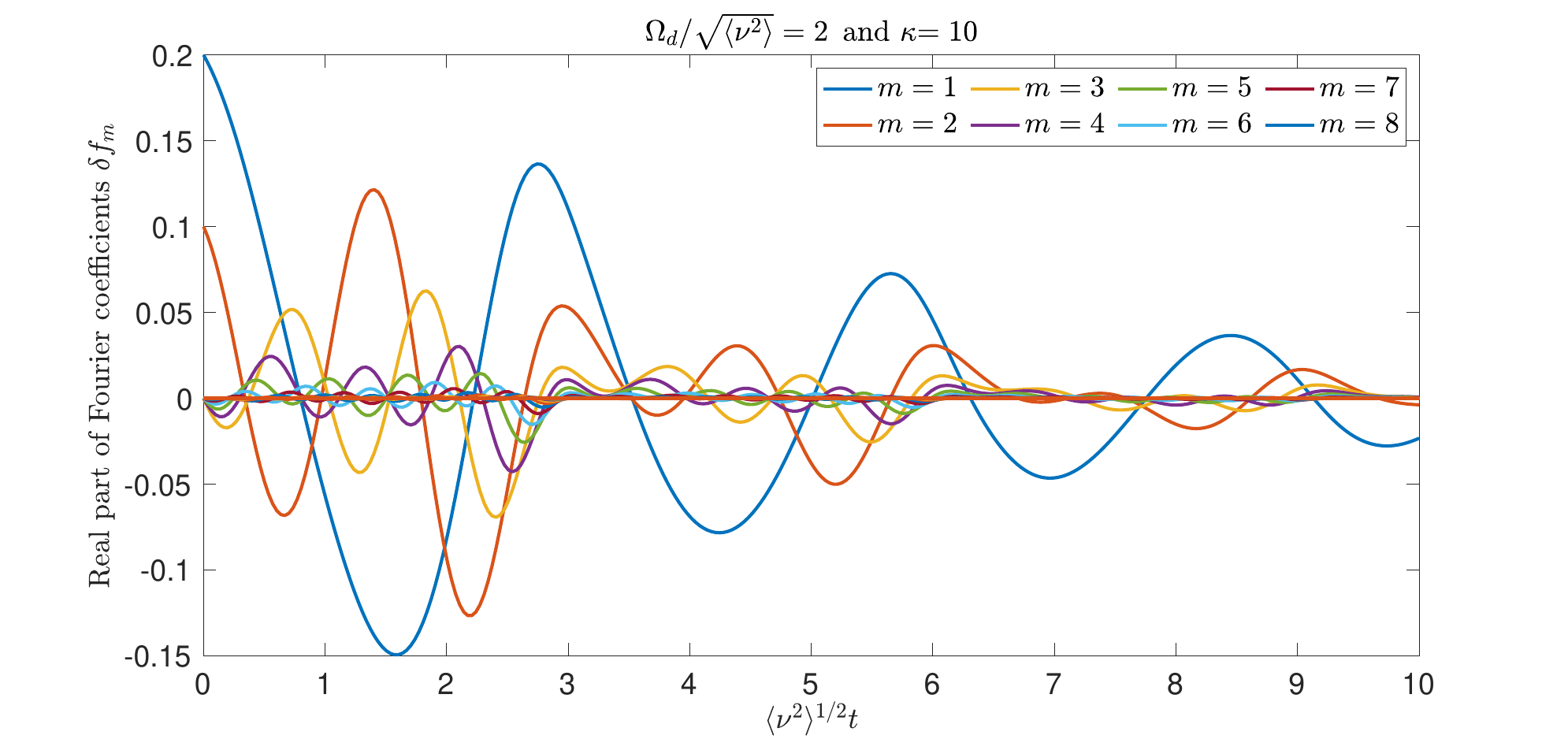}
         \caption{}
      \label{fig:4b}
     \end{subfigure}
     \hfill
      \begin{subfigure}[b]{0.8\textwidth}
         \centering
         \includegraphics[width=\textwidth]{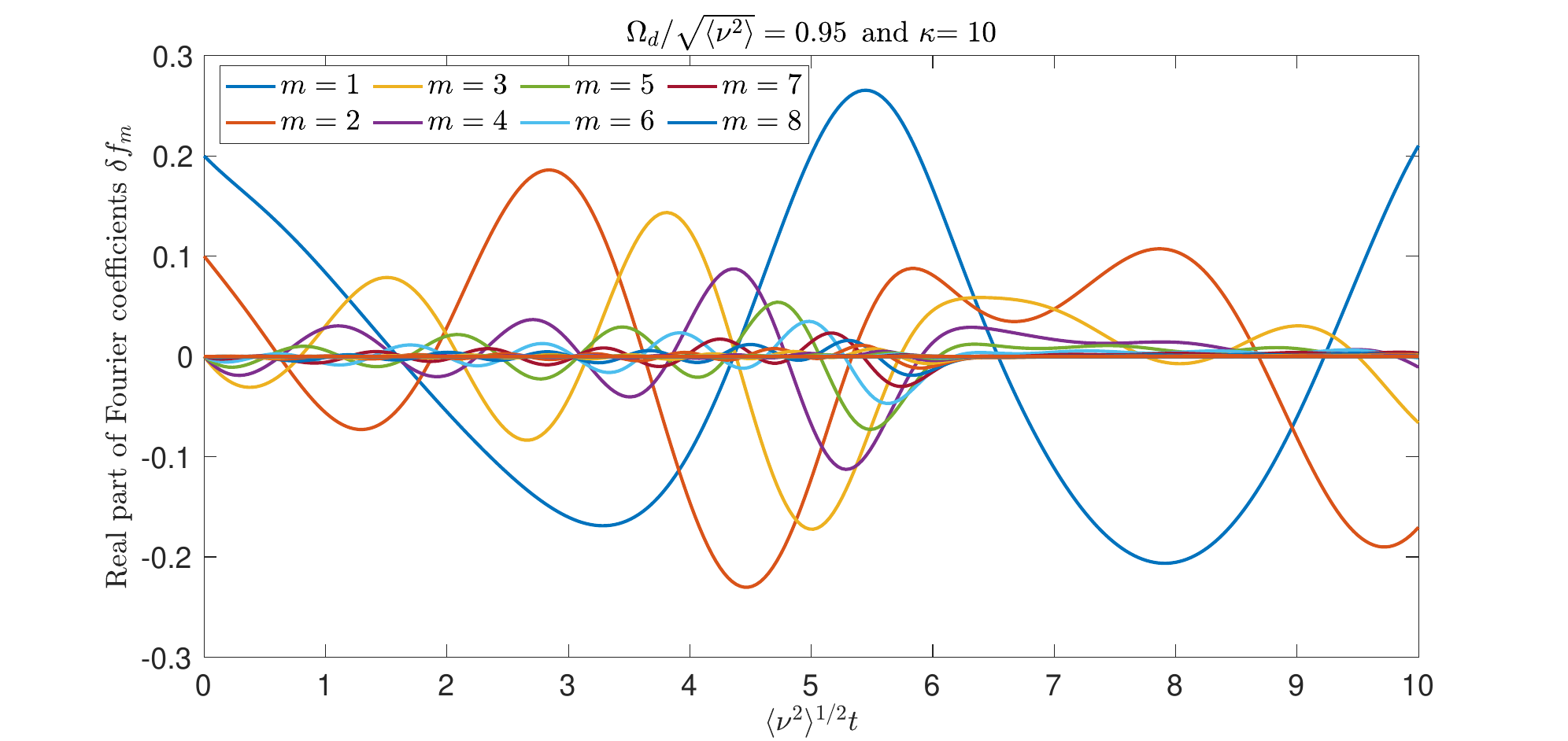}
         \caption{}
         \label{fig:4c}
     \end{subfigure}
        \caption{Same as Figure (\ref{fig:3}) with the time evolution of the Fourier coefficients $\delta f_m$ for stable parameters (panel a), marginally stable parameters (panel b) and unstable parameters (panel c). The $\kappa$ parameter is equal to 10 for all three panels, which corresponds to a loss region localized along 20\% of the drift orbit.}
        \label{fig:4}
\end{figure}

Figure~(\ref{fig:4}) presents a similar analysis to that shown in Figure~(\ref{fig:3}), depicting the time evolution of the Fourier coefficients $\delta f_m$ for three cases: stable parameters (panel a), marginally stable parameters (panel b), and unstable parameters (panel c). In this example, the parameter $\kappa$ is fixed at 10 across all three panels, corresponding to a loss region localized over 20 \% of the drift orbit. In the stable regime, the initial modes exhibit gradual, independent damping, as the coupling is too weak to facilitate interaction between modes. In the marginally stable case, damping remains present, but the decay of modes is significantly slower. This is due to coupling between the $m=1$ and $m=2$ modes, which allows energy exchange and partial coherence. Panel (c) of Figure~(\ref{fig:4}) illustrates the system’s behavior just beyond the marginal stability threshold. In this regime, the initial perturbations persist over several drift periods, rather than decaying rapidly. These results indicate that, in both the marginally stable and weakly unstable regimes of the dissipative kinetic system described by Equation~\ref{EQ:oscillators2}, certain modes of the distribution function can resist damping—and in some cases, even grow—over timescales comparable to a few drift periods.

\begin{figure}[H]
     \centering
     \begin{subfigure}[b]{0.49\textwidth}
         \centering
         \includegraphics[width=\textwidth]{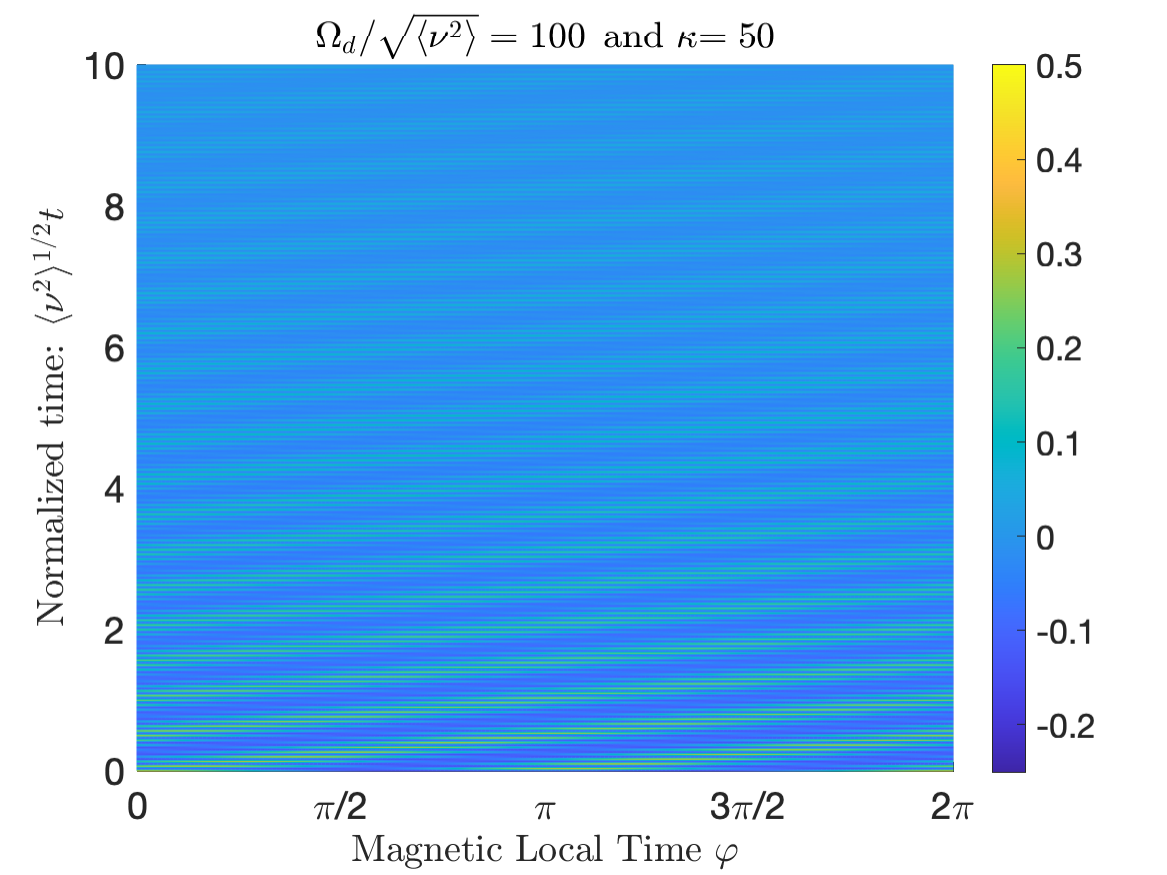}
         \caption{}
         \label{fig:5a}
     \end{subfigure}
     \hfill
     \begin{subfigure}[b]{0.49\textwidth}
         \centering
         \includegraphics[width=\textwidth]{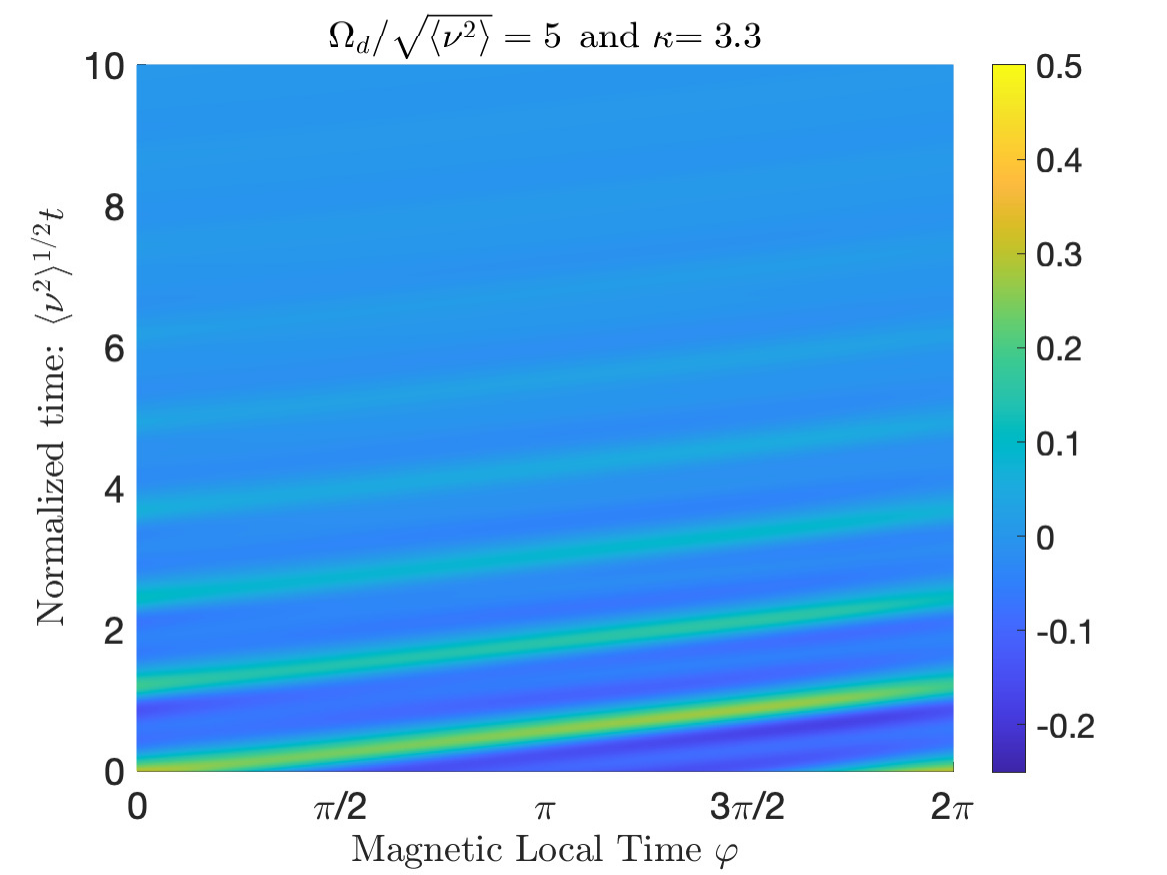}
         \caption{}
         \label{fig:5b}
     \end{subfigure}
     \hfill
     \begin{subfigure}[b]{0.49\textwidth}
         \centering
         \includegraphics[width=\textwidth]{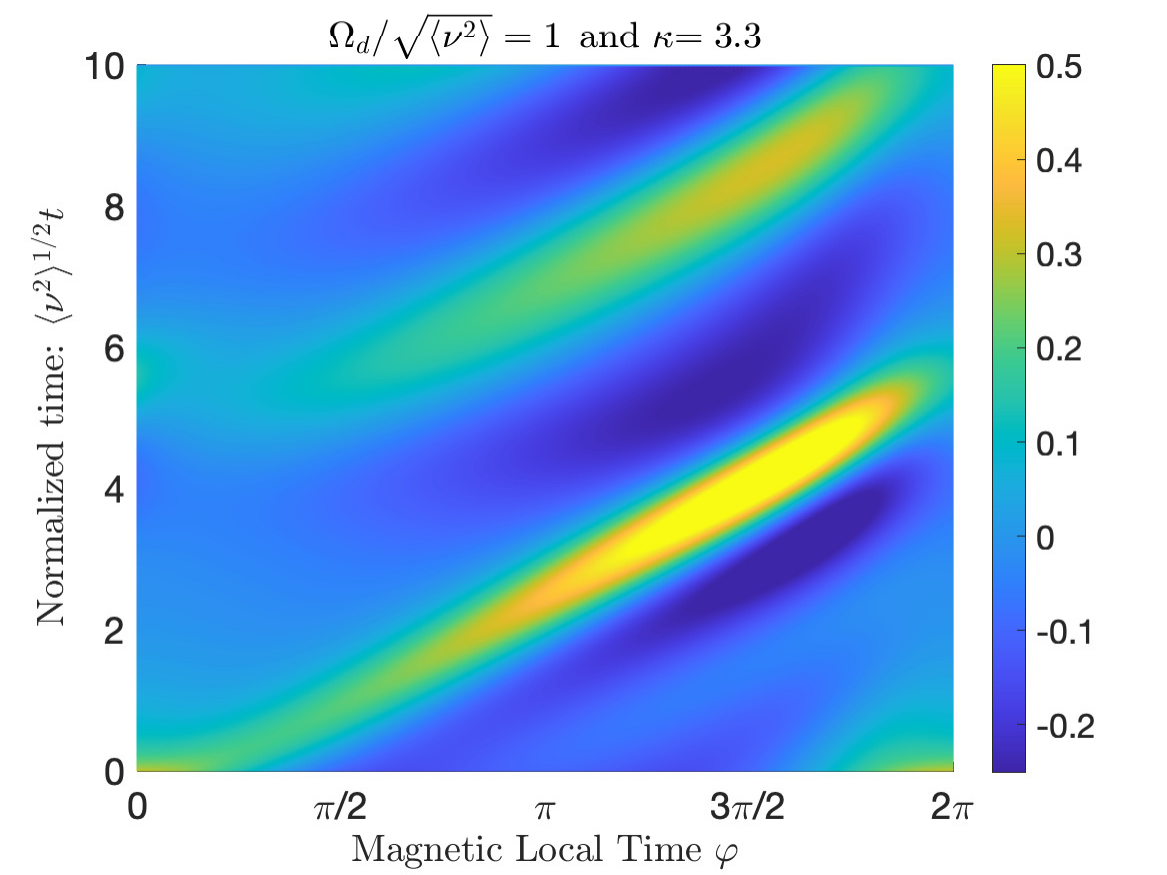}
         \caption{}
      \label{fig:5c}
     \end{subfigure}
     \hfill
      \begin{subfigure}[b]{0.49\textwidth}
         \centering
         \includegraphics[width=\textwidth]{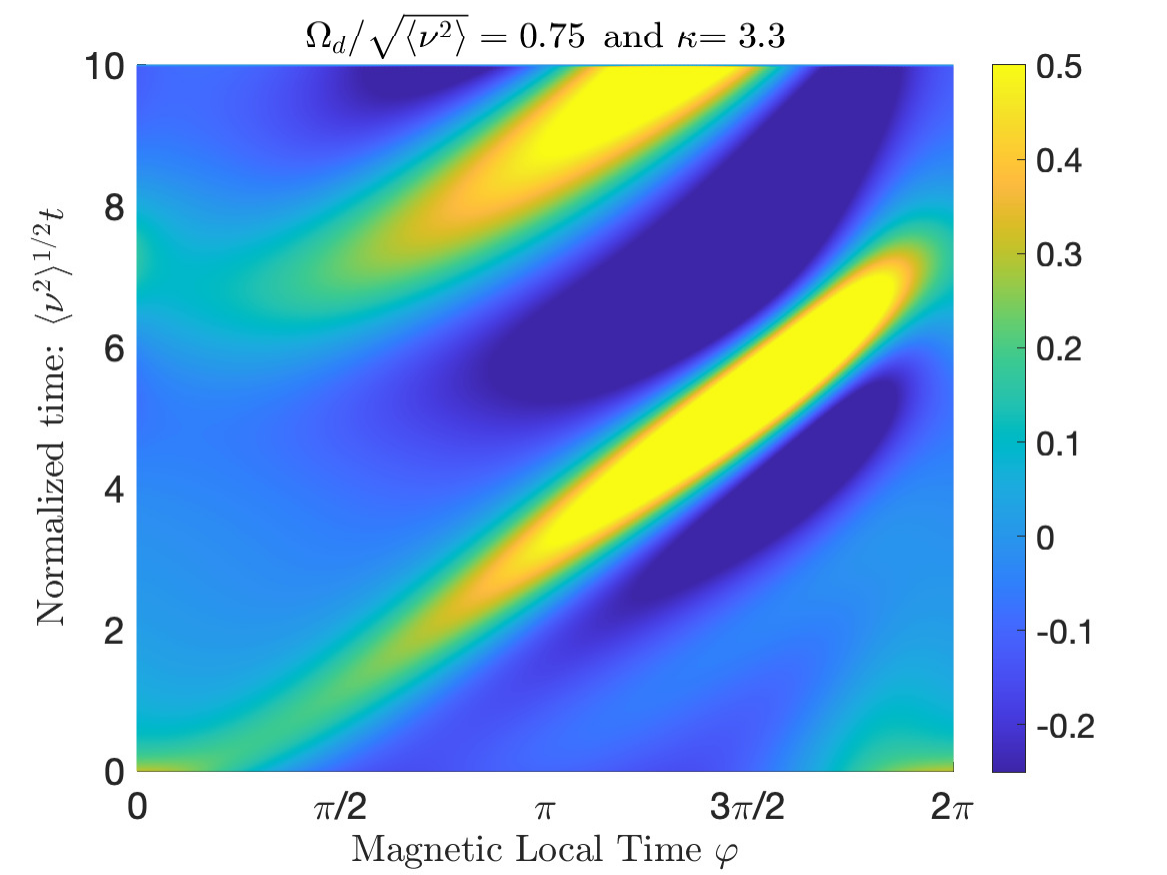}
         \caption{}
         \label{fig:5d}
     \end{subfigure}
    \caption{Surface plot of the perturbed distribution function $\delta f$ as a function of the magnetic local time $\varphi$ and normalized time for the {undriven case}, i.e. $\eta_m=0$ for all $m$ values. \textbf{(a)}  The loss region is highly localized ($\kappa = 50$) but inefficient at absorbing particles $\Omega_d/\langle \nu^2 \rangle^{1/2}=100$, leading to phase-mixing of the initial perturbation. \textbf{(b)} The loss region covers $50\%$ of the drift orbit, with a loss rate that is slow relative to the drift period. \textbf{(c)} The system is at marginal stability, allowing the initial perturbation to reappear at later times and different magnetic local times. \textbf{(d)} The system is linearly unstable, causing the perturbed distribution to grow to levels comparable to the background distribution, i.e., $\delta f \simeq f_0$.}
        \label{fig:5}
\end{figure}

Figure (\ref{fig:3}) and (\ref{fig:4}) are consistent with the stability analysis presented in Section (\ref{section:stability}). But in order to determine the signature of stable and unstable Fourier modes in the distribution function, we need to reconstruct the MLT dependent distribution functions and look at its evolution as a function of time. Figure (\ref{fig:5}) shows surface plot of the perturbed distribution function $\delta f=\sum_m \delta f_m e^{im\varphi}$ as a function of the magnetic local time $\varphi$ and normalized time for the undriven case, i.e. $\eta_m=0$ for all $m$ values. As reminder, setting $\eta_m=0$ for all $m$ values indicates either the absence of a poloidal electric field, or of a radial gradient in the distribution function $f_0$.  Panel \textbf{(a)} of Figure (\ref{fig:5}) shows when the loss region is highly localized ($\kappa = 50$) but inefficient at absorbing particles $\Omega_d/\langle \nu^2 \rangle^{1/2}=100$, leading to phase-mixing of the initial perturbation. Panel \textbf{(b)} of Figure (\ref{fig:5}) is when the loss region covers $50\%$ of the drift orbit, but with a loss rate that is slow relative to the drift period. In this instance, the initial perturbation are once more slowly decaying in time. Panel \textbf{(c)} shows when system is at marginal stability, allowing the initial perturbation to reappear at later times and different magnetic local times. Panel \textbf{(d)} is when the system is linearly unstable, causing the perturbed distribution to grow to levels comparable to the background distribution, i.e., $\delta f \simeq f_0$. 

\begin{figure}[H]
     \centering
     \begin{subfigure}[b]{0.49\textwidth}
         \centering
         \includegraphics[width=\textwidth]{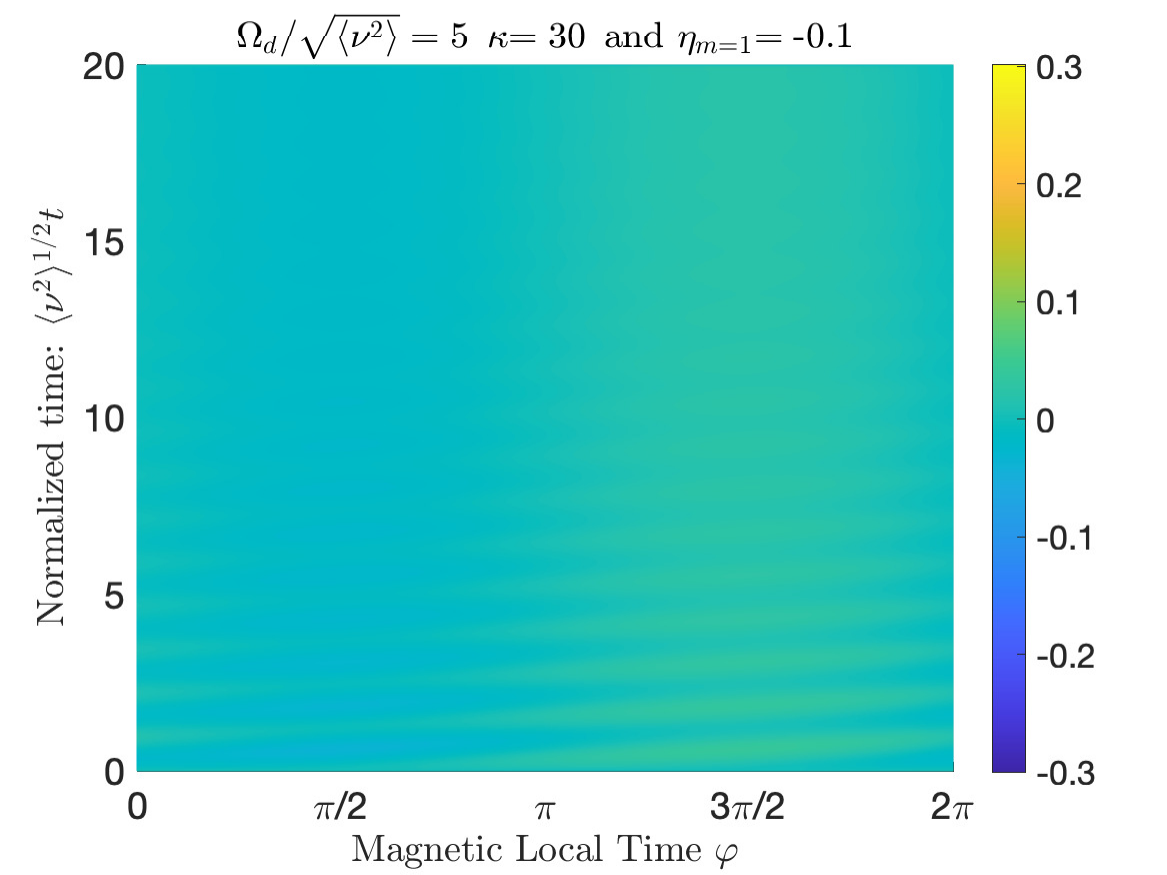}
         \caption{}
         \label{fig:6a}
     \end{subfigure}
     \hfill
     \begin{subfigure}[b]{0.49\textwidth}
         \centering
         \includegraphics[width=\textwidth]{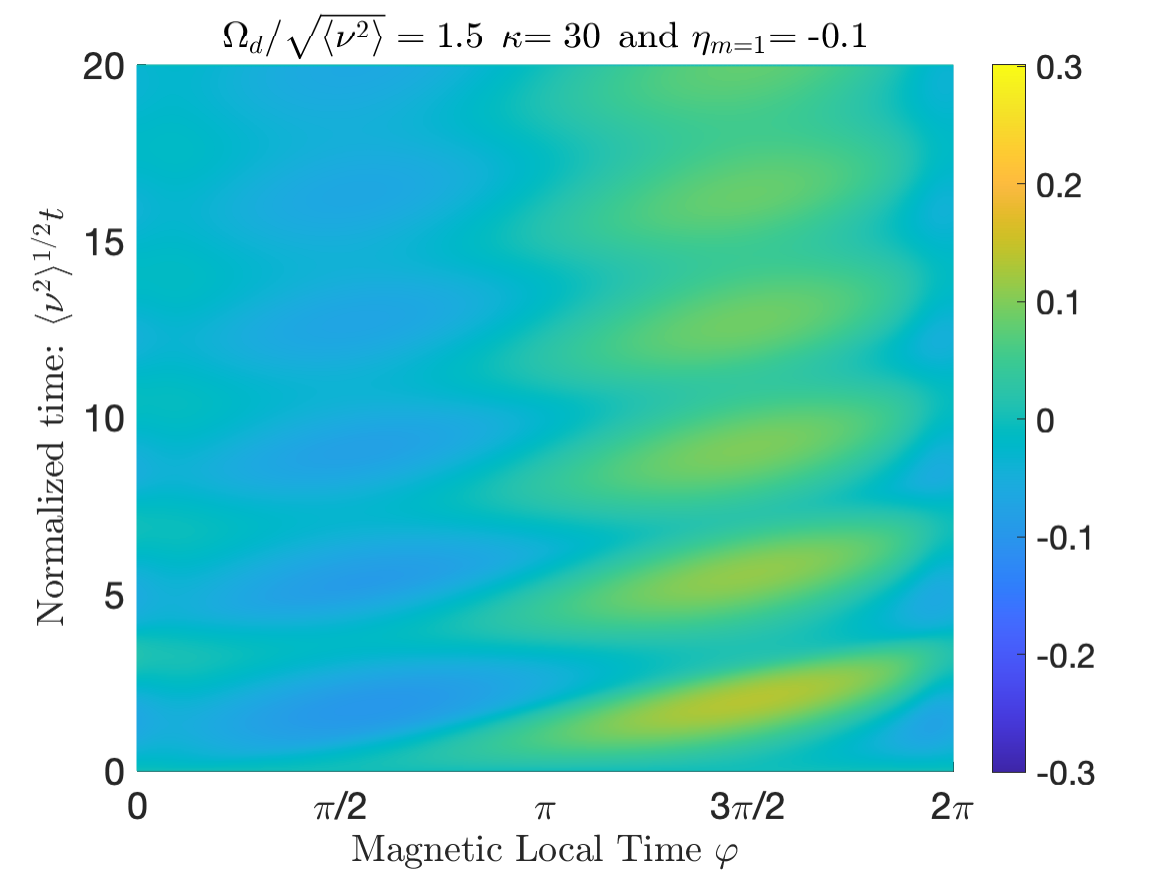}
         \caption{}
         \label{fig:6b}
     \end{subfigure}
     \hfill
     \begin{subfigure}[b]{0.49\textwidth}
         \centering
         \includegraphics[width=\textwidth]{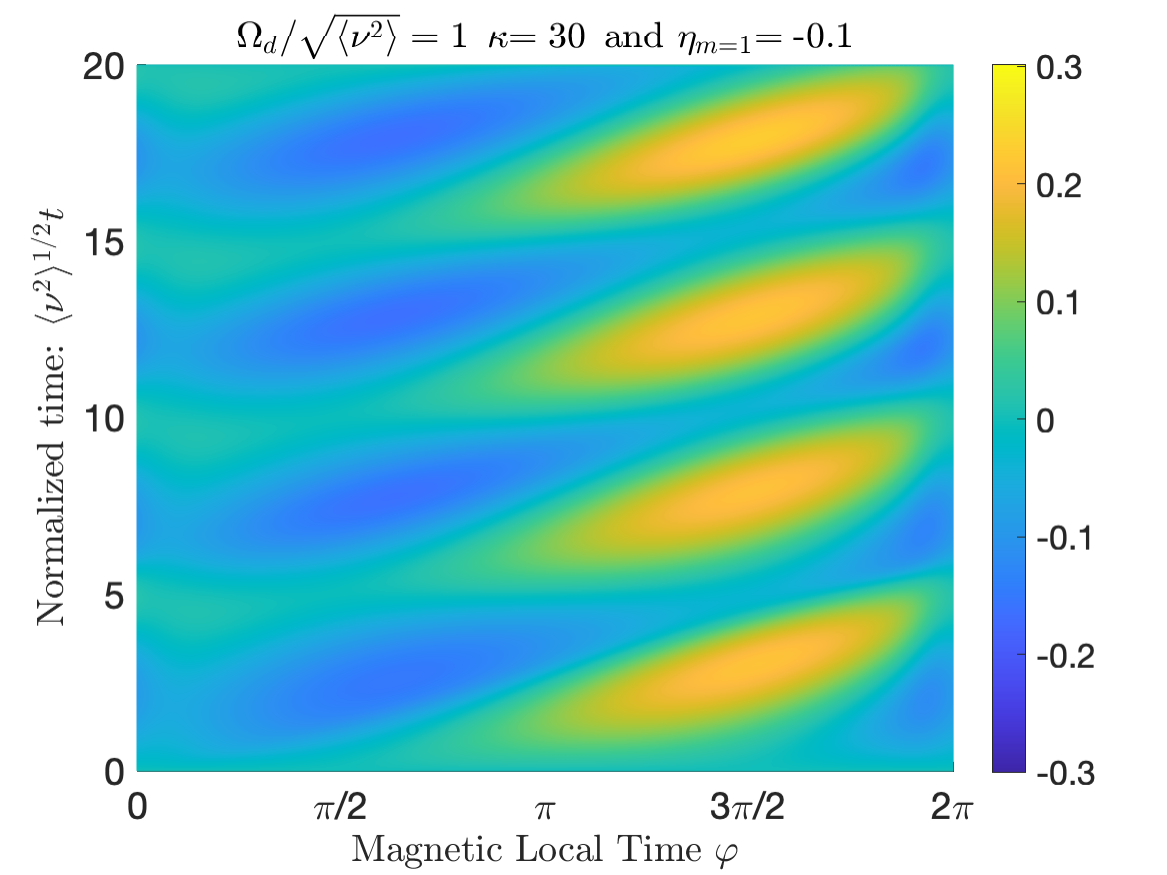}
         \caption{}
      \label{fig:6c}
     \end{subfigure}
     \hfill
      \begin{subfigure}[b]{0.49\textwidth}
         \centering
         \includegraphics[width=\textwidth]{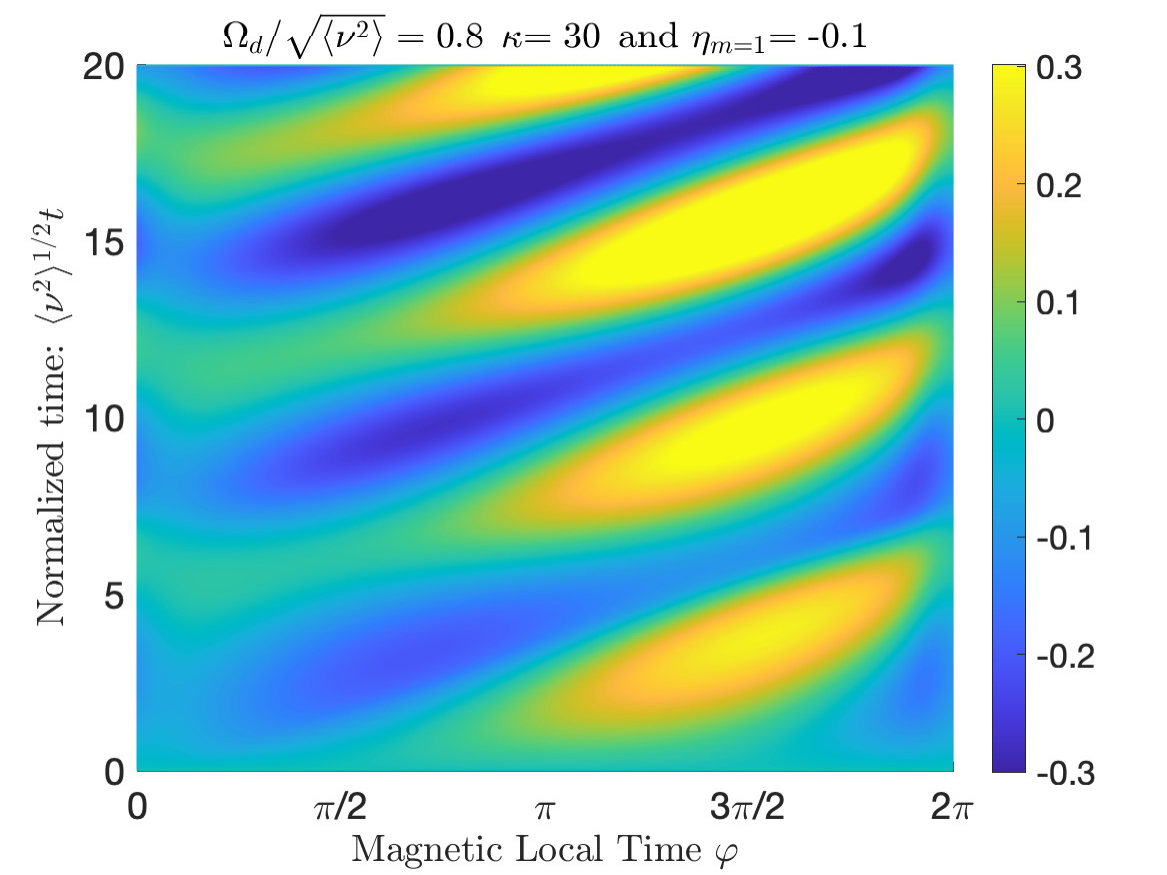}
         \caption{}
         \label{fig:6d}
     \end{subfigure}
  \caption{Surface plot of the perturbed distribution function $\delta f$ as a function of the magnetic local time $\varphi$ and normalized time for the \textbf{driven case}, i.e. $\eta_{m}=-0.1$ for $m=1$ and $\eta_m=0$ for $m\neq 1$. The  Fourier coeffficients $\delta f_m$ are initialized at zero at $t=0$. All panels are for 15\% MLT localized loss regions ($\kappa=30$). \textbf{(a)}  Even though the loss region is highly localized the system is linearly stable with $\Omega_d/\langle \nu^2 \rangle^{1/2}=5$, leading to phase-mixing and decaying of the perturbation driven by the electric field.  \textbf{(b)} The loss rate is fast compared to the drift period with $\Omega_d/\langle \nu^2 \rangle^{1/2}=1.5$ but the system is still stable, and linearly coupled mode eventually decay. \textbf{(c)} The system is at marginal stability, with $\Omega_d/\langle \nu^2 \rangle^{1/2}=1$, allowing the initial perturbation to reappear at later times and different magnetic local times. \textbf{(d)} The system is linearly unstable, $\Omega_d/\langle \nu^2 \rangle^{1/2}=0.8$, causing the electric field perturbation of the distribution function to grow at MLTs away from the loss region.}
        \label{fig:6}
\end{figure}

The extension to cases where the system is driven can also be solved numerically. This corresponds to situations in which the distribution function at a given drift shell $L$ possesses a radial gradient, and the radial $E \times B$ drift is non-zero. Figure~(\ref{fig:6}) presents surface plots of the perturbed distribution function $\delta f$ as a function of magnetic local time $\varphi$ and normalized time for the driven case, specifically with $\eta_{m} = -0.1$ for $m=1$ and $\eta_m = 0$ for $m \neq 1$. The Fourier coefficients $\delta f_m$ are initialized to zero at $t=0$. All panels correspond to a case with 15\% MLT-localized loss regions ($\kappa = 30$). Panel~\textbf{(a)} shows that despite the highly localized loss region, the system is linearly stable with $\Omega_d/\langle \nu^2 \rangle^{1/2} = 5$, leading to phase mixing and decay of the perturbation driven by the electric field. Panel~\textbf{(b)} represents a scenario where the loss rate is fast compared to the drift period, with $\Omega_d/\langle \nu^2 \rangle^{1/2} = 1.5$, yet the system remains stable, and the linearly coupled modes eventually decay. In Panel~\textbf{(c)}, the system is at marginal stability with $\Omega_d/\langle \nu^2 \rangle^{1/2} = 1$, allowing the initial perturbation to reappear at later times and at different magnetic local times. Finally, Panel~\textbf{(d)} illustrates a linearly unstable case with $\Omega_d/\langle \nu^2 \rangle^{1/2} = 0.8$, where the driven perturbation in the distribution function grows at magnetic local times away from the loss region.

The stability analysis and the figures discussed thus far demonstrate that when particle absorption is sufficiently fast and the loss region is sufficiently localized, the distribution function can exhibit growth or stabilization of certain Fourier modes at the expense of others. However, in situ spacecraft measurements record particle fluxes in the spacecraft frame as it traverses the magnetosphere, sampling different magnetic local time (MLT) locations. To facilitate comparison with observations—and to assess the synchronization behavior across different Fourier modes—we plot the full distribution function as a function of MLT over several drift periods. 

\begin{figure}[H]
     \centering
     \begin{subfigure}[b]{0.6\textwidth}
         \centering
         \includegraphics[width=\textwidth]{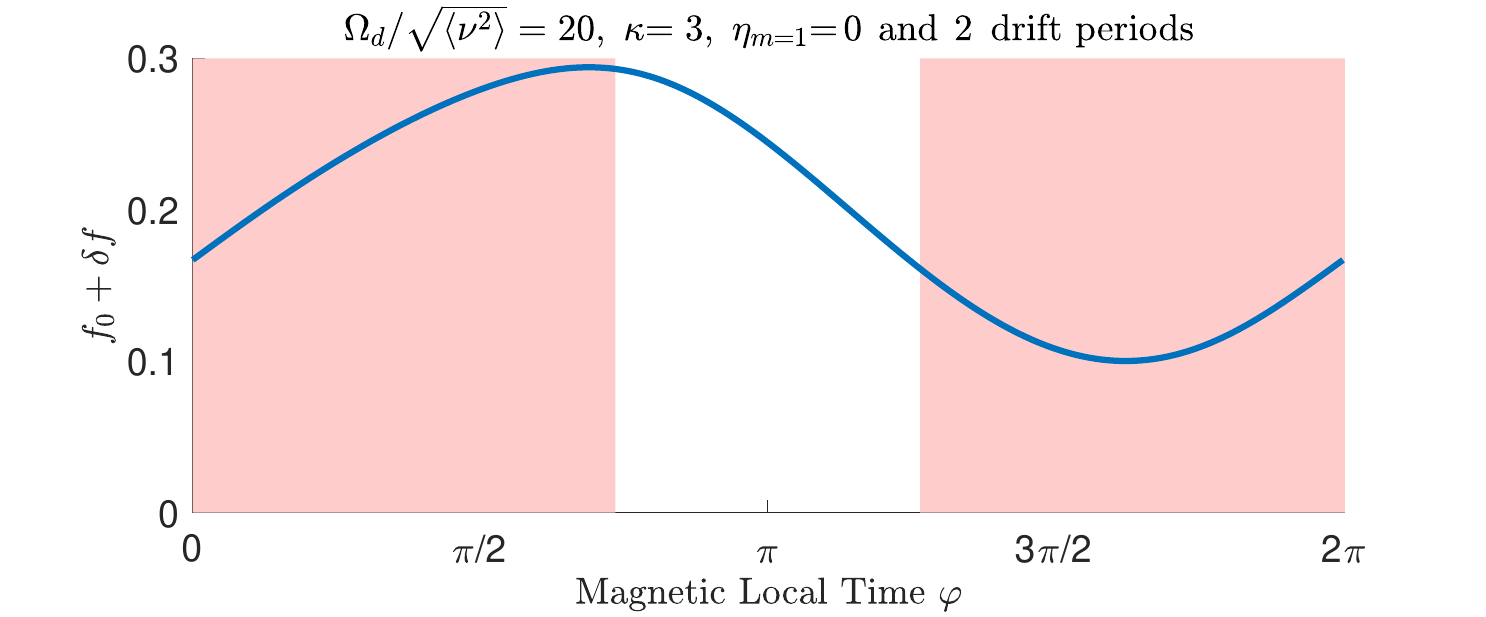}
         \caption{}
         \label{fig:7a}
     \end{subfigure}
     \hfill
     \begin{subfigure}[b]{0.6\textwidth}
         \centering
         \includegraphics[width=\textwidth]{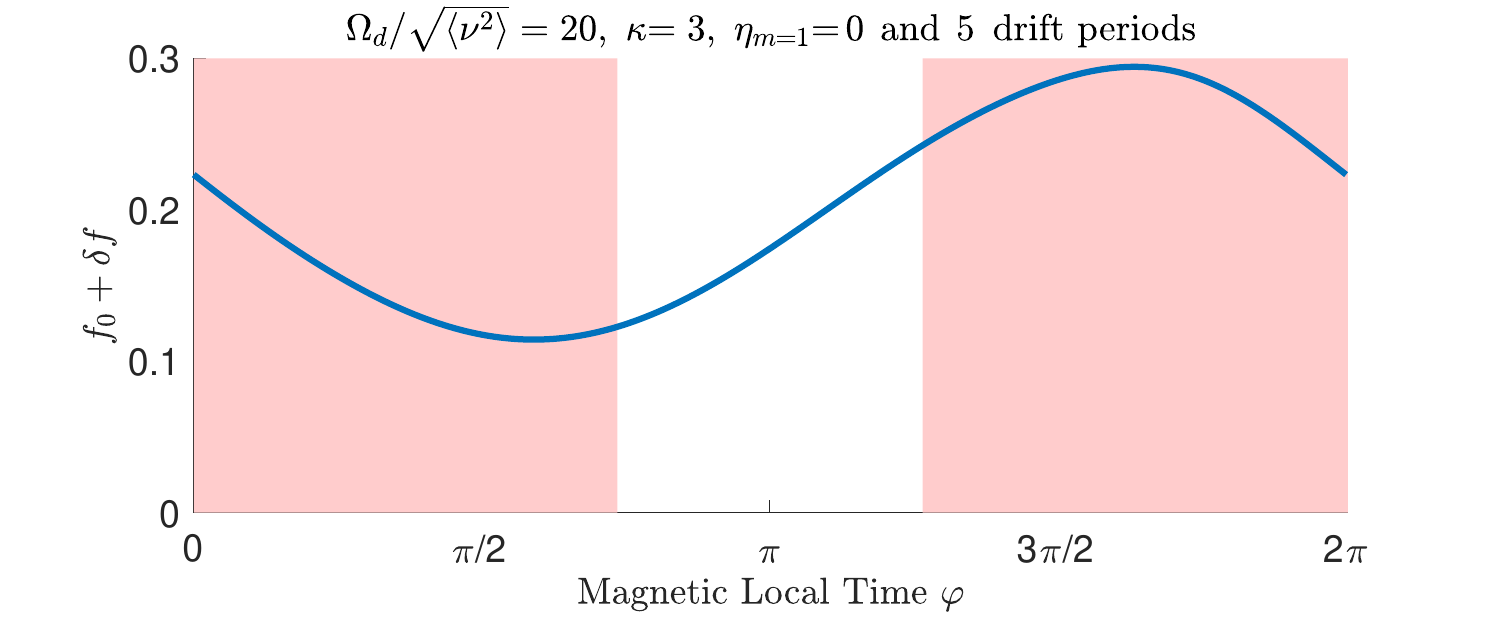}
         \caption{}
         \label{fig:7b}
     \end{subfigure}
     \hfill
     \begin{subfigure}[b]{0.6\textwidth}
         \centering
         \includegraphics[width=\textwidth]{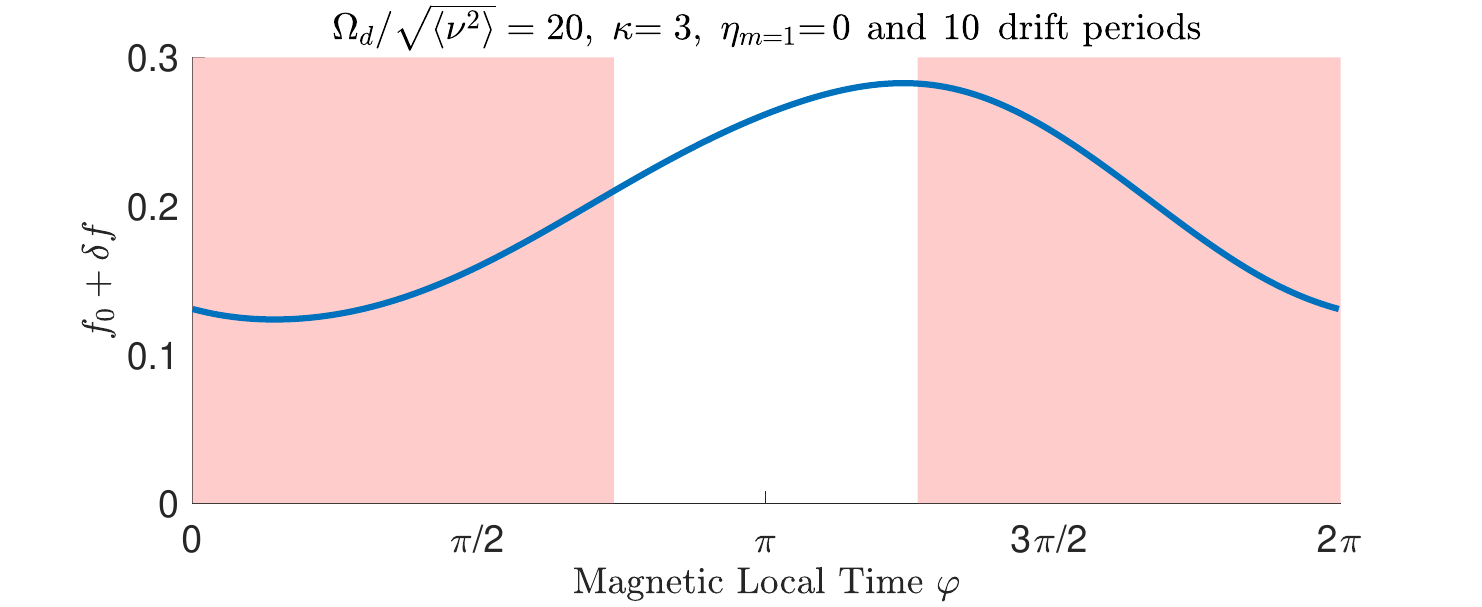}
         \caption{}
      \label{fig:7c}
     \end{subfigure}
     \hfill
     \begin{subfigure}[b]{0.6\textwidth}
         \centering
         \includegraphics[width=\textwidth]{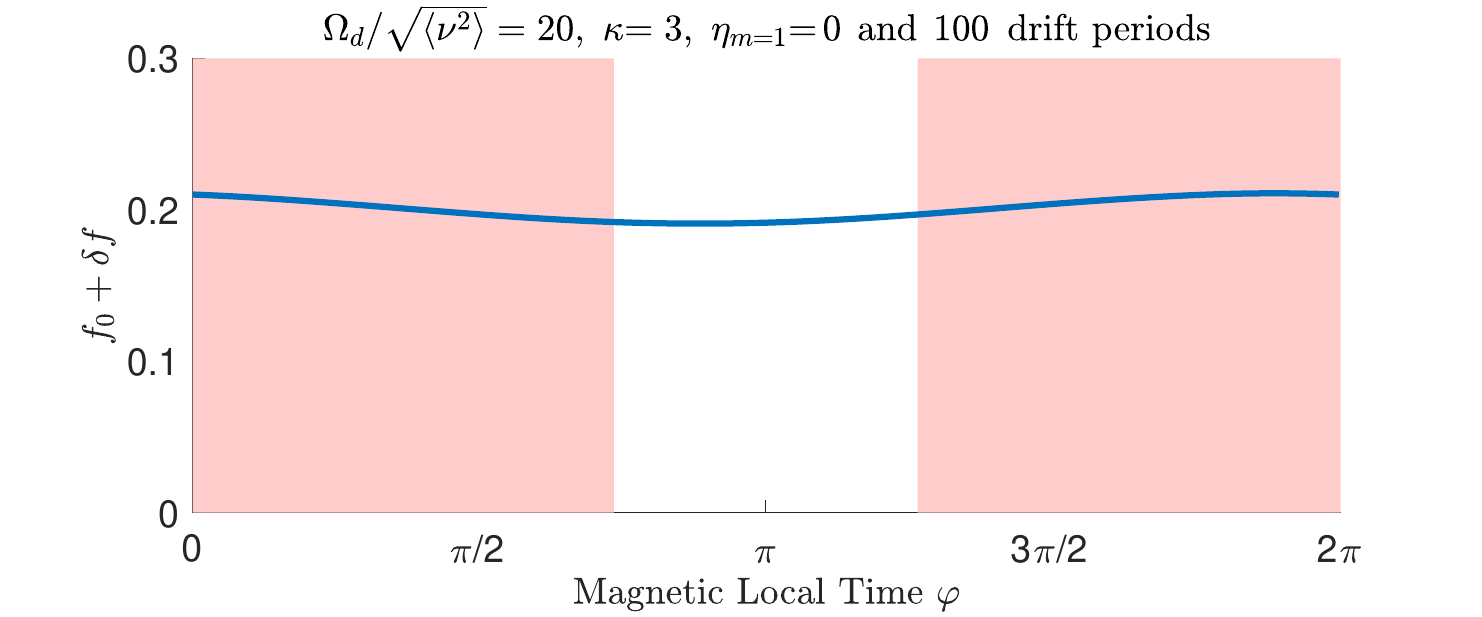}
         \caption{}
      \label{fig:7d}
     \end{subfigure}
  \caption{Total distribution function $f_0+\delta f$ as a function of the magnetic local time after 2 \textbf{(a)}  , 5 \textbf{(b)}, 10 \textbf{(c)}  and 100 \textbf{(d)}  drift periods for the undriven case, i.e. $\eta_{m}=0$. The  coeffficient $\delta f_{m=1}$ is initialized at 0.1 at $t=0$. The loss region defined as within 2 standard deviation from it's center is shaded in pink. \textbf{(a)}  The system is linearly stable with $\Omega_d/\langle \nu^2 \rangle^{1/2}=20$. and the $m=1$ mode of the distribution function is weakly affected by the losses.}
        \label{fig:7}
\end{figure}

In Figure (\ref{fig:7}) we show the total distribution function $f_0+\delta f$ as a function of the magnetic local time after 2 \textbf{(a)}  , 5 \textbf{(b)}, 10 \textbf{(c)}  and 100 \textbf{(d)}  drift periods for the undriven case, i.e. $\eta_{m}=0$.  The system is linearly stable with $\Omega_d/\langle \nu^2 \rangle^{1/2}=20$ and the $m=1$ mode of the distribution function is weakly affected by the losses. The  coefficient $\delta f_{m=1}$ is initialized at 0.1 at $t=0$. The loss region defined as within 2 standard deviation from it's center is shaded in pink. \textbf{(a)} In this particular example, the initial perturbation, which is essentially a drift echo, oscillates with frequency $\Omega_d$, but gradually damps away. Losses are far too slow to sustain synchronization, and the initial $m=1$ mode decays entirely, leaving the background distribution unaffected. 

\begin{figure}[H]
     \centering
     \begin{subfigure}[b]{0.8\textwidth}
         \centering
         \includegraphics[width=\textwidth]{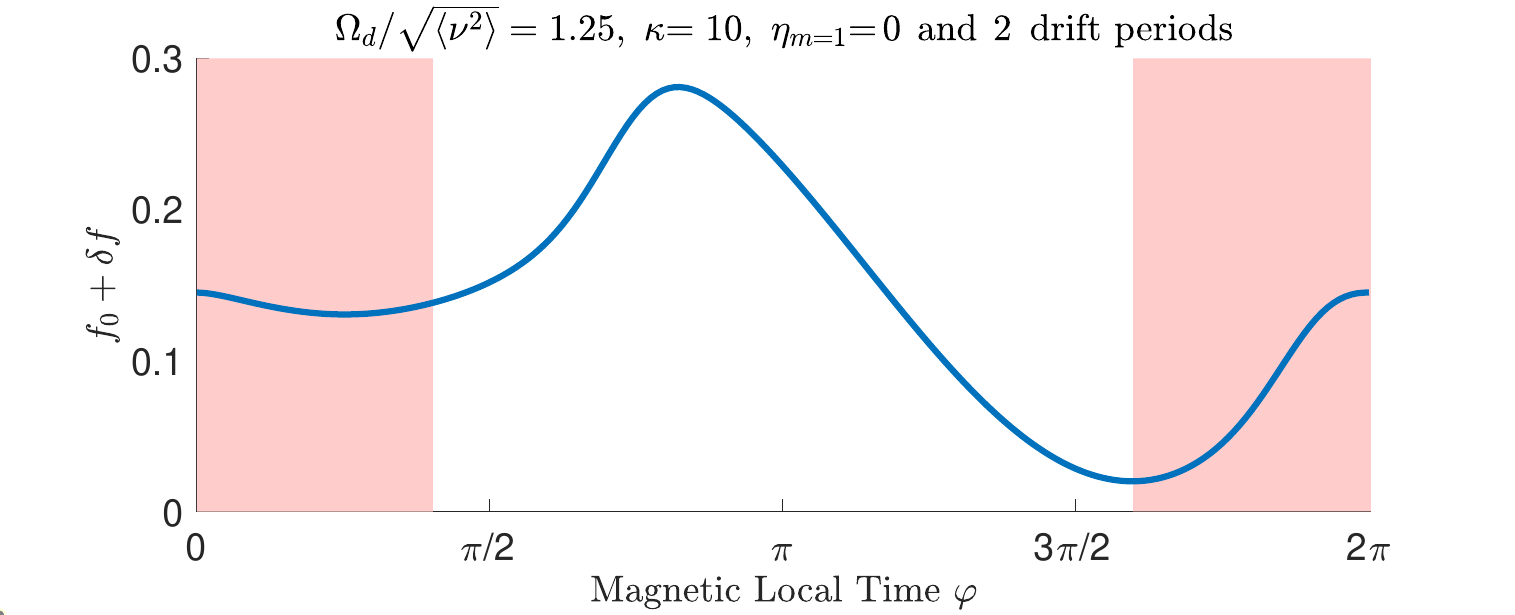}
         \caption{}
         \label{fig:11a}
     \end{subfigure}
     \hfill
     \begin{subfigure}[b]{0.8\textwidth}
         \centering
         \includegraphics[width=\textwidth]{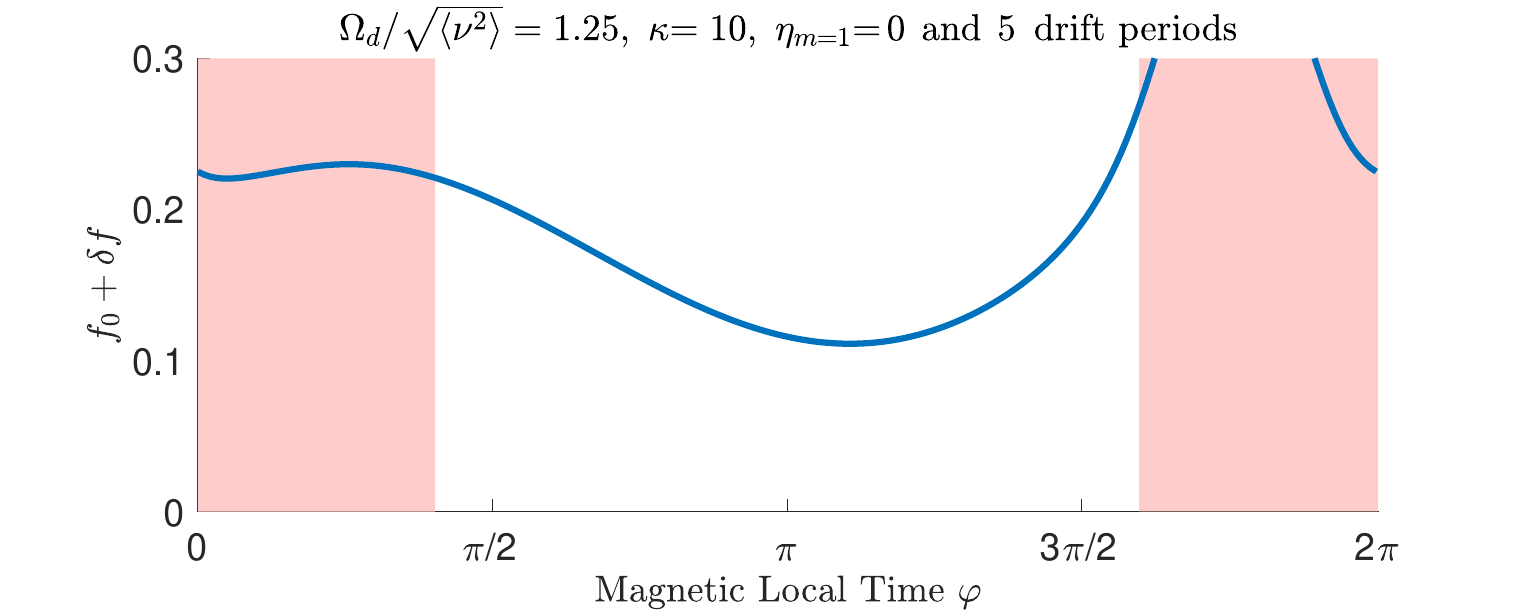}
         \caption{}
         \label{fig:11b}
     \end{subfigure}
     \hfill
     \begin{subfigure}[b]{0.8\textwidth}
         \centering
         \includegraphics[width=\textwidth]{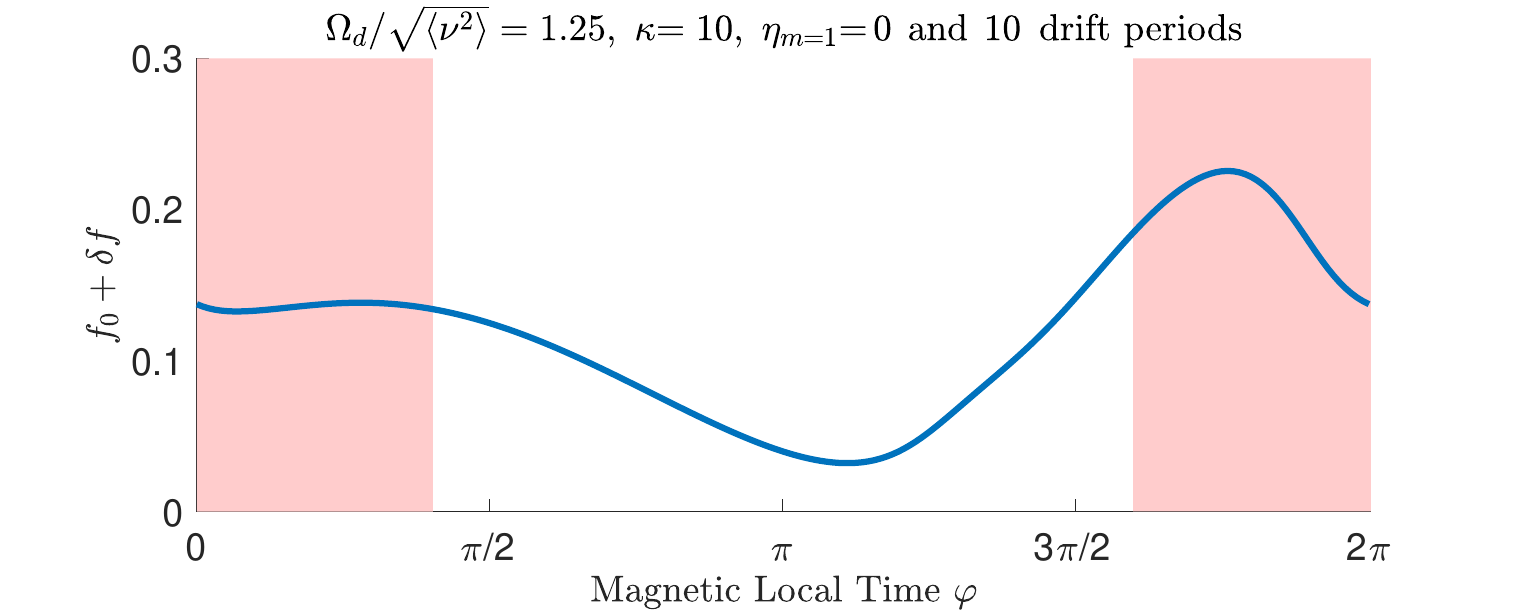}
         \caption{}
      \label{fig:11c}
     \end{subfigure}
       \caption{Same as for Figure (\ref{fig:7}) but for a system much closer to marginal stability with $\Omega_d/\langle \nu^2 \rangle^{1/2}=1.25$ and $\kappa=10$.}
        \label{fig:11}
\end{figure}

In Figure (\ref{fig:11}) the parameters are chosen closer to marginal stability with $\Omega_d/\langle \nu^2 \rangle^{1/2}=1.25$ and $\kappa=10$ with an initial perturbation $\delta f_{m=1}\neq 0$. Synchronization between the different Fourier modes appear as an enhancement of the distribution function on timescales that are comparable to a few drift periods. Figure (\ref{fig:11}) indicates that even though the system is dissipative, the synchronization gives the appearance of a refilling of the phase-space density. Thus, when MLT losses are sufficiently localized and fast, i.e., comparable to the drift period, the distribution function can appear to refill faster than it decays.

\begin{figure}[H]
     \centering
     \begin{subfigure}[b]{0.8\textwidth}
         \centering
         \includegraphics[width=\textwidth]{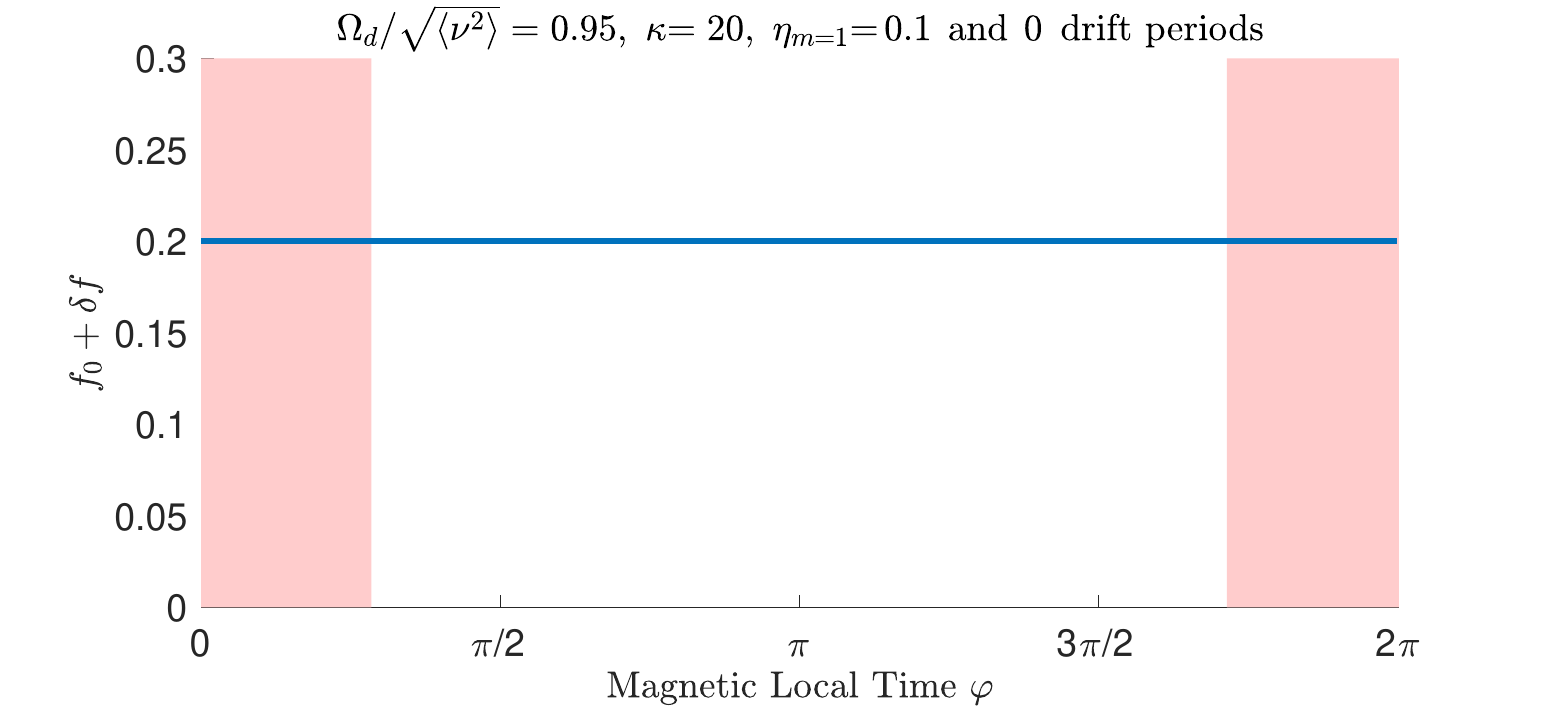}
         \caption{}
         \label{fig:12a}
     \end{subfigure}
     \hfill
     \begin{subfigure}[b]{0.8\textwidth}
         \centering
         \includegraphics[width=\textwidth]{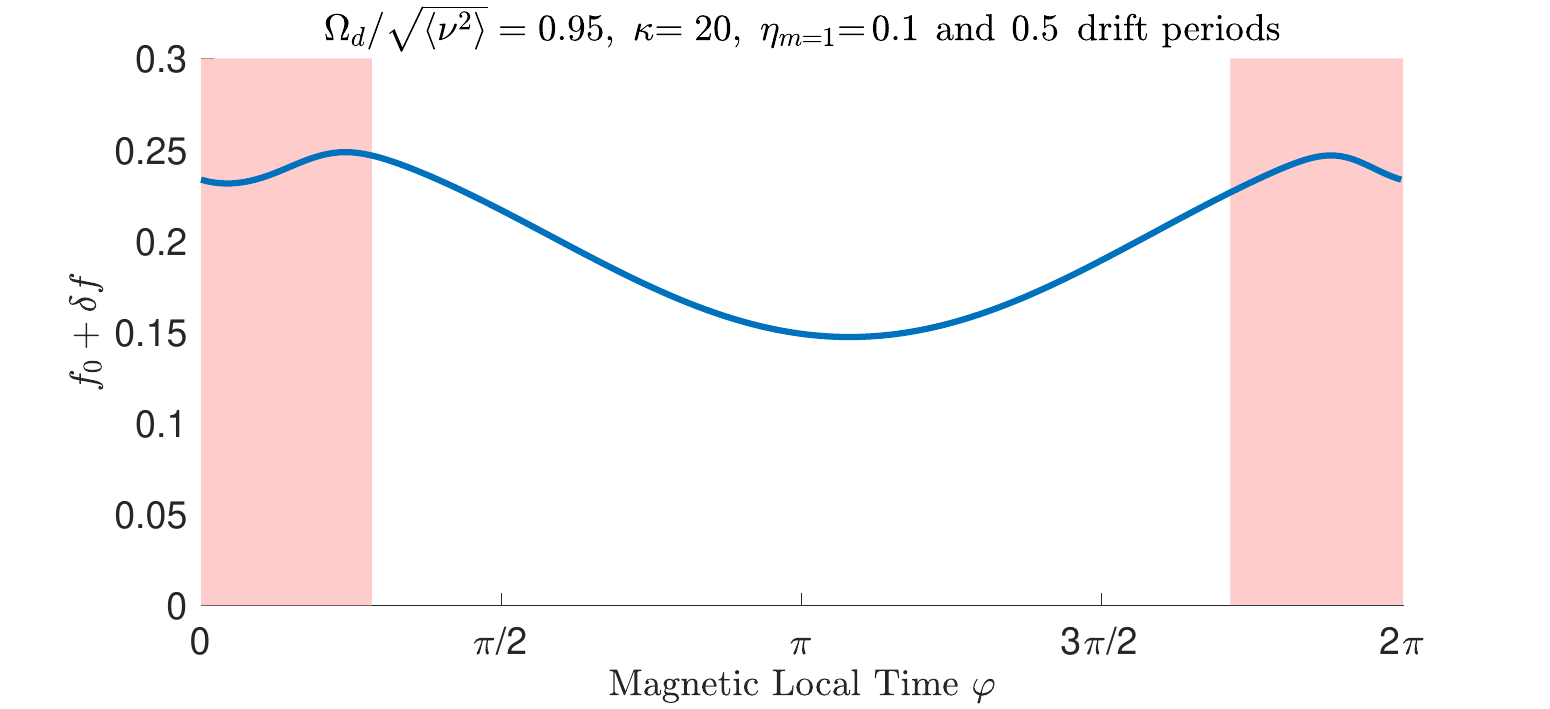}
         \caption{}
         \label{fig:12b}
     \end{subfigure}
     \hfill
     \begin{subfigure}[b]{0.8\textwidth}
         \centering
         \includegraphics[width=\textwidth]{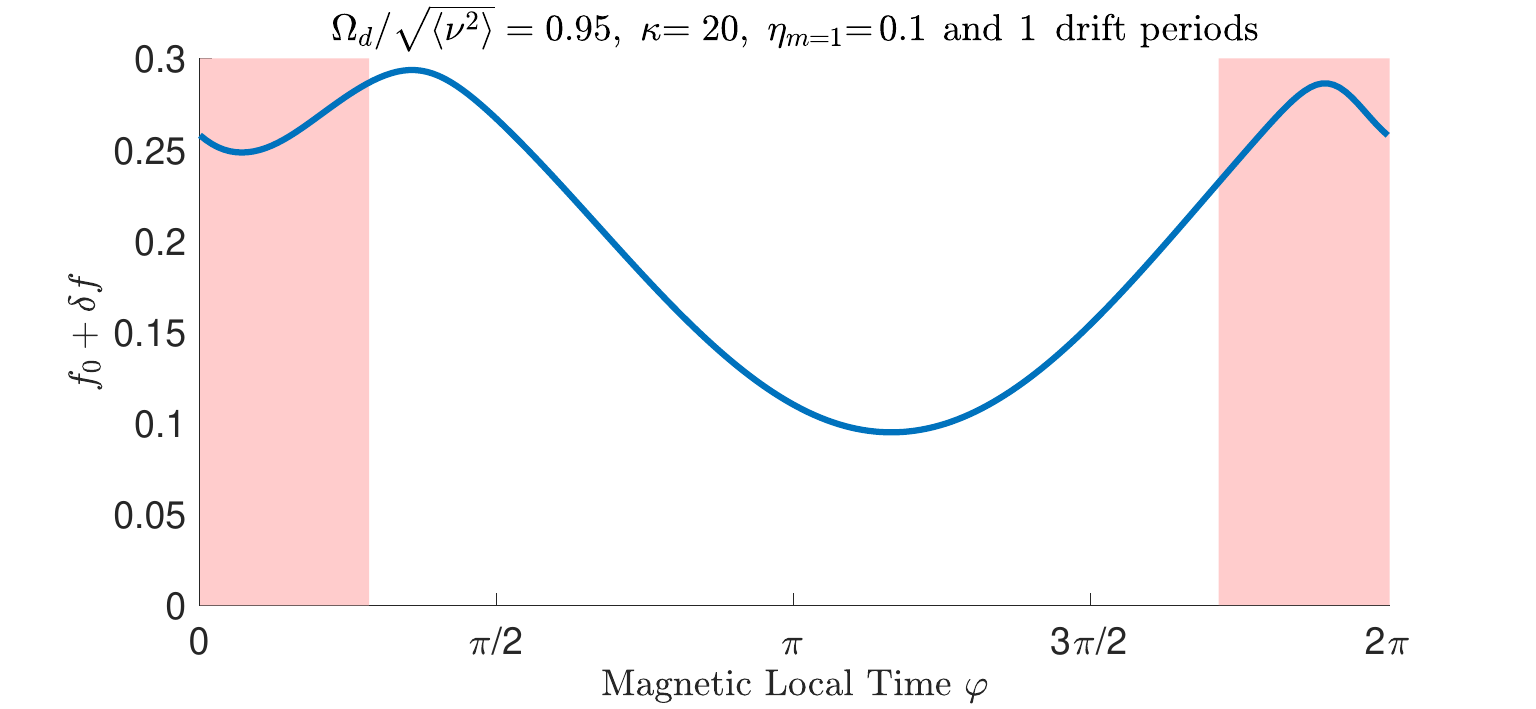}
         \caption{}
      \label{fig:12c}
     \end{subfigure}
       \caption{Same as for Figure (\ref{fig:11}) but for a driven linearly unstable regime with $\Omega_d/\langle \nu^2 \rangle^{1/2}=0.95$ and $\kappa=20$. There is no initial perturbation of the distribution function at $t=0$. But for $t>0$, the $m=1$ driving mode is switched on, resulting in a perturbation of the distribution function. Since the system is linearly unstable, the mode $m=1$ in the distribution function does not damp secularly but rather grows as it exchanges with $m\neq 1$ modes. A spacecraft sweeping the MLT downstream of the loss region would first see an initial depletion followed by an enhancement.}
        \label{fig:12_Microsignature}
\end{figure}

In Figure (\ref{fig:12_Microsignature}), the system is driven and linearly unstable, with $\Omega_d/\langle \nu^2 \rangle^{1/2} = 0.95$, $\kappa = 20$ and $\delta f_m=0$ for all $m$ values. There is no initial perturbation of the distribution function at $t = 0$, but for $t > 0$, the $m = 1$ driving mode is triggered, introducing a perturbation into the distribution function. Due to the linear instability, the $m = 1$ mode does not decay secularly but instead grows through nonlinear interactions with modes where $m \neq 1$.

A spacecraft sweeping through magnetic local time (MLT) downstream of the loss region would observe an initial depletion in the distribution function, followed by a localized enhancement. Such signatures are more likely to be observed in particle populations with slow drift periods. When accounting for the corotation electric field, electrons in gas giants exhibit longer drift periods than protons with comparable first adiabatic invariants, since the gradient-$B$ drift for electrons opposes the direction of corotation. As a result, the synchronization of perturbed Fourier modes in the distribution function—associated with the linearly unstable regime—is more likely to be observable in energetic electrons rather than protons. We, however, note a significant caveat with our results before a more complete comparison with observational measurements of microsignatures can be made. The coupled Equations (\ref{EQ:oscillators2}) are solved for a single drift shell, that is for a fixed $L$ values. Since MLT localized loss region with large absorption rates ($\Omega_d/\sqrt{\langle \nu^2\rangle}<1$) result in a hole of the distribution function, they also lead to a large gradient around the moon radial location. And one must therefore solve (\ref{EQ:oscillators2}) for various drift-shells to determine how radial gradients shape the dynamical evolution of the hole created by the absorption region. Our theoretical framework accounts for both quasi-linear radial diffusion and fast timescales losses, and thus, a solving jointly the time evolution for $f_0$ and all values of $\delta f_m$ on fast timescales is for the first time theoretically possible. We therefore conclude this section by emphasizing that the apparent refilling of the distribution function can potentially be understood as a phase-space synchronization process, since it operates on timescales much faster than quasi-linear radial diffusion.


\subsection{Correspondence with the Kuramoto model}
In the previous section, we have shown that the drift-kinetic distribution function of magnetically trapped particles interacting with MLT-localized absorbing regions can exhibit synchronized behavior and develop phase-space enhancements, despite the presence of irreversible losses. This behavior is particularly striking given the system’s inherently dissipative nature, with moons acting as sinks for particle populations. Mathematically, it is shown that when the source region is very localized ($\kappa >1$) and the absorption rate is comparable to the azimuthal drift period ($\Omega_d/\sqrt{\langle\nu^2 \rangle} \leq 1$), Fourier modes in the distribution function decomposed along magnetic local times are linearly instable. If we think of a collection of magnetically trapped particles as a fluid in phase space, then localized enhancements (associated with injections) or depletions (due to interactions with moons) give rise to magnetic local time (MLT)-localized fluctuations in the distribution function—akin to the beating of a drum resonating along specific harmonics. However, we can move beyond this analogy. In the following, we demonstrate that the emergence of additional Fourier modes with azimuthal wave numbers $m$ in the distribution function can be more rigorously understood through the lens of a textbook model for synchronization in physically coupled systems: the Kuramoto model \cite{Acebron05}.

The Kuramoto model describes a large population of coupled limit-cycle oscillators whose natural frequencies $\omega_i$ are drawn from some prescribed distribution $g(\omega)$. If the coupling strength between the oscillators exceeds a certain threshold, the system exhibits a phase transition: some of the oscillators spontaneously synchronize, while others remain incoherent \cite{STROGATZ20001}. In our particular problem, the oscillators are the Fourier modes $\delta f_m$ of the MLT decomposed distribution function. The natural frequencies for each Fourier modes are the drift frequencies $m\Omega_d$ due to the combination of magnetic gradient drift and corotation drift, and the coupling frequencies are a function of the ratio of the Bessel function $\psi_m < 1$ with magnitude determined by the parameter $\kappa$. 

In order to show the correspondence with the Kuramoto model we consider the main equation of our problem once again:
\begin{equation}
\frac{d a_m}{dt} + (i\omega_m + \sigma) a_m = -2\sigma \sum_{m' \neq m} \psi_{m'-m} a_{m'}
\label{eq:original}
\end{equation}
Our aim is to derive equations for the amplitude and phase of each mode. We first write the complex amplitude in polar form:
\begin{equation}
a_m(t) = R_m(t) e^{i\theta_m(t)}
\end{equation}
where \( R_m(t) \in \mathbb{R}^+ \) is the amplitude and \( \theta_m(t) \in \mathbb{R} \) is the phase.
If we compute the time derivative of \( a_m \), we find:
\begin{align}
\frac{d a_m}{dt} &= \frac{d}{dt} \left(R_m e^{i\theta_m}\right) = \dot{R}_m e^{i\theta_m} + i R_m \dot{\theta}_m e^{i\theta_m} \\
&= e^{i\theta_m} \left( \dot{R}_m + i R_m \dot{\theta}_m \right)
\end{align}
Substituting into equation~\eqref{eq:original} and dividing both sides by \( e^{i\theta_m} \) we find:
\begin{equation}
\label{EQ:B5}
\dot{R}_m + i R_m \dot{\theta}_m + i\omega_m R_m + \sigma R_m
= -2\sigma \sum_{m' \neq m} \psi_{m'-m} R_{m'} e^{i(\theta_{m'} - \theta_m)}
\end{equation}
The next step is to extract an evolution equation for the amplitude $R_m$ and the phase $\theta_m$. In order to do so, we separate the real and imaginary parts. The real part of Equation (\ref{EQ:B5}) gives the Amplitude equation:
\begin{equation}
\dot{R}_m + \sigma R_m = -2\sigma \sum_{m' \neq m} \psi_{m'-m} R_{m'} \cos(\theta_{m'} - \theta_m),
\end{equation}
whereas the imaginary part of Equation (\ref{EQ:B5}), gives the phase equation:
\begin{equation}
R_m \dot{\theta}_m + \omega_m R_m = 2\sigma \sum_{m' \neq m} \psi_{m'-m} R_{m'} \sin(\theta_{m'} - \theta_m),
\end{equation}
We divide both imaginary and real parts of Equation (\ref{EQ:B5}) by \( R_m \) (assuming \( R_m \ne 0 \)) to find:
\begin{equation}
\dot{\theta}_m = -\omega_m + 2\sigma \sum_{m' \neq m} \psi_{m'-m} \frac{R_{m'}}{R_m} \sin(\theta_{m'} - \theta_m)
\end{equation}

The final system of equations is given by: 
\begin{align}
\label{EQ:Kuramoto}
\dot{R}_m &= -\sigma R_m - 2\sigma \sum_{m' \neq m} \psi_{m'-m} R_{m'} \cos(\theta_{m'} - \theta_m) \\
\dot{\theta}_m &= -\omega_m + 2\sigma \sum_{m' \neq m} \psi_{m'-m} \frac{R_{m'}}{R_m} \sin(\theta_{m'} - \theta_m)
\end{align}
The equation governing the phase $\theta_m$ corresponds to the classical Kuramoto equation, but with a time-dependent coupling term given by $K_m = 2\sigma  \psi_{m’-m} R_{m’}/R_m$ \cite{Acebron05} and non-identical oscillating frequencies, i.e., $\omega_m=m\omega_{m=1}$. The Kuramoto model has been widely applied to describe synchronization phenomena in diverse systems, including networks of cardiac pacemaker cells, flashing fireflies, laser arrays, and superconducting Josephson junctions \cite{STROGATZ20001}.

As discussed in Section \ref{section:stability}, modeling an absorption region over a very localized portion of the drift orbit necessitates computing a large number $N$ of Fourier modes $a_m$. For instance, if $N = 30$, we must solve the Kuramoto equation for $N$ phases and $N$ amplitudes. Analyzing this generalized Kuramoto model—with $N$ oscillators—from the perspectives of bifurcation and chaos theory can shed light on the diverse responses of the distribution function as a function of the parameter $\kappa$. While a detailed bifurcation analysis lies beyond the scope of this work, it is worth noting that the Kuramoto model is known to exhibit chaotic behavior and quasiperiodic dynamics under certain conditions \cite{Maistrenko05}.

The set of equations (\ref{EQ:Kuramoto}) thus describes the coupled evolution of both the amplitude $R_m$ and phase $\theta_m$ of each mode $m$, extending the classical Kuramoto framework by incorporating amplitude dynamics. This connection between a drift-kinetic system undergoing localized dissipation and the Kuramoto model offers an alternative explanation for the apparent refilling of the distribution function of magnetically trapped particles colocated with moons on Jupiter and Saturn. In other words, the microsignatures observed in gas giants may arise, not from radial diffusion, which is too slow to operate on timescales comparable to the drift period, but from synchronized phase-space dynamics analogous to those found in the rhythmic beating of heart cells.

\section{Conclusion}\label{Sec:Conclusion}
We have developed a theoretical framework to account for spatially localized losses of magnetically trapped particles in planetary magnetospheres. While the model is general enough to capture phenomena such as magnetopause shadowing at Earth, our focus has been on highly localized and efficient absorption regions—specifically, the influence of moons in the gas giant environments. In situ measurements over the past 50 years have shown that particle fluxes transiting through moons in gas giant systems undergo energy, pitch-angle, and drift-speed-dependent absorption \cite{vanAllen1980Mimas}. Observations of the resulting electron flux depletions in the moons’ wake have consistently revealed an apparent refilling of the affected phase-space regions on timescales comparable to, or even shorter than, a single drift period. Such observed depletions followed by apparent refilling are commonly referred to as \textit{microsignatures} and the physical mechanism originally proposed to explain this refilling—first quantified by \citet{vanAllen1980Mimas}—has been radial diffusion \cite{Parker60, Falthammar65}. 

However, in this communication, we have argued that current quasi-linear models of radial diffusion—originally developed in the early days of the space era \cite{Falthammar65, Parker60} and later extended to incorporate more realistic drivers of radial transport \cite{Elkington99, Fei06, Cunningham16, Lejosne20, Osmane25}—are fundamentally inconsistent with the observed refilling of fluxes on timescales comparable to just a few drift periods. We therefore contend that either (i) a new class of radial diffusion models must be developed that can operate on fast timescales, necessarily involving higher-order effects and large-amplitude fluctuation drivers \cite{osmane2023radial, Art24} \footnote{\citet{Art24} demonstrate that intense monochromatic ULF waves can accelerate radial electron transport and mimic the effects of electron injections into the lower L-shells. In the context of a transport equation, this process would be represented by a drift term. This theoretical argument gains support from a recent study by \citet{Camporeale22}, which shows that a drift-diffusion equation provides a more accurate description of storm-time electron flux dynamics. While such a mechanism—modeled via a drift transport coefficient—could, in principle, be generalized to the magnetospheres of gas giants, the origin of similarly large-amplitude fields violating the third adiabatic invariant and coinciding with moon drift shells remains, to the best of our knowledge, unknown.}, or (ii) the explanation for microsignatures lies in non-diffusive radial transport processes that have, until now, been largely unaccounted for. We have chosen to pursue the second option—focusing on non-diffusive radial transport processes as a possible explanation for microsignatures—while reserving the development of fast-timescale radial diffusion models for future studies.

By allowing for a violation of phase-space density conservation in the drift-kinetic distribution function, we have shown that localized and efficient loss mechanisms, mimicking moon–magnetosphere interactions acting on guiding center orbits, can offer a new approach to tackling the problem of microsignatures. When the distribution function is decomposed into azimuthal Fourier modes, we find that linearly unstable behavior arises in the presence of loss regions that are both strongly localized in magnetic local time and efficient at removing particles on timescales comparable to their azimuthal drift period. In this regime, the observed refilling of the distribution function cannot be attributed to radial diffusion, which fundamentally operates on timescales longer than a drift period. Instead, we recover apparent refilling as a result of synchronization dynamics among the Fourier modes of the distribution function—a non-diffusive mechanism that can explain the persistence and evolution of microsignature structures. The possible interpretation of microsignatures as a manifestation of synchronization is further supported by our demonstration that the drift-kinetic equation with MLT-localized sources is mathematically analogous to a generalized Kuramoto model—the canonical framework for describing synchronization phenomena across a wide range of physical systems \cite{Acebron05, STROGATZ20001}.

Our approach offers new tools and perspectives for understanding spatially localized sink regions in planetary magnetospheres. In particular, the model can be applied to quantify the impact of magnetopause shadowing over timescales comparable to a few azimuthal drift orbits in Earth’s radiation belts \cite{Turner12, George22b}. Nonetheless, several limitations of the current model must be addressed in future developments. First, localized absorbing moons will result in large radial gradients of the distribution function. In this study, we focused on demonstrating the impact on a single drift shells. However, a complete and more consistent comparison with observational measurements, requires solving the full system of equations for every MLT and for radial distances above and below absorbing region's radial distance. Additionally, we do not treat Maxwell’s equations self-consistently with the drift-kinetic equations, and electromagnetic fluctuations are neglected. The former implies that energetic particles are treated as passive tracers, while the latter assumes that the dominant perturbations are electrostatic. These assumptions are appropriate for relativistic particles in the radiation belts of Jupiter or Saturn, but break down for lower-energy populations such as ring current particles in Earth’s magnetosphere. The impact of kinetic scale electromagnetic perturbations can be added as diffusion along pitch-angle and energy to the drift kinetic equation, as is commonly done for global magnetospheric models \cite{Allanson}, but we nonetheless expect these processes to be too slow to explain the microsignature structures on timescales comparable to less than a drift period. Second, our analysis is restricted to equatorially trapped particles. As a result, the instability criteria derived for the Fourier modes apply only to particles with $90^\circ$ pitch angles, i.e., those satisfying $\dot{v}_\parallel \simeq 0$. To extend the model to particles interacting with moons located at higher magnetic latitudes, the full distribution function must retain its dependence on $v_\parallel$, and curvature drift effects must be included. Accounting for the bounce motion introduces a new timescale for the problem, i.e. the bounce motion, which comes with it's own set of resonant interactions with electromagnetic fluctuations \cite{Brizard22} and which can result in faster radial diffusion \cite{Ozturk07}. Third, the current model lacks a detailed representation of the absorption frequency that accounts for pitch-angle and energy dependence, as well as charge exchange processes. A more accurate treatment of these factors is essential for improving predictive capabilities in realistic planetary magnetospheres.

In summary, our results show that localized, efficient sinks in magnetized plasmas can synchronize the phase-space structure of energetic particles, producing apparent refilling on timescales far shorter than quasi-linear diffusion predicts. By mapping the drift-kinetic dynamics to a generalized Kuramoto model, we establish a direct link between plasma transport and universal synchronization phenomena. This reframes the long-standing puzzle of microsignatures not as a diffusive process but as an emergent, non-diffusive phase-space synchronization. Beyond gas giant magnetospheres, our framework suggests that synchronization-driven transport could play a role in other weakly collisional plasma environments, from planetary radiation belts to laboratory fusion devices. Future work should test these predictions against in-situ measurements from missions such as Cassini and Juno, extend the model to multi-shell geometries and pitch-angle dependence, and incorporate self-consistent coupling to electromagnetic fluctuations. More broadly, our findings invite a re-examination of the foundations of transport theory in plasmas with spatially localized sinks, highlighting synchronization as a fundamental yet underexplored mechanism in non-equilibrium physics.

\begin{acknowledgments}
\addtocontents{toc}{}  
\noindent AO gratefully acknowledges many valuable discussions with Yixin Hao on the subject of planetary radiation belts. ER is supported by DLR with the JUICE/PEP contract 50 QJ 2303. PK has been supported through NASA Cassini Data Analysis (CDAP) grants 80NSSC20K0478 and 80NSSC21K0534.
\end{acknowledgments}

.

\appendix
\section{Revisiting the radial diffusion explanation for microsignatures}
In this appendix, we present the explanation of microsignatures in terms of radial diffusion, as described and first quantified by \citet{vanAllen1980Mimas}. The analytical solution derived by  \citet{vanAllen1980Mimas} remains to this day the one used to determine radial diffusion coefficients in outer planetary magnetospheres (see, e.g., \citet{Roussos07}). We therefore find it useful to revisit the explanation of microsignatures in terms of radial diffusion to assess its limitations and, as shown below, to generalize it to cases where the radius of the absorption region, hereafter denoted by the parameter  $b$, cannot be considered much smaller than the the planetary radius, or when the radial diffusion dependence on the normalized radial distance can not be treated as a constant on spatial scales comparable to the absorption region \footnote{Even though absorbing moons in gas giants have very small sizes compared to the planetary radius, rings, which can also be the site of absorption, have characteristic sizes comparable to the respective planetary radius \cite{Rousso18rings}.}.  

The role of radial diffusion for the refilling holes in electron fluxes sustained by moons can be described as follows. Magnetically trapped electrons and protons transit through different regions of the magnetosphere according to their guiding center trajectories. As the particles flux transit through moons they can experience absorption at different rates, depending on their energy, pitch-angle and net azimuthal drift speed which is a function of the co-rotation electric field, the sign of their charge and the first adiabatic invariant. The depleted fluxes now possess a radial gradient colocated with the moon and as they continue their azimuthal drift motion, spacecraft measurements, demonstrate that electron fluxes experience what seems to be a refilling on timescales comparable or smaller than a single drift period.  

If we assume that radial diffusion primarily shapes the absorption signature profile, then this profile should be described by the radial diffusion equation \cite{Falthammar65, Lejosne19, Osmane21a, osmane2023radial}: 

\begin{equation}
\label{Eq:radial_diffusion}
    \frac{\partial f}{\partial t}=L^2\frac{\partial}{\partial L}\left(\frac{D_{LL}}{L^2} \frac{\partial f}{\partial L}\right),
\end{equation}

\noindent where $D_{LL}$ is the radial diffusion equation, and $L$ is proportional to the inverse of the third adiabatic invariant, i.e. the magnetic flux. In the case where the fluctuations violating the third invariant are purely electrostatic, the variable $L$ is identitical the normalised radial distance $r/R_p$, where $R_p$ is the planetary radius. 

It should also be emphasized that, within the framework of quasi-linear radial diffusion, that is, the framework that led to the development of Equation (\ref{Eq:radial_diffusion}) from the late 1950s and early 1960s \cite{Parker60, Falthammar65} to contemporary radial diffusion models \cite{Lejosne19}, that the distribution function $f$ represents only the drift-averaged component of the particle distribution. Components of the distribution function that vary on timescales comparable to the drift period are not captured by a diffusion equation \cite{osmane2023radial}. Thus, as argued in the introduction, using Equation~(\ref{Eq:radial_diffusion}) to quantify the refilling of absorption signatures over timescales of a few drift periods, or even shorter, introduces inconsistencies with the very framework that permits radial diffusion to occur. But for now we ignore such theoretical considerations and go ahead and describe the steps required to recover the equation given by \citet{vanAllen1980Mimas}. 

\subsection{\citet{vanAllen1980Mimas} solution for small absorption regions with $b/R_p\ll 1$}
If the absorption region is very small in size compared to planetary radius, and the losses downstream of the moon are very deep, then the radial diffusion Equation (\ref{Eq:radial_diffusion}) can be simplified to: 
\begin{equation}
\label{Eq:radial_diffusion2}
    \frac{\partial f}{\partial t}\simeq D_{LL} \frac{\partial^2 f}{\partial L^2},
\end{equation}
The reason for this simplification is that the gradient term $L^2\frac{\partial f}{\partial L} \frac{\partial (D_{LL}/L^2)}{\partial L}$ can be ignored because hole profiles in $f$ such as the one asummed from moon absorption lead to much larger second derivative than first derivatives. Moreover, if the moon is small in size, $D_{LL}$ is effectively constant within the moon shadow. 

Solution for the diffusion Equation (\ref{Eq:radial_diffusion2}) should be calculated in the semi-infinite domain $L\in[0, \infty)$, but we proceed according to \citet{vanAllen1980Mimas} and compute it in the infinite domain $L\in(-\infty, \infty)$ and assume $D_{LL}$ to be constant. The self-similar solution to Equation (\ref{Eq:radial_diffusion2}) is given by: 

\begin{equation}
    f(L, t)= \frac{1}{\sqrt{4\pi D_{LL}t}}\exp{\left(-\frac{L^2}{4D_{LL}t}\right)}
\end{equation}
and if one takes the initial profile for a hole centred at $L_0$ and with diameter $2b/R_p$ in the electron distribution function given by 
\begin{equation}
\label{Eq:radial_diffusion_initial_condition}
    f(L, t=0) = 
    \begin{cases}
        0, & \text{if } L_0 - \frac{b}{R_p} < L < L_0 + \frac{b}{R_p} \\
        1, & \text{otherwise}
    \end{cases}
\end{equation}
the solution is given by convolving the self-similar solution  $G(L, t; L', 0)$ solution to the diffusion equation, with the initial condition: 
\begin{eqnarray}
      f(L,t)&=& \int_{-\infty}^{\infty} d L' G(L, t; L', 0)f(L', t=0) \nonumber \\
      &=&\int_{-\infty}^{\infty} d L' \frac{1}{\sqrt{4\pi D_{LL}t}}\exp{\left(-\frac{(L-L')^2}{4D_{LL}t}\right)} f(L', t=0) \nonumber \\
      &=&\left(\int_{-\infty}^{L_0-\frac{b}{R_p}}d L'+\int^{\infty}_{L_0+\frac{b}{R_p}}d L'\right)\frac{1}{\sqrt{4\pi D_{LL}t}}\exp{\left(-\frac{(L-L')^2}{4D_{LL}t}\right)}  \nonumber \\
      &=& 1 - \frac{1}{2} \left[ \text{erf} \left( \frac{L - L_0 + \frac{b}{R_p}}{2\sqrt{D_{LL} t}} \right) - \text{erf} \left( \frac{L - L_0 - \frac{b}{R_p}}{2\sqrt{D_{LL} t}} \right) \right]
\end{eqnarray}

In the above derivation, the radial distance is normalized by the planetary radius $R_p$, whereas the diffusion coefficients in \citet{vanAllen1980Mimas} has units of km$^2$ per seconds. The $D_{LL}$ used in the above derivation has units of $s^{-1}$. Thus, by setting $L_0=0$ and writing $\tau=4 D_{LL} t R_p^2/b^2$, the following solution is identical to the one appearing in Figure 5 of \citet{vanAllen1980Mimas} and to Equation (3) of \citet{Roussos07}: 
\begin{eqnarray}
\label{EQ:Van_Allen_Solution}
     f(L,t)=1 - \frac{1}{2}   \left[ \text{erf} \left( \frac{1 + \frac{L R_p}{b}}{\sqrt{\tau}} \right) +\text{erf} \left( \frac{1 - \frac{L R_p}{b}}{\sqrt{\tau}} \right) \right].
\end{eqnarray}
In the next section, we relax the assumption of small loss regions, find a more general solution for the refilling of the phase-space density and compare the solution to that of \citet{vanAllen1980Mimas}.

\subsection{General solution for arbitrarily sized absorption regions}
In this section, we solve the radial diffusion for the case where the dominant fluctuations are electrostatic but where the absorption region can be comparable to the planetary radius, i.e. $b \simeq R_p$. While the moons are small in size, and satisfy the limit $b \ll R_p$, rings can be widespread in the equatorial plane, and if one wanted to account for the refilling absorption regions downstream of rings, one would need to account for values of $b \geq R_p$ \cite{Rousso18rings}.

In such instance, the radial diffusion coefficients, expressed in terms of the electrostatic fluctuations $\delta E_{\varphi,m}$ is given by \cite{Osmane25}: 
\begin{equation}
    \label{Eq:DLL_correct}
    D_{LL}=L^4 \frac{\langle\delta E_{\varphi, m}^2\rangle}{B_p^2 R_p^2}
\end{equation}
and thus, the radial diffusion is spatially dependent and with units of one over seconds. In the following we solve the exact radial diffusion equation, and then take a limit where the absorption region is very small compared to the planetary radius, i.e. $b\ll R_p$.

Since $D_{LL}$ scales as $L^4$ for the case of a pure electrostatic fluctuations, expressing the radial diffusion equation in terms of the inverse of the radial distance, i.e. $x=1/L$, allows us to solve the classical diffusion equation instead: 
\begin{equation}
\label{Eq:radial_diffusion3}
    \frac{\partial f}{\partial t}=L^2\frac{\partial}{\partial L}\left(\frac{D_{LL}}{L^2} \frac{\partial f}{\partial L}\right) \Longrightarrow  \frac{\partial f(x)}{\partial t}=\frac{\langle\delta E_{\varphi, m}^2\rangle}{B_p^2 R_p^2} \frac{\partial^2 f(x)}{\partial x^2}
\end{equation}

We now proceed to solve the radial diffusion equation in terms of $x$, instead of $L$. We also normalise time according to $\tau=\frac{\langle\delta E_{\varphi, m}^2\rangle}{B_p^2 R_p^2} t$. We can enforce an absorbing inner boundary with an image solution, such that for $L=0$, or equivalently for $x=\infty$, the distribution function cancels, but we skip this step since (1) the absorption are so small that the effects of the boundaries can be ignored, and (2) we want to compare the general solution to that of \citet{vanAllen1980Mimas} which was computed for an infinite domain. Using Green's function solutions or equivalently, the self-similar solution, for the radial diffusion equation, as shown previously in \citet{Osmane21a}, the propagator for the inverse of the radial distance is given by: 
\begin{equation}
    G(x, \tau; x') = \frac{1}{\sqrt{4\pi \tau}}
    \exp\left(-\frac{(x - x')^2}{4\tau} \right).
\end{equation} 
The initial condition expressed in terms of $x$ is now:
\begin{equation}
    {f}(x, \tau=0) =
    \begin{cases}
        0, & x_1 < x < x_2, \\
        1, & \text{otherwise}.
    \end{cases}
\end{equation}

where \( x_1 \) and \( x_2 \) are given by:

\begin{equation}
    x_1 = \frac{1}{L_0 + \frac{b}{R_p}}, \quad x_2 = \frac{1}{L_0 - \frac{b}{R_p}}.
\end{equation}

The solution for $\tau>0$ can therefore be obtained via convolution:

\begin{equation}
    f(x, \tau) = \int_{-\infty}^\infty G(x, \tau; x') {f}_0(x') \, dx'.
\end{equation}
Using standard integrals for Gaussian function, the final solution in \( x \)-space is:
\begin{equation}
    f(x, \tau) = 1 - \frac{1}{2} \left[
    \text{erf} \left( \frac{x_2 - x}{2\sqrt{\tau}} \right)
    - \text{erf} \left( \frac{x_1 - x}{2\sqrt{\tau}} \right)
    \right].
\end{equation}
Since $x = \frac{1}{L}$, we obtain the final solution in terms of \( L \): 

\begin{eqnarray}
\label{EQ:correct_solution}
     f(L, \tau) &=& 1 - \frac{1}{2} \Bigg{[}
    \text{erf} \left( \frac{\frac{1}{L_0 - \frac{b}{R_p}} - \frac{1}{L}}{2\sqrt{\tau}} \right)
    - \text{erf} \left( \frac{\frac{1}{L_0 + \frac{b}{R_p}} - \frac{1}{L}}{2\sqrt{\tau}} \right) \Bigg{]}.  
\end{eqnarray}
In order to compare the above solution to the one derived by \citet{vanAllen1980Mimas}, we do the following transformation for the normalised time variable: 
\begin{equation}
    \tau\longrightarrow 4L^4 \frac{\langle\delta E_{\varphi, m}^2\rangle}{B_p^2 R_p^2}t \frac{R_p^2}{b^2}= 4 D_{LL} t  \frac{R_p^2}{b^2}
\end{equation}
which allows us to write Equation (\ref{EQ:correct_solution}) as
\begin{eqnarray}
\label{EQ:correct_solution2}
     f(L, \tau) &=& 1 - \frac{1}{2} \Bigg{[}
    \text{erf} \left( \frac{\frac{L^2 R_p/b}{L_0 - \frac{b}{R_p}} - L\frac{R_p}{b}}{\sqrt{\tau}} \right)
    - \text{erf} \left( \frac{\frac{L^2 R_p/b}{L_0 + \frac{b}{R_p}} - L\frac{R_p}{b}}{\sqrt{\tau}} \right) \Bigg{]}.  \nonumber \\ 
    &=&1 - \frac{1}{2} \Bigg{[}
    \text{erf} \left(\frac{{L^2 R_p/b-LL_0R_p/b+L}}{(L_0 - \frac{b}{R_p})\sqrt{\tau}}\right)
    - \text{erf} \left(\frac{{L^2 R_p/b-LL_0R_p/b-L}}{(L_0 + \frac{b}{R_p})\sqrt{\tau}}\right) \Bigg{]}. \nonumber \\
   &=&1 - \frac{1}{2} \Bigg{[}
    \text{erf} \left(\frac{1+\frac{(L-L_0)R_p}{b}}{(L_0/L - \frac{b}{LR_p})\sqrt{\tau}}\right)
    + \text{erf} \left(\frac{1-{\frac{(L-L_0)R_p}{b}}}{(L_0/L + \frac{b}{LR_p})\sqrt{\tau}}\right) \Bigg{]}.
\end{eqnarray}
We now want to answer how the above solution compares to the Equation (\ref{EQ:Van_Allen_Solution}) with $L_0\neq 0$, and which we rename below as: 
\begin{eqnarray}
\label{EQ:Van_Allen_SolutionL0}
     f_{\text{Van Allen}}(L,t)=1 - \frac{1}{2}   \left[ \text{erf} \left( \frac{1 + \frac{(L-L_0) R_p}{b}}{\sqrt{\tau}} \right) +\text{erf} \left( \frac{1 - \frac{(L-L_0) R_p}{b}}{\sqrt{\tau}} \right) \right].
\end{eqnarray}
A comparison of Equation (\ref{EQ:correct_solution2}) with (\ref{EQ:Van_Allen_SolutionL0}), both written for $\tau=4D_{LL}R_p^2 t/b^2$, only differ in the denominator of the arguments of the error functions. The terms in the denomitors $(L_0/L\pm b/LR_p)$ in the general solution for the radial diffusion equation (\ref{EQ:correct_solution2}) in the case of electrostatic fluctuations is negligeable when $b/R_p/L \ll 1$, and thus the solution of \citet{vanAllen1980Mimas} are appropriate in the limit of small absorption regions. The comparison between the general solution Equation (\ref{EQ:correct_solution2}) with (\ref{EQ:Van_Allen_SolutionL0}) is also shown in Figure (\ref{fig:12}), which demonstrates that when $b/R_p\ll 1$ (panel (a)), there is little difference in the refilling rate. Whereas when $b/R_p \simeq 1$, the refilling is asymmetric and associated with a drift of the hole towards higher $L$ regions, due to the nonlinear dependence of the radial diffusion coefficients. Since $D_{LL}\simeq L^4$, the absorption region at higher $L$ values fill up faster than at lower $L$ values, which is noticeable when the absorption region is not small.


\begin{figure}[H]
     \centering
         \includegraphics[width=0.45\textwidth]{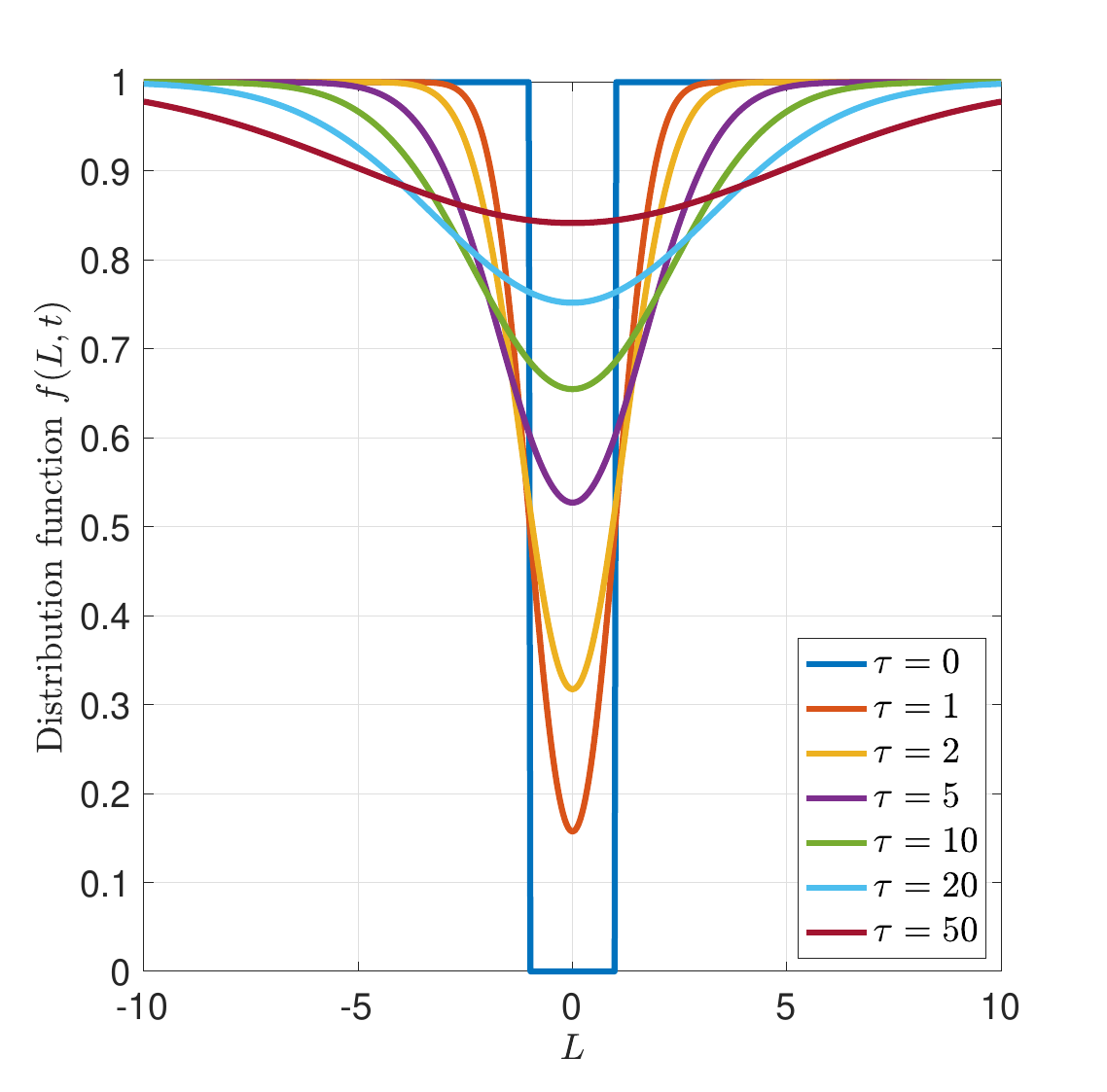}
           \caption{Refilling of the absorption holes for electrons as per Equation (\ref{EQ:Van_Allen_Solution}). The figure is identical in its content to Figure 5 in \citet{vanAllen1980Mimas}.}
        \label{fig:Van_Allen_solution}
\end{figure}

\begin{figure}[H]
     \centering
     \begin{subfigure}[b]{0.49\textwidth}
         \centering
         \includegraphics[width=\textwidth]{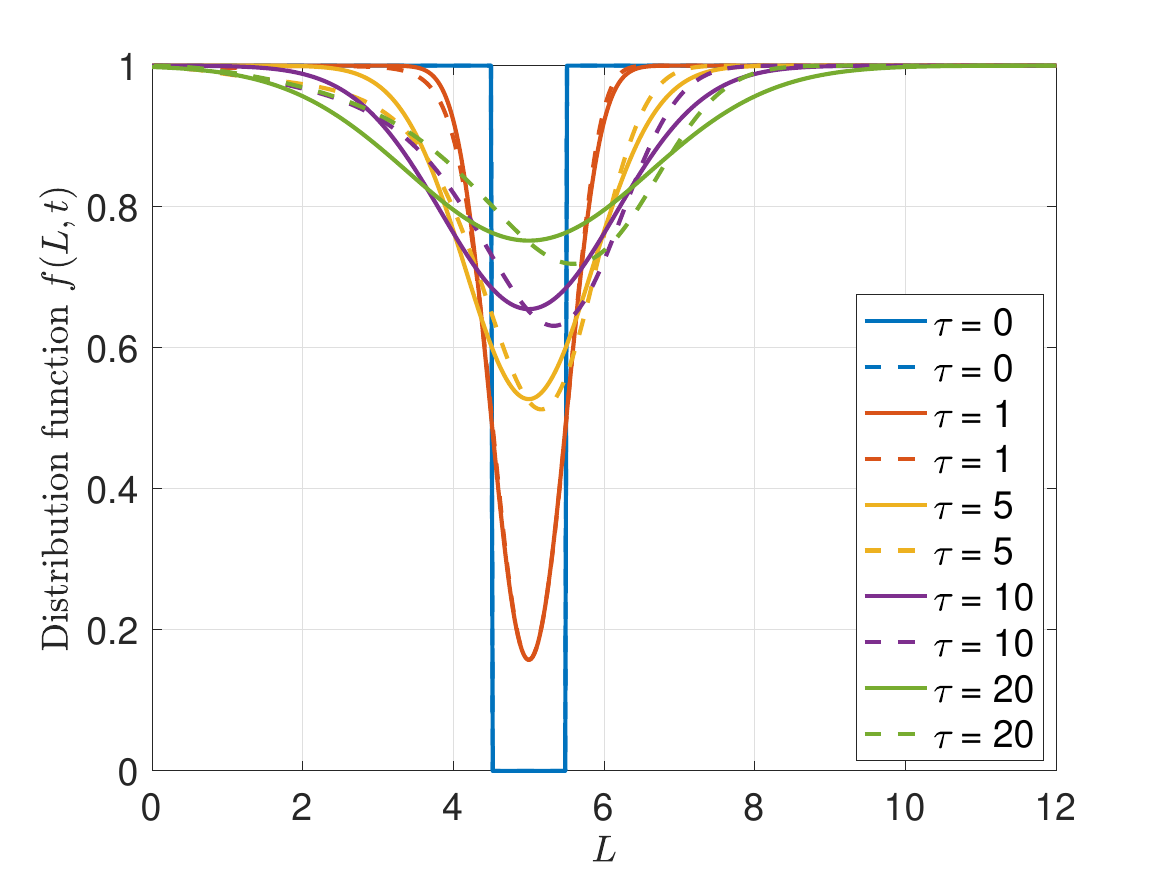}
         \caption{}
         \label{fig:appendix_a}
     \end{subfigure}
     \hfill
     \begin{subfigure}[b]{0.49\textwidth}
         \centering
         \includegraphics[width=\textwidth]{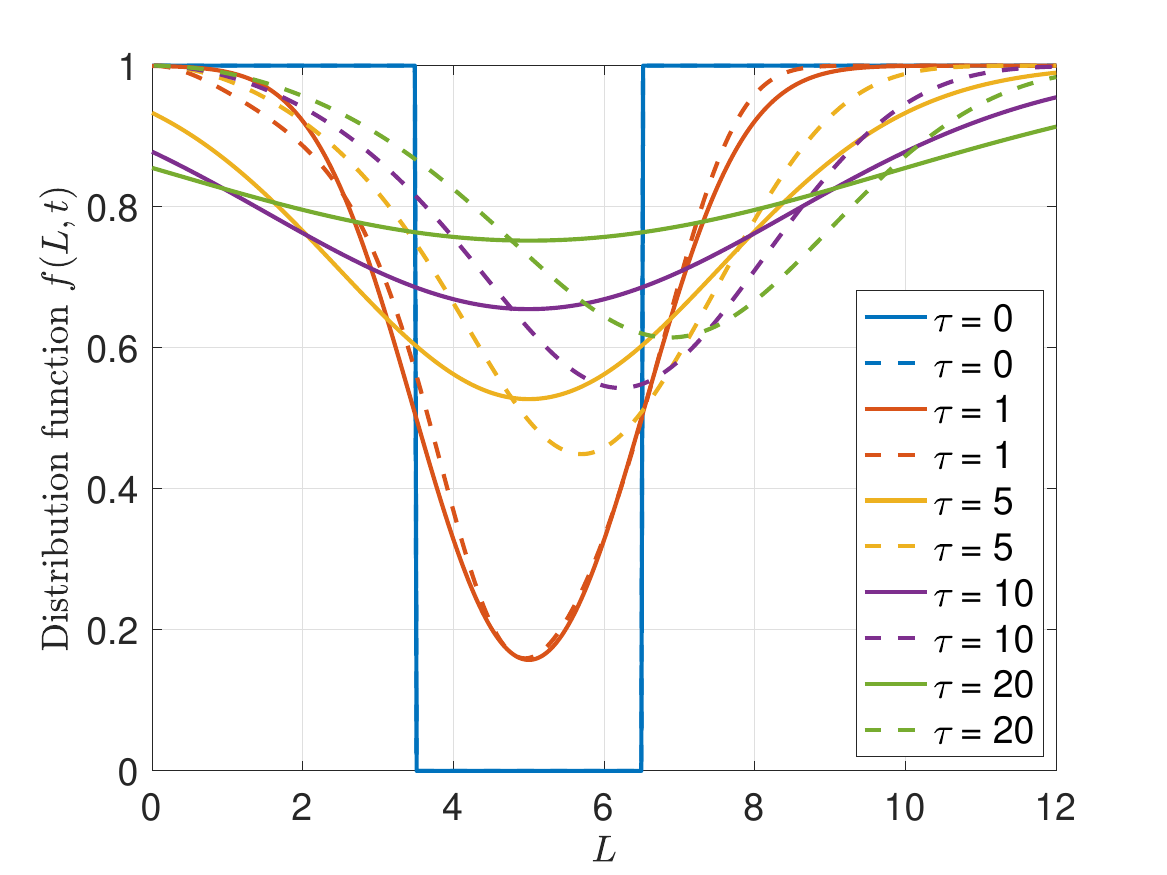}
         \caption{}
         \label{fig:appendix_b}
     \end{subfigure}
  \caption{Refilling of the distribution function for moons located at $L_0=5$ and with normalised radiuses (a) $b/R_p=0.5$ and (b) $b/R_p=1.5$. In each subplot, the solid lines represent the refilling according to the \citet{vanAllen1980Mimas} solution given by Equation (\ref{EQ:Van_Allen_SolutionL0}). The dashed lines are the general solution (\ref{EQ:correct_solution2}) for the radial diffusion equation for an arbitrary value of $b/R_p$.  When the moon is small, radial diffusion is symmetric, and the \citet{vanAllen1980Mimas} is quantitatively consistent with the general solution. However, the refilling becomes asymmetric for $b\geq 1 R_p$ and is accompanied by a drift of the holes to higher $L$ regions.}
        \label{fig:12}
\end{figure}


\bibliography{apssamp}

\end{document}